\documentclass[9pt,twocolumn,twoside]{pnas-new}

\templatetype{pnasresearcharticle} %
\usepackage{caption}
\usepackage{subcaption}
\usepackage{multirow}
\usepackage{multicol}
\usepackage{color}
\usepackage{csquotes}
\usepackage{tcolorbox}
\newcommand{\ps}[1]{\textcolor{orange}{[Punyajoy: #1]}}
\newcommand{\kg}[1]{\textcolor{green}{[Kiran: #1]}}

\newcommand{\updated}[1]{\textcolor{blue}{[Updated: #1]}}

\newcommand{\appendixnewname}{SI text}
\title{On the rise of fear speech in online social media}

\author[a]{Punyajoy Saha}
\author[b]{Kiran Garimella} 
\author[a]{Narla Komal Kalyan}
\author[a]{Saurabh Kumar Pandey}
\author[a]{Pauras Mangesh Meher}
\author[a]{Binny Mathew}
\author[a]{Animesh Mukherjee}

\affil[a]{Department of Computer Science and Engineering, Indian Institute of Technology, Kharagpur, India -- 721302}
\affil[b]{School of Communication and Information, Rutgers University, NJ 08901}

\leadauthor{Saha} 

\significancestatement{
Existential fear has always been a concern across human history, and even transcends to the rest of the animal world. This fear is so deeply ingrained that even the slightest `knock' to it could spark a violent conflict among different groups. Here we demonstrate how social media platforms are used to extensively mediate elements of existential fear as \textit{fear speech} posts. Their non-toxic and argumentative nature make them appealing to even benign users who in turn contribute to their wide prevalence by resharing, liking and replying to them. Remarkably, this prevalence is far stronger than the more well-known hate speech posts. Our work necessitates consolidated moderation efforts and awareness campaigns to mitigate the harmful effects of fear speech.

}

\authorcontributions{P.S., K.G., B.M. and A.M. designed research;
P.S., S.K.P., N.K.K. and P.M.M performed research and contributed analytical tools; 
P.S., K.G., B.M. and A.M. wrote the paper.}
\authordeclaration{The authors declare no conflict of interest.}
\correspondingauthor{\textsuperscript{2}To whom correspondence should be addressed. E-mail: punyajoys@iitkgp.ac.in}


\begin{abstract}
Recently, social media platforms are heavily moderated to prevent the spread of online \textit{hate speech}~\cite{Hatespee37:online} which is usually fertile in toxic words and is directed toward an individual or a community. Owing to such heavy moderation newer and more subtle techniques are being deployed. One of the most striking among these is \textit{fear speech}~\cite{buyse2014words}. Fear speech, as the name suggests, attempts to incite fear about a target community ~\cite{TheRadic56:online}. Although subtle, it might be highly effective often pushing communities toward a physical conflict~\cite{buyse2014words}. Therefore, understanding their prevalence in social media is of paramount importance. This article presents a large-scale study to understand the prevalence of 400K fear speech over 700K hate speech posts collected from \texttt{Gab.com}. Remarkably, users posting large number of fear speech accrue more followers and occupy more central positions in the social network compared to users posting large number of hate speech. They are also able to reach out to the benign users far more effectively than hate speech users through the replies, re-posts and mentions. This has connections to the fact that unlike hate speech, fear speech has almost zero toxic content which makes it look plausible. Moreover, while topics of fear speech posts mostly portray a community as a perpetrator using a (fake) chain of argumentation, hate speech topics hurl direct multi-target insults thus pointing to why general users could be more gullible to fear speech. Our findings transcend even to other platforms (Twitter and Facebook) and thus necessitates using sophisticated moderation policies and mass awareness to combat fear speech. %

\end{abstract}

\dates{This manuscript was compiled on \today}
\doi{\url{www.pnas.org/cgi/doi/10.1073/pnas.XXXXXXXXXX}}

\begin{document}

\maketitle
\thispagestyle{firststyle}
\ifthenelse{\boolean{shortarticle}}{\ifthenelse{\boolean{singlecolumn}}{\abscontentformatted}{\abscontent}}{}

\dropcap{C}ontent moderation plays an important role in removing harmful and irrelevant posts (spam), thereby keeping the platforms safe. Social media companies like Facebook\footnote{https://transparency.fb.com/en-gb/policies/community-standards/hate-speech/} and Twitter\footnote{ttps://help.twitter.com/en/rules-and-policies/hateful-conduct-policy} have detailed guidelines as what is considered hateful in their platforms. These companies use such guidelines 
to appoint manual and automatic moderators to delete hateful posts/suspend the hateful users\footnote{https://about.fb.com/news/2021/02/update-on-our-progress-on-ai-and-hate-speech-detection/}. Subsequently, the research community has started putting consolidated efforts to automate and, thereby, scale up this moderation creating better datasets and machine learning models to accurately detect hate speech. The datasets span across different platforms including Twitter~\cite{basile2019semeval,waseem2016hateful}, Gab~\cite{kennedy2018gab}, Reddit~\cite{hada2021ruddit} etc. Further, the models also range from simple ones like mSVM~\cite{macavaney2019hate} to complex AI architectures like transformers~\cite{aluru2020deep}.

While these advances are indeed encouraging, newer and more subtle forms of harmful content are inflicting the online world which most often go unnoticed. One such form of malicious content is \textit{fear speech} which involves spreading \textit{fear} about one or more target communities in the online, and eventually, the physical world. 
In this context, we note that existential fear can bias peaceful people toward extremism. In a controlled experiment~\cite{cohen06}, a group of Iranian students were found to support doctrines related to the understanding of the value of human life as opposed to a jihadist call for suicide bombing. However, when they were frightened about death, they subscribed toward the bomber, even expressing desire to become a martyr themselves. From time to time, mortality salience polarizes an individual or a group to stick firmly to their own beliefs while demonizing others with opposing beliefs. This arises from the fear of endangerment of their own clan. For instance, while the fear that was generated due to the 9/11 incident was real, it also made Americans more vulnerable to psychological manipulation. In this context, Florette Cohen notes that “fear tactics have been used by politicians for years to sway votes”.\footnote{https://www.psychologicalscience.org/news/releases/the-political-effects-of-existential-fear.html} In a survey~\cite{cohen04} conducted by Cohen and her colleagues, the authors asked the participants to think about the fear of death and then gave them statements from three fictitious political personalities. One of them was a charismatic who stressed on ingroup favoritism, the second, a technocrat presenting practical solutions to realistic problems and the third preaching democratic values. When primed with the fear of death, the support for the fictional charismatic leader went up by eightfolds. With the advent of social media it has become easier to propel the prevalence of such fear tactics.

In real life, elements of fear are often found associated with events of violence. The posts of the alleged attacker who shot worshippers at the Pittsburgh synagogue in October 2018, portrayed the HIAS\footnote{\url{https://www.hias.org/}, a non profit organisation that provides aids to the refugees.} as an organisation supporting refugee invasion~\cite{TheRadic56:online}. Similarly, the shooter of the Christchurch event in 2019 released a manifesto - `Great Replacement'. This manifesto contained elements of fear in the form of `non-whites' replacing `whites' in the future~\cite{Christch52:online}. A recent mass shooting in Buffalo  shooting~\cite{Buffalom54:online} also denotes another such racially motivated attack.  Such association is also well grounded in the literature of inter-group conflicts~\cite{buyse2014words}. 

Fear is also used by politicians and media figures. Politicians in the United States~\cite{Trumpand26:online} and European nations~\cite{Wilderst27:online} portray immigration as an invasion and asylum seekers as dangerous. A viral poster in The Brexit campaign - \textit{`Breaking point'}, shows non-whites as invaders and as a danger to the British resources~\cite{reid2019buses}. Media figures like Tucker Carlson often cite low birth rates among Americans as a threat to cultural identity. Previous work on `fear speech'~\cite{saha2021short} also find similar themes in public political WhatsApp groups in India during the general elections of 2019. 

One of the representative messages from our study reproduced below shows the intricate structure of fear speech.

\begin{tcolorbox}[colback=gray!20!white,colframe=black]
  Hundreds of South Americans are marching through Mexico, aiming to cross the US Border and demand asylum in the US. No one in Mexico is stopping them. This is a national security threat and should be dealt with by force if necessary. What else is our military good for if they can’t stop an invading force?	

\end{tcolorbox}

Note that this message has no toxic words and is weaved into a series of arguments citing evidences, establishing a case of nationwide fear and finally inciting users to take an action. Such views often resonate with the opinions of the `common' audience and they in turn contribute to spread the message deeper and farther into the network. 

The central objective of this article is to investigate the prevalence of fear speech \hyperref[sec:anno-guide]{See this section}) in a loosely moderated social media platform like \texttt{Gab.com}. Since no known dataset is available for such a study, we devise an algorithmic pipeline to first build a dataset of 400,000 fear speech posts to be contrasted with another 700,000 hate speech posts. Based on the analysis of this dataset, the central result that we arrive at demonstrates how users posting a large number of fear speech are successful in garnering significantly more followers compared to the users posting large number of hate speech  (\hyperref[sec:followers]{See this section})
The former are also more effective in reaching out to the general users through re-posts, replies and mentions (\hyperref[sec:normal-users]{See this section}). We elucidate that this is because of the non-toxic and argumentative nature of the posts that make them look more plausible and thus widely accepted. Some such prevailing arguments in fear speech correspond to violence by the Muslim community (10\% of all fear speech posts), Jews controlling media and culture (10\% of all fear speech posts), white genocide in South Africa (7\% of all fear speech posts), etc. In contrast to this, the traditional hate speech (posts mostly correspond to hurling insulting remarks or calling for deportation of the target community (\hyperref[sec:topics]{See this section} for popular topics and \hyperref[sec:anno-guide]{this section} for definitions). The seemingly benign nature makes fear speech more credulous to the users than hate speech facilitating its increased prevalence in the network.   

We stress that such forms of highly destructive speech should not go unnoticed and call for more sophisticated moderation mechanisms along with mass awareness. We believe that this article can lay the foundation stone for such an initiative.

\section*{Dataset}

There is no data available in the literature that allows for the study of prevalence of fear speech (See Materials and Methods section) in social media. Therefore we had to setup an end-to-end pipeline to build our dataset. We make use of the Gab platform for data collection. Gab is a social media platform alternative to Twitter and was launched in May, 2016. It has 100,000 estimated active and 4 million total users.\footnote{\url{https://en.wikipedia.org/wiki/Gab_(social_network)} as of March 2021} Unlike Twitter, it has a `lax' moderation policy for harmful content and presents itself as a champion of `freedom of speech'. It came under scrutiny in the Pittsburgh shooting case where the sole suspect posted a message on Gab indicating an immediate intent to cause harm before the shooting and also had a history of antisemitic posts~\cite{Screwthe78:online}. Recently, Gab was one of the platforms used to plan the storming of the United States Capitol on January 6, 2021~\cite{SocialMe40:online}. Given these facts, we reasoned that Gab should be a breeding ground for the type of data we wanted for our investigation.

The site allows anyone to read and write posts up to 3,000 characters called ``gabs''. In Gab, posts can be re-posted, quoted, and used as replies to other posts. Similar to Twitter, Gab also supports mentions and hashtags and users can follow one another. We started off with a huge dump already crawled from Gab in a previous study~\cite{10.1145/3178876.3186162}. This contains all the posts and their metadata from October 2016 to July 2018. In total there are 21 million posts. Further it has the re-post and reply information for each post. In addition the dump also hosts user bios and the follower/followee information per month. In total, there are $\sim$ 280,000 users having at least one post\footnote{Out of these users only 2\% have a following/follower ratio higher than 10~\cite{schaffer2009twitter} and around 1.2\% users post more than 3 messages per day~\cite{HowManyT7:online}. Thus the number of users that can be classified as bots based on the above two measures is negligible.}

\if{0}\begin{table}
\centering
\caption{Dataset Details}
\label{tab1}
\begin{tabular}{|c|c|}
\hline
\textbf{}              & \textbf{Dataset Details}                    \\ \hline
Timeframe                            & 2016-2018                       \\ \hline
Total number of posts                 & 21,200,212                      \\ \hline
Number of unique users                & 279,961 \\ \hline
\end{tabular}
\end{table}
\fi

In order to prepare the dataset for our study, we annotate 10K posts from Gab using a hybrid set of annotators. A group of four expert annotators and 103 crowd workers from Amazon Mechanical Turk were chosen based on a rigorous test of their annotation performance (see Methods for details). The task was to mark each post as (a) fear speech, (b) hate speech or (c) normal. Further, a post could have both fear and hate components and thus these were annotated with multiple labels. 

The annotators were asked to strictly adhere to the operational definitions of fear speech and hate speech as follows -- \textit{fear speech} is an expression aimed at instilling (existential) fear of a target group based on attributes such as race, religion, ethnic origin, sexual orientation, disability, or gender~\cite{saha2021short}, and \textit{hate speech} is a language used to express hatred towards a targeted individual or group or is intended to be derogatory, to humiliate, or to insult the members of the group, based on attributes such as race, religion, ethnic origin, sexual orientation, disability, or gender~\cite{mathew2020hatexplain}. In addition the crowd workers were given a number of examples and multiple rounds of training to enable them to perform the annotations as accurately as possible. The annotation went in 24 rounds with a small number of samples annotated in each round so as to reduce the overall mental toll faced by the annotators. In each round the sample posts were chosen based on the presence of a set of manually prepared keywords which increases the possibility of gathering such samples that are susceptible to be fear or hate speech. All rounds of annotation were closely monitored by the experts and corrective steps were taken as and when necessary (see Methods for full details of the annotation process). 

After this elaborate process we arrived at a dataset consisting of $\sim$ 10,000 annotated posts. Out of these, around 1800 were fear speech and 4000 were hate speech. The inter-annotator agreement values were as follows -- Krippendorff's $\alpha=0.30$ and Fleiss $\kappa=0.34$. These agreement values are comparable with such complex tasks in similar domain and settings~\cite{del2017hate,saha2021short,ousidhoum-etal-2019-multilingual}. Some examples of fear vs hate speech are noted in Table~\ref{table:samples_initial}. Note the use of various arguments in the fear speech posts such as the target community (a) replacing indigenous population (first instance), (b) being a physical danger to the society (second instance), and (iii) causing cultural threat (third instance).

\begin{table*}[!htbp]
\centering
\caption{This table shows the example of fear and hate speech in the Gab dataset.}
\begin{tabular}{p{8cm}|p{8cm}}
\hline
\textbf{Fear speech} & \textbf{Hate speech}\\ \hline
Germany is no longer German. German media celebrates school where 80\% of class is non-German \#GabFam \#Politics \#Europe \#Merkel \#Relocation \#Muslims \#BanIslam \#Invasion \#StopRelocation \#WhiteGenocide & You are a camel piss drinking goat f**ing imbecile now get off my timeline you disgusting piece of s**t.\\\hline
TILL White people won't protest for their SAFETY. Hell, it's not just Whites. Asian \& Middle Eastern shopkeepers are frequent victims.Young Black Males are a DANGER to society. SOME are ok, but we don't know who is who. We need PROTECTION \& the RIGHT NOT to race mix! & I hear Botswana is lovely in the spring. All ni**ers should go there. And stay. \\\hline
Jewish poison pouring out of our media and Hollywood is destroying Christianity & Because Jews are lying pigs. I'm really thinking this is a genetic thing..\\\hline
\end{tabular}
\label{table:samples_initial}
\end{table*}

\subsection*{Scaled up dataset}
Our objective was to study the large-scale prevalence of fear speech and compare the same with that of hate speech. Therefore we needed to scale up the annotated data. One easy way to achieve this would be to use standard toxicity classifiers over the whole dataset. We verify if this is possible using one of the state-of-the-art tools --- the Perspective API~\cite{Perspect94:online}. If we pass our base dataset through this classifier we observe that the average toxicity score of the fear speech posts returned by the API is 0.51 as opposed to 0.69 for the hate speech posts. This difference is also statistically significant with $p < 1e^{-6}$ as per the M-W U test~\cite{shier2004statistics}. The normal posts have a toxicity score of 0.47 and is very close to that of the fear speech post. Thus distinguishing fear speech from normal speech using such classifiers would be very difficult if not impossible. Hence we develop a sophisticated BERT based architecture to perform multi-label classification of an input post. We train and test the model using the base dataset and obtain a macro-F1 score of 0.63 (see Methods for the detailed description of the model). We next ran this model to classify all the 2 million posts in our dataset. For fear speech if we consider only those machine generated labels as correct where the confidence of the classifier is $>0.7$, our results are $>70$\% accurate (confirmed by a second round of expert annotation of a small number of samples). For hate speech, a similar accuracy is obtained if the decision confidence of the classifier is $>0.9$. We manually observe that increasing the threshold further did not improve the score further. Therefore we empirically fix these two decision confidence levels to finally obtain a scaled up dataset comprising $\sim$ 400K fear speech and $\sim$ 700K hate speech posts (see Methods for more details). All our analysis that follow in this article are performed on this dataset.%

\section*{Prevalence of fear speech}
The prevalence of a particular entity in any social network can be directly attributed to its users and fear speech is no exception. Therefore the first task is to identify users who have a strong propensity to post fear speech. Similarly, for comparison we also need to select users posting hate speech. 

\begin{table*}[!htbp]
\centering
\caption{Examples of fear speech which are popular in terms of likes/replies/reposts. The bold number per row shows the engagement factor based on which the specific post is cited.}
\label{table:samples_engaged}

\begin{tabular}{p{10cm}|c|c|c}

\hline
\textbf{Post} & \textbf{Likes} & \textbf{Reposts} & \textbf{Replies}\\ \hline
It’s the future. I was promised flying cars and cured cancer. Instead I got “hate speech,” a third world invasion, and an internet controlled by the ADL and SPLC. I’ll be damned if I let this be the “future” my kids grow up in. & \textbf{920} & \textbf{361} & 41 \\\hline
PUBLIC SERVICE ANNOUNCEMENT FROM IDENTITY EVROPA San Francisco is a dangerous sanctuary city where the law does not apply to illegal invaders. Enter at your own risk! & \textbf{593} & 205 & 16 \\\hline
This family lost a mother. She was killed by a Sudanese migrant in church yesterday in Antioch, Tennessee. Media silence is deafening.\#MelanieSmith  & \textbf{625} & \textbf{268} & 0 \\\hline
80K whites dead in South Africa in an ongoing genocide = Silence. 30 dead in a highly suspicious unconfirmed G-S attack in the Middle East = World War 3 & \textbf{588} & \textbf{282} & 15 \\\hline
It's not too late. A Charlottesville 2 could feature a memorial for Heather Heyer blaming antifa for jostling a land whale with an explosive heart. That'd probably get attention, and it'd probably be hard to separate from the message that the Alt-Right were innocent victims just trying to speak before jewish domestic terrorists started killing Whites.	& 0 & 0 & \textbf{77} \\\hline
I had an uber driver telling me this recently, after me going on about niggers (We in oz have a SMALL population of niggers per capita) he finally came clean they slashed his seats with knives and they SMELL particularly bad, Again this is in Australia where I see Africans maybe 10 a year. So far one tried to rob me and my uber driver had that happen! Imagine the US!! & 9 & 1 & \textbf{85} \\\hline
\end{tabular}
\end{table*}

\subsection*{User selection}
Out of 280K users we observe that as high as 9200 users have posted at least 10 fear/hate speech posts. However, we were interested in the extreme behaviour, i.e., we wanted to identify those users who have extreme propensity to post fear speech or hate speech. To this purpose, we find users falling in the top 10 percentile in terms of the number of fear speech or hate speech posted by them. We remove those common users that belong to both these sets\footnote{We perform a separate analysis on this set of users in \appendixnewname~in section \ref{sec:common_users}.}. We end up with 479 extreme fear speech (ExFear) and 483 extreme hate speech (ExHate) users\footnote{Out of this set, 476 users matched in term of propensity score based matching. See \appendixnewname~in section \ref{sec:matching} for more details.}. The choice of these set of users is motivated by the fact that they would be the central actors responsible for the prevalence of fear/hate speech.

\subsection*{User characterisation}
In this section, we compare the ExFear users with the ExHate users.  In total, ExFear users posted 2.6 millon posts, out of which 104k were fear speech and 26k were hate speech. Similarly, ExHate users posted 2 million posts out of which 184k were hate speech and 18k were fear speech. We consider three different aspects -- their position in the social network, their overall reach of the normal users and temporal trends. %

\subsubsection*{Position in the social network}
\label{sec:followers}

We construct the social network based on all the follower-followee relationship among the users till the end of the timeline (i.e., June 2018) ~\cite{mishra2019abusive}. %
This network consists of 279,961 nodes and 1960,869 edges. 

The first quantities that we compare are the number of \textit{followers} and \textit{followings} for each type of user. The plot in Figure~\ref{fig:follower_following} shows that both the number of followers and the followings of ExFear users are larger than that of the ExHate users. The results are statistically significant with $p<0.0001$ (M-W U test).

\begin{figure}[h]
     \centering
     \begin{subfigure}[h]{0.45\linewidth}
         \centering
         \includegraphics[width=\textwidth]{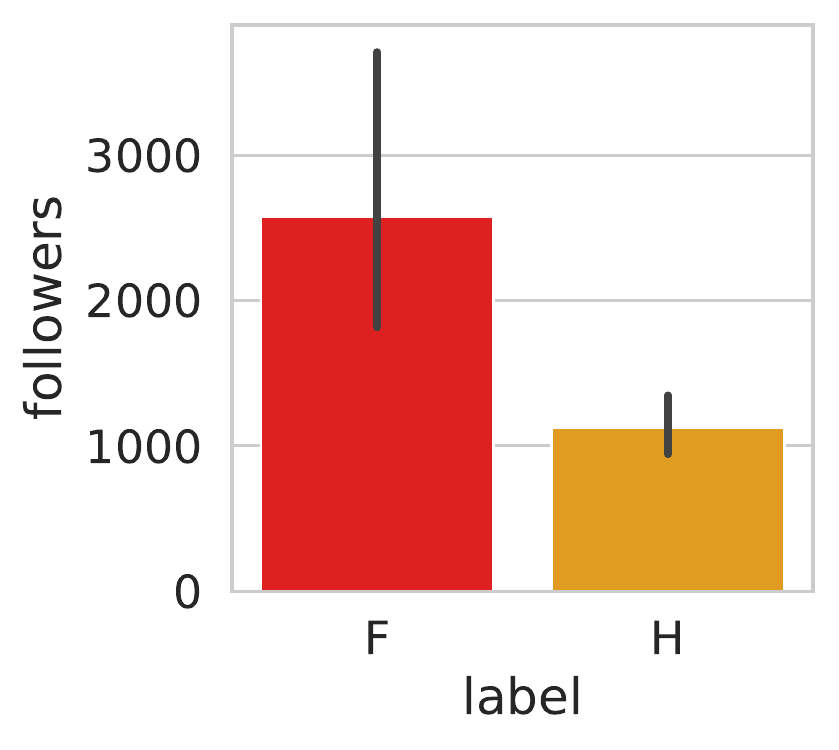}
         \caption{Average number of followers for different groups of users.}
         \label{fig:followers}
     \end{subfigure}
     \hfill
     \begin{subfigure}[h]{0.45\linewidth}
         \centering
         \includegraphics[width=\textwidth]{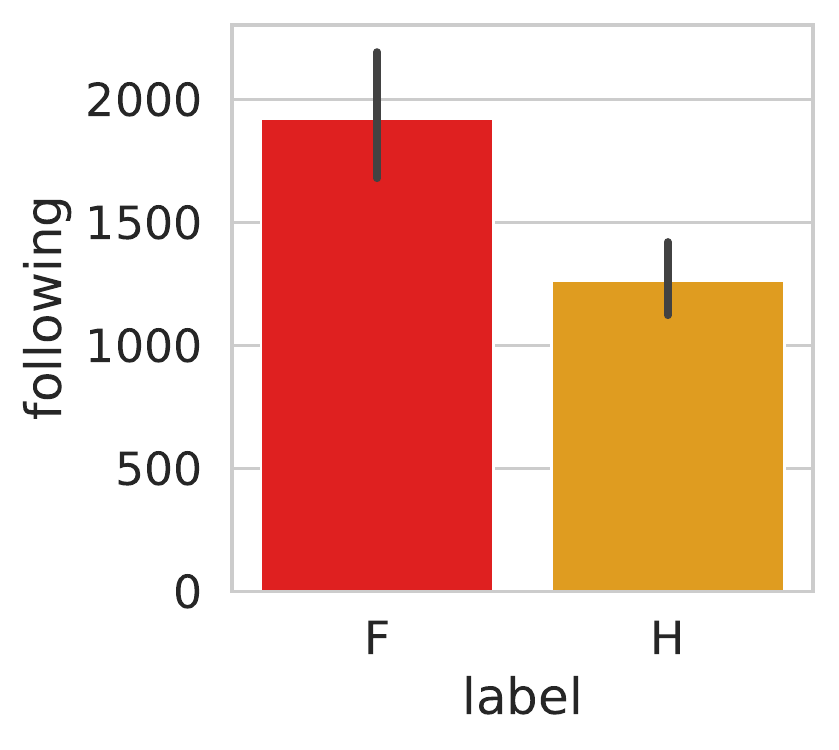}
         \caption{Average number of followings for different groups of users.}
         \label{fig:following}
     \end{subfigure}
    \caption{Plots denoting follower-following properties for ExFear (F) and ExHate (H) users.The results are significant at $p<0.0001$ using Mann-Whitney U test.}%
    \label{fig:follower_following}
\end{figure}

Next we compute the \textit{betweenness} and the \textit{eigen-vector} centrality of the nodes from the undirected version of this network. These metrics are known to express the positional importance of the nodes; while eigenvector centrality indicates the influence of the nodes, betweenness centrality indicates the degree to which a node stands in between other nodes. From Figure~\ref{fig:eigen_vector} we observe that, in terms of eigenvector centrality, the ExFear users are more central compared to the ExHate users. Once again the results are statistically significant with $p < 0.001$ (M-W U test). The observations remain same for the betweenness centrality with the ExFear users far more central compared to the ExHate users (see Figure \ref{fig:betweenness}).

These results together show that the ExFear users are far more strategically placed in the network compared to the ExHate users. Such advantageous positions of the ExFear users is a natural source for higher prevalence of fear speech in the network.

\begin{figure}[h]
     \centering
     \begin{subfigure}[h]{0.45\linewidth}
         \centering
         \includegraphics[width=\textwidth]{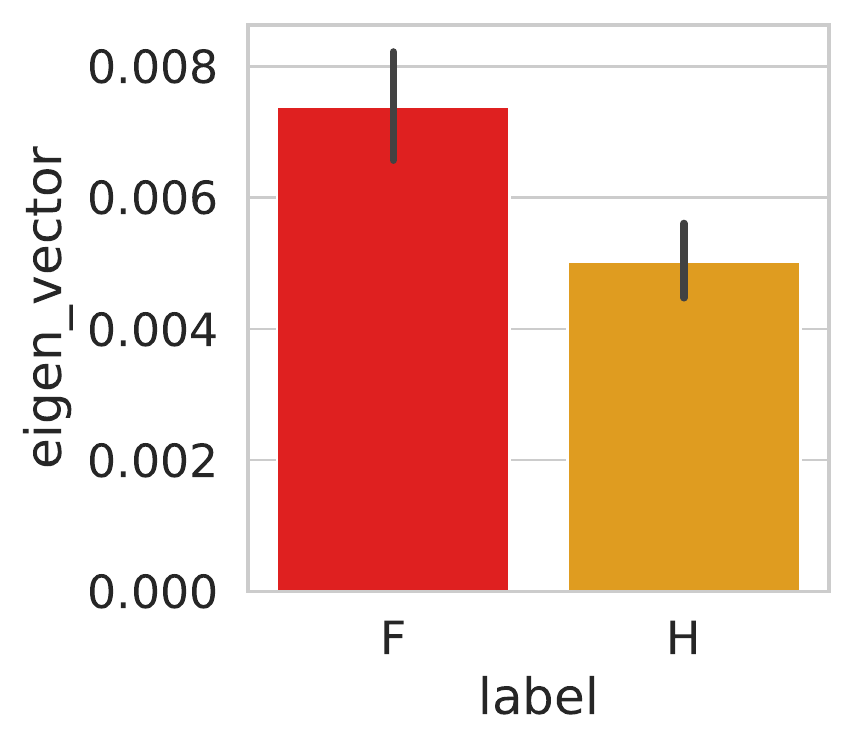}
         \caption{Eigenvector centrality of users.}
         \label{fig:eigen_vector}
     \end{subfigure}
     \hfill
     \begin{subfigure}[h]{0.45\linewidth}
         \centering
         \includegraphics[width=\textwidth]{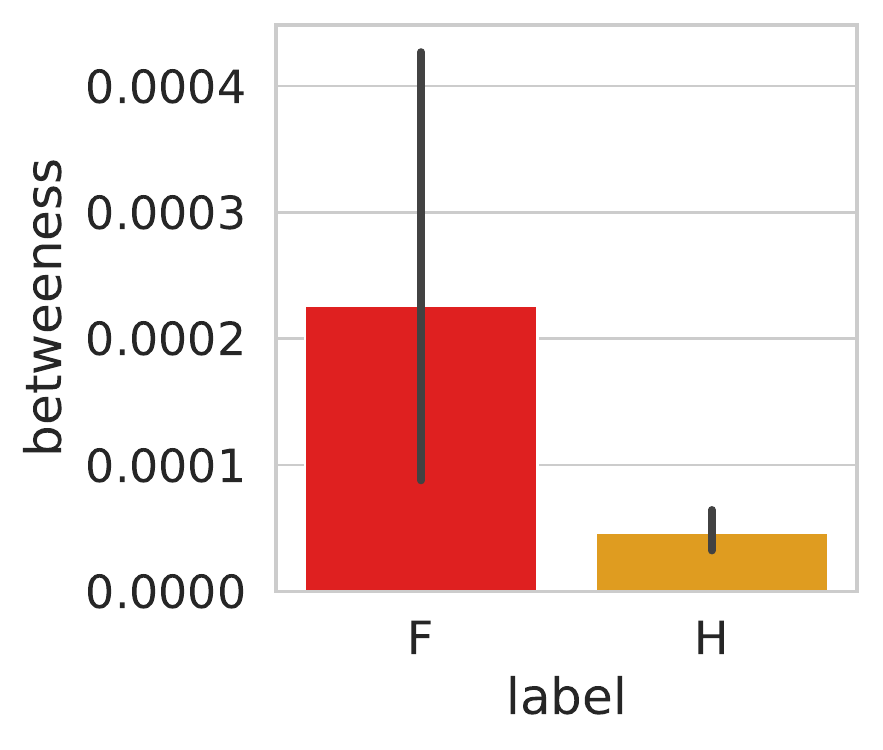}
         \caption{Betweenness centrality of users.}
         \label{fig:betweenness}
     \end{subfigure}
    \caption{Centrality measures of ExFear (F) and ExHate (H) users. The results are significant at $p<0.0001$ using Mann-Whitney U test.}
    \label{fig:network_metrics}
\end{figure}

\subsubsection*{Reach of the normal users}
\label{sec:normal-users}
We label users who never post fear or hate speech (decision confidence of the model $<0.5$) as \textit{normal users}. Here we investigate how the ExFear and ExHate users interact with the normal users. First, we find that the average percentage of normal followers out of all followers for ExFear users (21\%) is higher than for ExHate users (18\%). This difference is statistically significant with $p < 1e^{-6}$ (M-W U test). 

The number of posts made by ExFear and that by ExHate users are both of the tune of two million each, i.e., their posting activity is quite similar. Therefore, we plot the number of normal users re-posting the posts of ExFear vs ExHate users. The results in Figure~\ref{fig:normal_reposts} show that larger number of normal users re-post the posts of ExFear users compared to that of ExHate users ($p < 1e^{-6}$ (M-W U test)). Further, the total number of reposts by normal users to the posts made by ExFear users is larger than posts of ExHate users (see Figure~\ref{fig:total_normal_reposts}). The same trend persists for both mentions and replies. ExFear users mention more number of normal users in their posts (Figure~\ref{fig:normal_mentions}) compared to ExHate users. Moreover, the total number of posts by the former having normal users mentioned is also higher (Figure~\ref{fig:total_normal_mentions}). The number of normal users replying to the posts of ExFear uses is higher than that of ExHate users (Figure~\ref{fig:normal_replies}). Finally the number of replies obtained from the normal users by the ExFear users is larger than that of ExHate users (Figure~\ref{fig:total_normal_replies}). Hence, we show that ExFear users impacts the normal users more. 

We also analyse the impact on normal users based on the posts they receive. Since, it is not possible to know if someone received a post directly, we assumed that an user `A' would receive a post from a particular user `B'  if he/she is following that user ‘B’. We consider the top 500 normal users based on their number of posts. An additional constraint was that they should have at least one ExFear and one ExHate user in their following. This resulted in 179 users. We find that these users receive around 1.5\% fear speech and 2\% hate speech posts from their followings. Surprisingly, although the percentage of fear speech received by them is less, they end up reposting the fear speech almost \textit{four} times more (average of 1.10 posts) compared to hate speech (average of 0.28 posts). The results are statistically significant ($p<0.001$, M-W U test).

In our final experiment in this section we go a step forward to assess the perception about fear vs hate speech among in-the-wild users. We recruit human judges from Amazon Mechanical Turk for this experiment. We create a survey by posing pairs of fear and hate speech and ask human judges to select the post they believed in more. Each pair of posts is judged by \textit{nine} random judges. All these judges have high approval rates ($>95\%$) and high approved hits ($>1000$). We got the posts judged in three batches, with incremental number of post pairs in each batch -- precisely, 25, 30 and 45 pairs in the three successive batches. Each batch is further divided into pages of \textit{three} pairs each. We make sure that the judges in successive batches do not overlap to ensure diversity of opinions. In total, 246 unique judges participated in the task (68 in the first batch, 82 in the second and 96 in the third). For each pair, we select that post between fear and hate speech to be believable which receives majority of the votes. We find that out of 100 pairs, in as many as 69 pairs (i.e., $69\%$), fear speech posts were voted to be the more believable out of the two.

\begin{figure}[!htpb]
     \centering
     \begin{subfigure}[h]{0.45\linewidth}
         \centering
         \includegraphics[width=\textwidth]{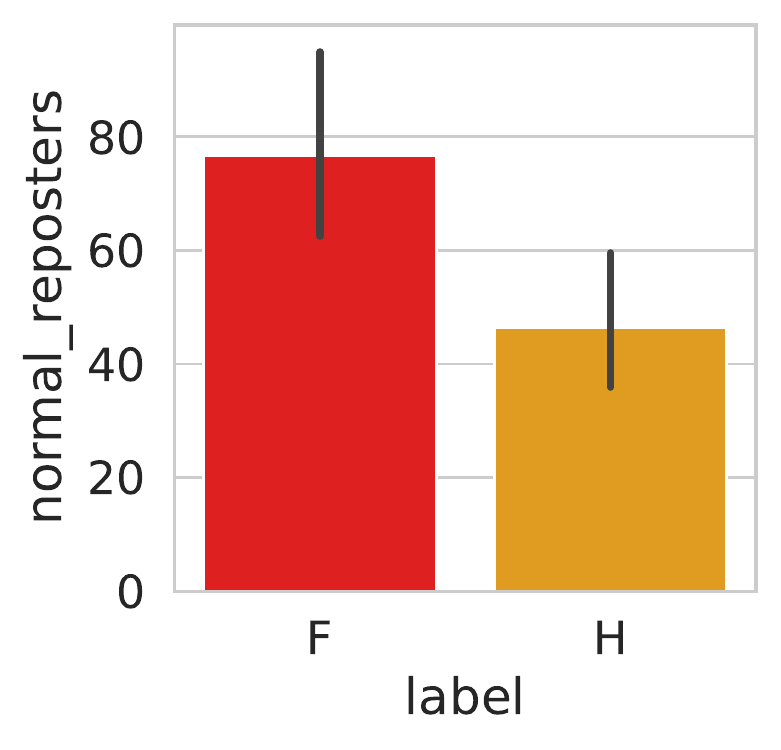}
         \caption{\# of normal reposters per user.}
         \label{fig:normal_reposts}
     \end{subfigure}
     \hfill
     \begin{subfigure}[h]{0.45\linewidth}
         \centering
         \includegraphics[width=\textwidth]{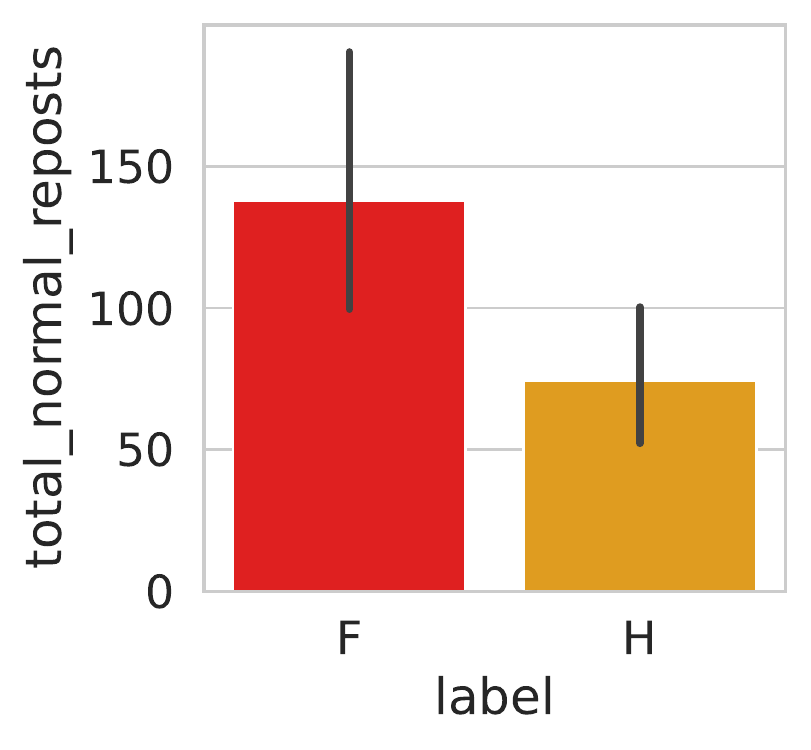}
         \caption{\# of reposts from normal users per user.}
         \label{fig:total_normal_reposts}
     \end{subfigure}
    \caption{Distribution of reposts from normal users for ExFear (F) and ExHate (H) users. The results are significant at $p<1e^{-6}$ using Mann-Whitney U test.}
    \label{fig:reposts}
\end{figure}

\begin{figure}[h]
     \centering
     \begin{subfigure}[h]{0.45\linewidth}
         \centering
         \includegraphics[width=\textwidth]{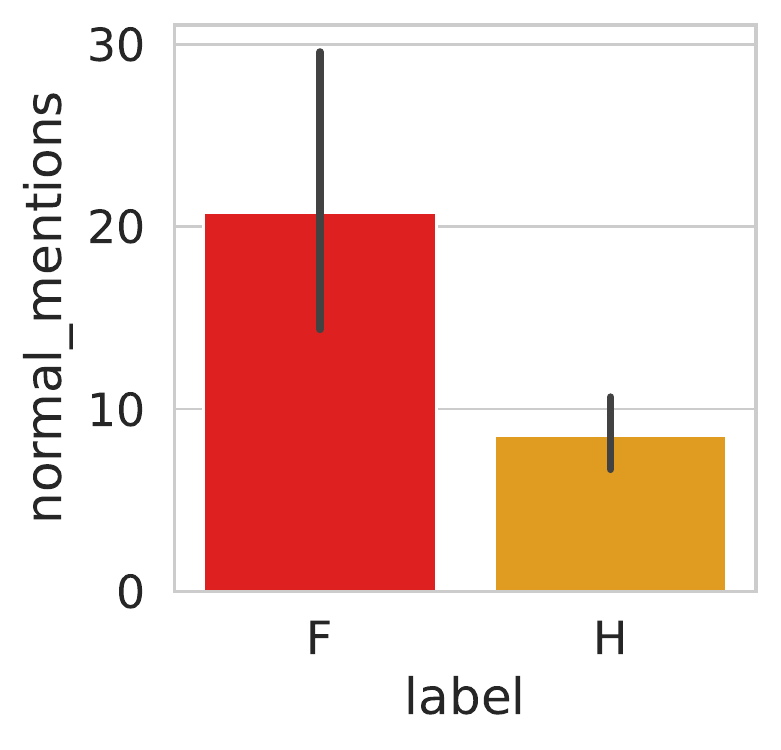}
         \caption{\# of normal mentioned users per user.}
         \label{fig:normal_mentions}
     \end{subfigure}
     \hfill
     \begin{subfigure}[h]{0.45\linewidth}
         \centering
         \includegraphics[width=\textwidth]{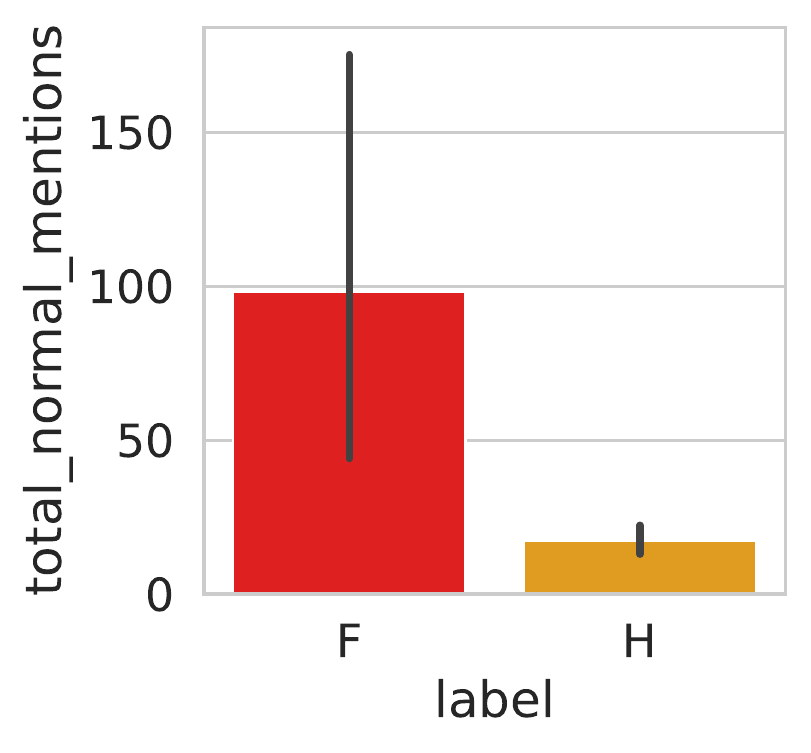}
         \caption{\# of normal mentions per user.}
         \label{fig:total_normal_mentions}
     \end{subfigure}
    \caption{Distribution of normal mentions for ExHate (H) and ExFear (F) users. The results are significant at $p<1e^{-6}$ using Mann-Whitney U test.}
    \label{fig:mentions}
\end{figure}

\begin{figure}[h]
     \centering
     \begin{subfigure}[h]{0.45\linewidth}
         \centering
         \includegraphics[width=\textwidth]{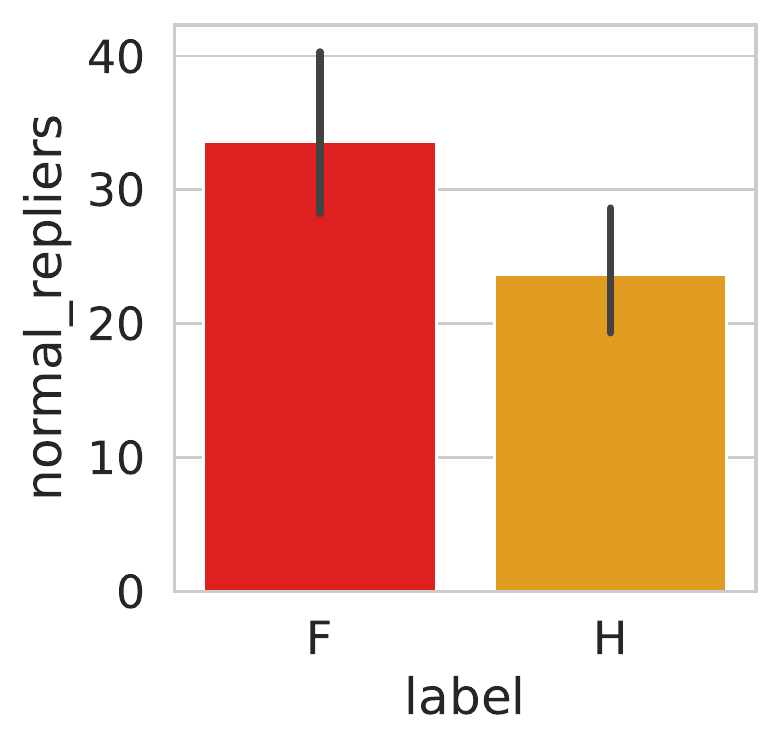}
         \caption{\# of normal repliers per user.}
         \label{fig:normal_replies}
     \end{subfigure}
     \hfill
     \begin{subfigure}[h]{0.45\linewidth}
         \centering
         \includegraphics[width=\textwidth]{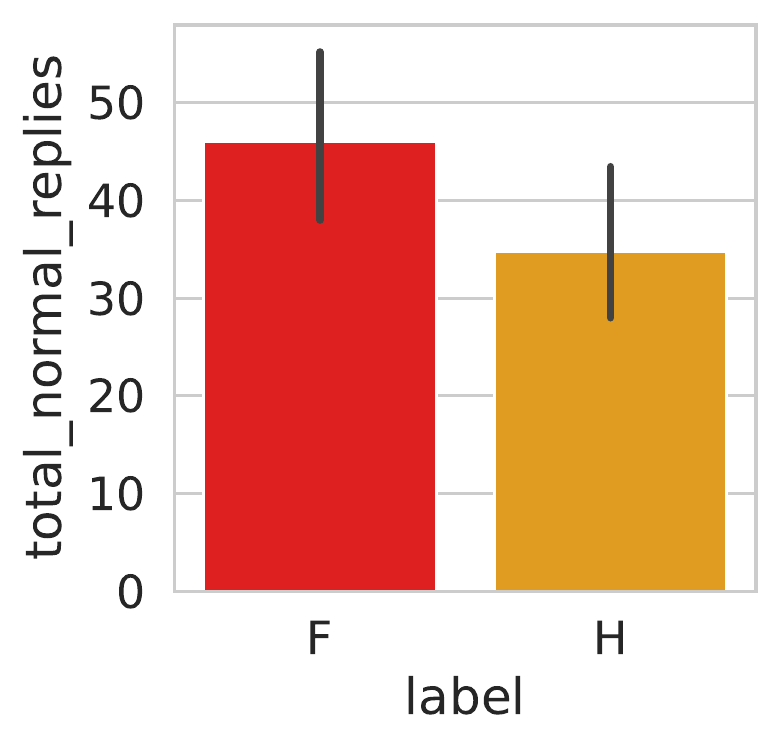}
         \caption{\# of replies from normal users per user.}
         \label{fig:total_normal_replies}
     \end{subfigure}
    \caption{Distributions of replies from normal users for ExFear (F) and ExHate (H) users.The results are significant at $p<1e^{-6}$ using Mann-Whitney U test.}
    \label{fig:replies}
\end{figure}

Overall, we observe that the ExFear users are far more well connected with the normal users compared to the ExHate users. Manual analysis reveals that the top reposted/replied/liked fear speech posts contain emotionally loaded language and/or urgent tone with the occasional usage of capital letters as shown in Table ~\ref{table:samples_engaged}. Often the posts pretend to narrate real incidents, foretell how bleak the future could be and cite (fake) statistics to make the content look realistic and convincing. We also obtain the top 10 normal users mentioned by ExFear users and find that they usually have a large number of followers ($\sim$ 1200) and followings ($\sim$ 1700)  but have less number of posts ($\sim$ 17). Manually analysing their profiles from Gab we find that their posts are generally on benign topics but they repost a lot of controversial topics which might be the reason why they get mentioned more by the ExFear users.

\subsubsection*{Temporal trends}

In this section we deep dive into the results obtained earlier to investigate the temporal evolution of different observables of interest. 

\begin{figure*}[ht]
     \centering
     \begin{subfigure}[h]{0.45\linewidth}
         \centering
         \includegraphics[width=0.7\textwidth]{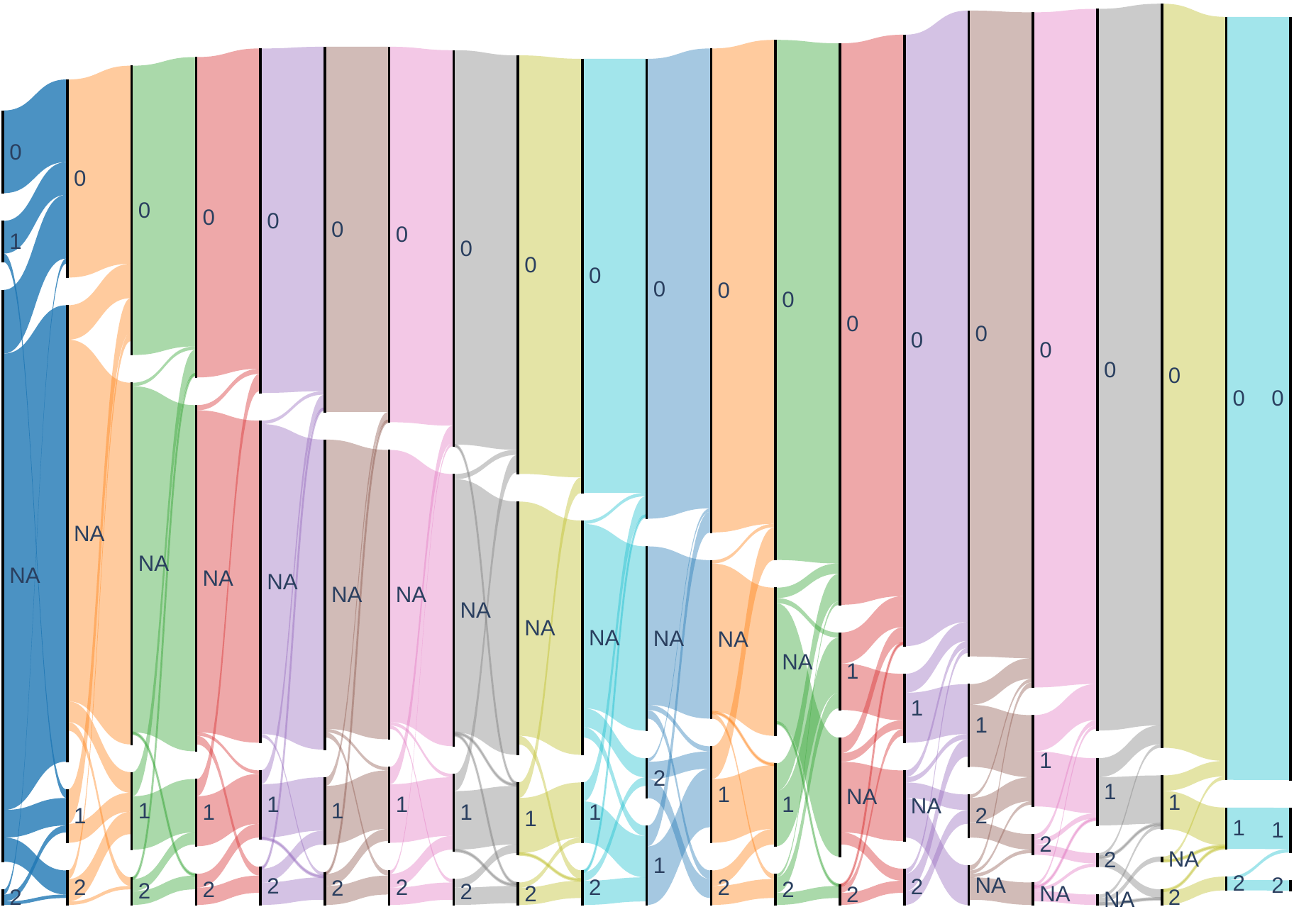}
         \caption{Extreme fear speech (ExFear) user movement}
         \label{fig:hate_temporal}
     \end{subfigure}
     \hfill
     \begin{subfigure}[h]{0.45\linewidth}
         \centering
         \includegraphics[width=0.7\textwidth]{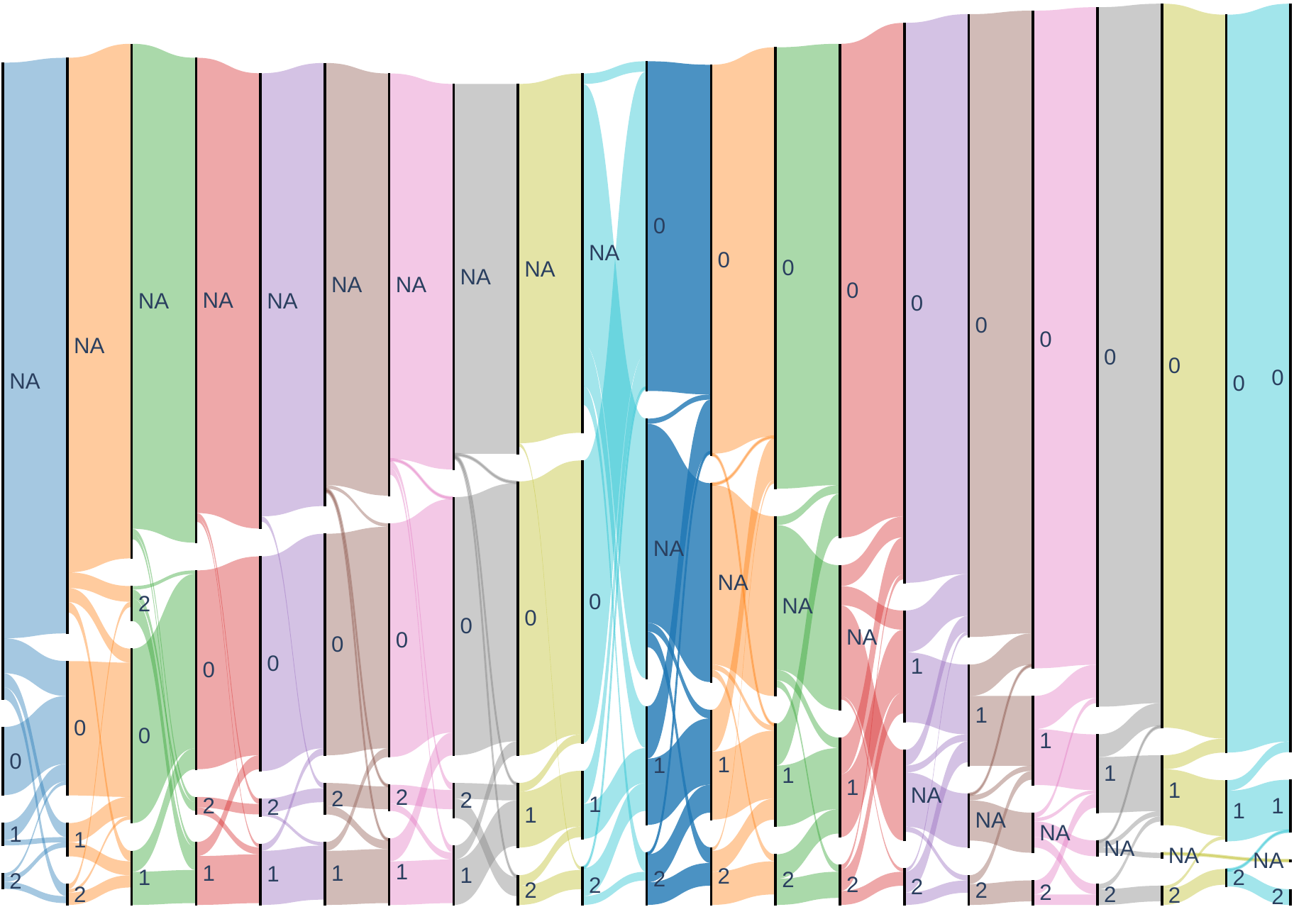}
         \caption{Extreme hate speech (ExHate) user movement}
         \label{fig:fear_temporal}
     \end{subfigure}
    \caption{Alluvial diagram to show the core transition for the users. The stubs represent dynamic graph state with the first stub indicating October 2016. A lower core value represents that a node is situated deeper in the network. `NA' denotes the set of users who are yet to join the networks each month from the total set of users. We show only the transitions among the three innermost cores for better visualisation. The dark blue band shows the month with the maximum influx for each graph. Maximum influx means during that month maximum number of users have jumped to an inner core with respect to their core.}
    \label{fig:temporal_followership}
\end{figure*}

As a first step we investigate how the ExFear and ExHate users move in the follower-followee network over time. To this purpose, for each month we construct an undirected follower-followee network and perform the standard $k$-core decomposition~\cite{batagelj2011fast}. Such a decomposition is known to segregate the network into `shells' with the innermost few shells containing the most influential nodes. We divide the nodes into 10 buckets in terms of the percentile ranks based on their $k$-core values, i.e., top 10 percentile nodes in the first bucket, next 10 percentile nodes in the second bucket and so on. Note that therefore the first bucket consists of the most influential nodes while the last contains the least influential ones. Next we observe how the users move from one bucket to the other over time since they had joined the network. The temporal movement of the ExFear and ExHate users across the different buckets over time are shown in Figure ~\ref{fig:temporal_followership}. Both the ExFear and ExHate users are predominantly in the outer shell of the network at the time of their joining. However, as time progresses they accelerate steadily to the inner shells with the maximum influx happening in Oct 2016 for ExFear users and in August 2017 for the ExHate users\footnote{Here the maximum influx is defined as the maximum users jumping to a inner core considering their current core}. The maximum influx for ExHate users coincides with the Unite the Right  rally, Charlottesville\footnote{\url{https://time.com/charlottesville-white-nationalist-rally-clashes/}} while for ExFear users we see a jump toward the initial time period. This possibly indicates that a fraction of the ExFear users have remained all through in the core of the network from the very beginning. %
On average, ExFear users take lesser time (2.83 months) to reach the innermost core of the network compared to the ExHate users (3.32 months). %

Next, we investigate the temporal evolution of the engagement to the posts made by ExFear or ExHate users.  When considering replies by normal users, we observe that while for the first 2-3 months the trajectories are similar, after January 2017, the replies to ExFear users keep increasing while replies to ExHate users suffer a dip. The replies to ExFear users have a sudden peak around June 2017. After this, the replies to ExFear users dip below ExHate users possibly due to the influence of external event in the form of Unite the Right rally, Charlottesville. This might also suggest that many normal users started to subscribe to the hateful notions. If we consider the reposts by normal users\footnote{We have the reposts information from August 2017 in the dataset.}, we find that here the distributions are similar with the peaks occurring at similar time (March 2018). Considering the normal users' mentions by the two groups of users temporally, we find a significant difference in the two curves. While ExHate users use very less mentions of normal users, ExFear users heavily use the same with the peak occurring (60 times per mentioned users) at December 2017. Manual analysis revealed many of the fear speech posts had a comment about a target community followed by mentions of several users, social media influencers and news media sources etc. For e.g. ``Muslims want to double the number of mosques in France <link> @MichelOsef @Isleofcarl @Bill\_Murray @SatanIsAllah @Brea @HEDGE @PigtownGrump @Zucotic @TaratheLeo @kingmack @Psnow @TwoPats @MaryJane @TupacZaday @Reef''.

Overall, this section demonstrates that fear speech has a significantly larger prevalence in the social network compared to hate speech (more analysis with larger set of users can be found in the SI text). %
In the next section we investigate the content structure of fear speech which undoubtedly plays the central role in its wider prevalence.

\section*{Content structure}
In this section, we investigate the differences in the content structure of the fear speech posts from those of the hate speech posts. These differences rooted in their content plays a key role in shaping their prevalence.   

\subsection*{Fear speech is presented as topical arguments}

We analyze the text present in the fear speech post and compare them with the hate speech posts using widely popular NLP tools as follows.

\subsubsection*{Topic modelling} 
\label{sec:topics}

We use the LDA model~\cite{hoffman2010online} to extract the topics in the fear and hate speech posts (More details in the \appendixnewname~section~\ref{sec:topic_modelling}). Next for each month, we plot their normalized distribution considering the total posts in that month. Overall we notice one very important difference between fear speech (see Figure~\ref{fig:topic_dist_fear}) and hate speech topics (see Figure~\ref{fig:topic_dist_hate}). Topics in the fear speech mostly portrayed other communities as perpetrators in a subtle and argumentative style, while topics in the hate speech were dehumanizing or insulting the target communities. 

Some of the illustrative examples of fear speech topics are -- `America needs to wake up' and `Ideology of Islam is dangerous' which are prevalent across all the months. Here the topic `America needs to wake up' makes implicit calls to Americans to see the atrocities by other communities. The topic `violence by Muslim communities' notes the various unconfirmed violent activities by the Muslim communities. It had a tiny share  initially (October \& November 2016) but increased to a significant ratio afterward. On the other hand, the topic -- `immigrants manipulating elections' was prominent during the initial time periods but died out after April 2017. Another interesting topic was `jews controlling media' - which points out how Jews control media platforms. Apart from that, illegal immigration as a problem was portrayed in topics like `illegal immigration in Europe', `illegal immigration in the USA' etc.

\begin{figure}[!h]
    \centering
         \includegraphics[width=\linewidth]{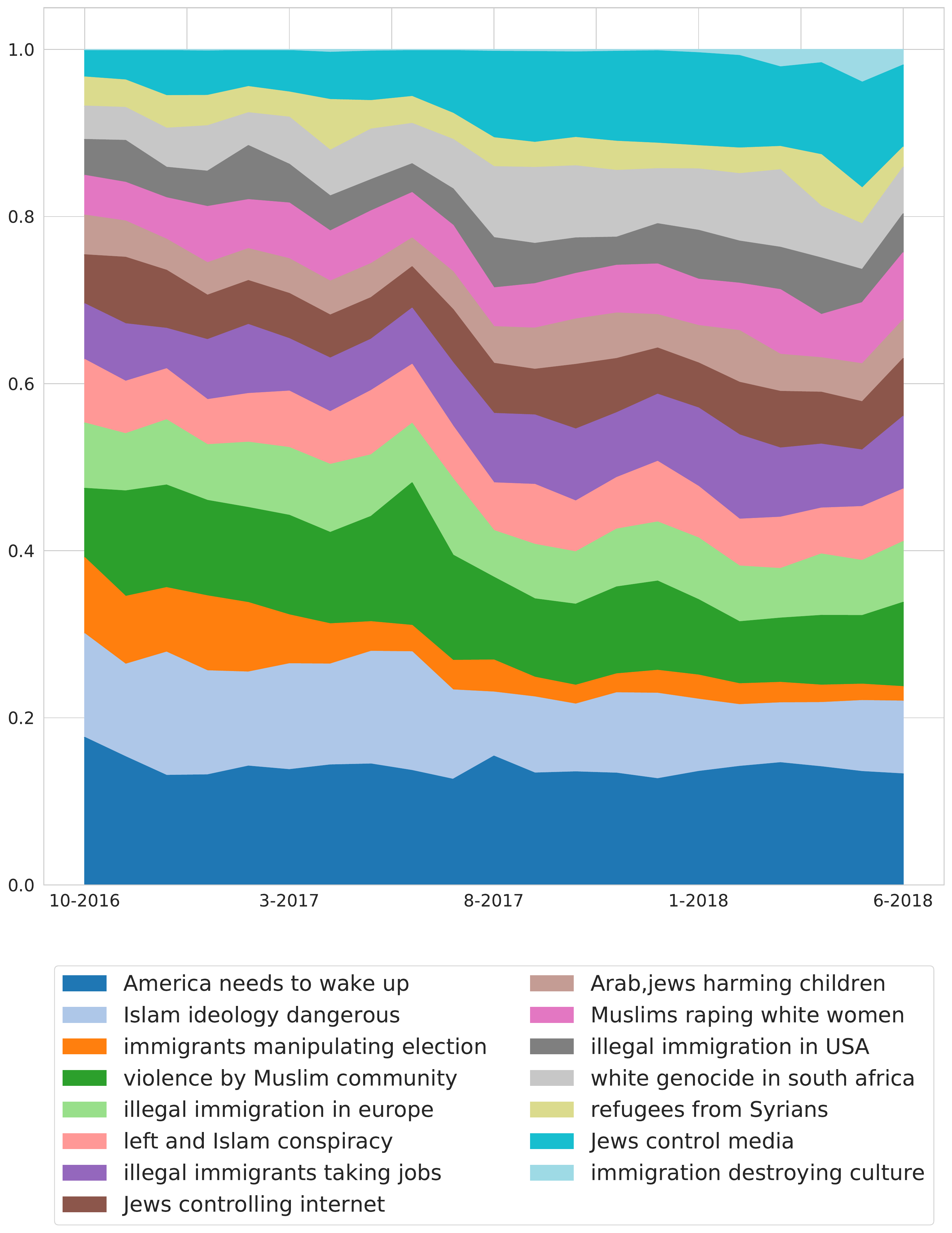}
    \caption{Top 10 topics and their normalised distribution per month for fear speech posts.}%
    \label{fig:topic_dist_fear}
\end{figure}

Among the hate speech topics, three of the most consistent ones are `multi-target insults' --- where a single hate post targeted multiple communities, `women being projected as prostitutes', and hate against voters from different demography. Other topics like insults of Muslims and Canadians occur rarely and have smaller ratios. Insults of the Jewish community rose after August 2017. This might be an effect of the influx of a lot of new users during that time period. The topic, which has posts targeting both homosexuals and Muslims reduced after March 2017 since it possibly merged with the multi-target insults. The ratio of posts under topics like `support for Nazi', `insulting and blaming Africans'  increased significantly after August 2017. %

\begin{figure}
    \centering
         \includegraphics[width=\linewidth]{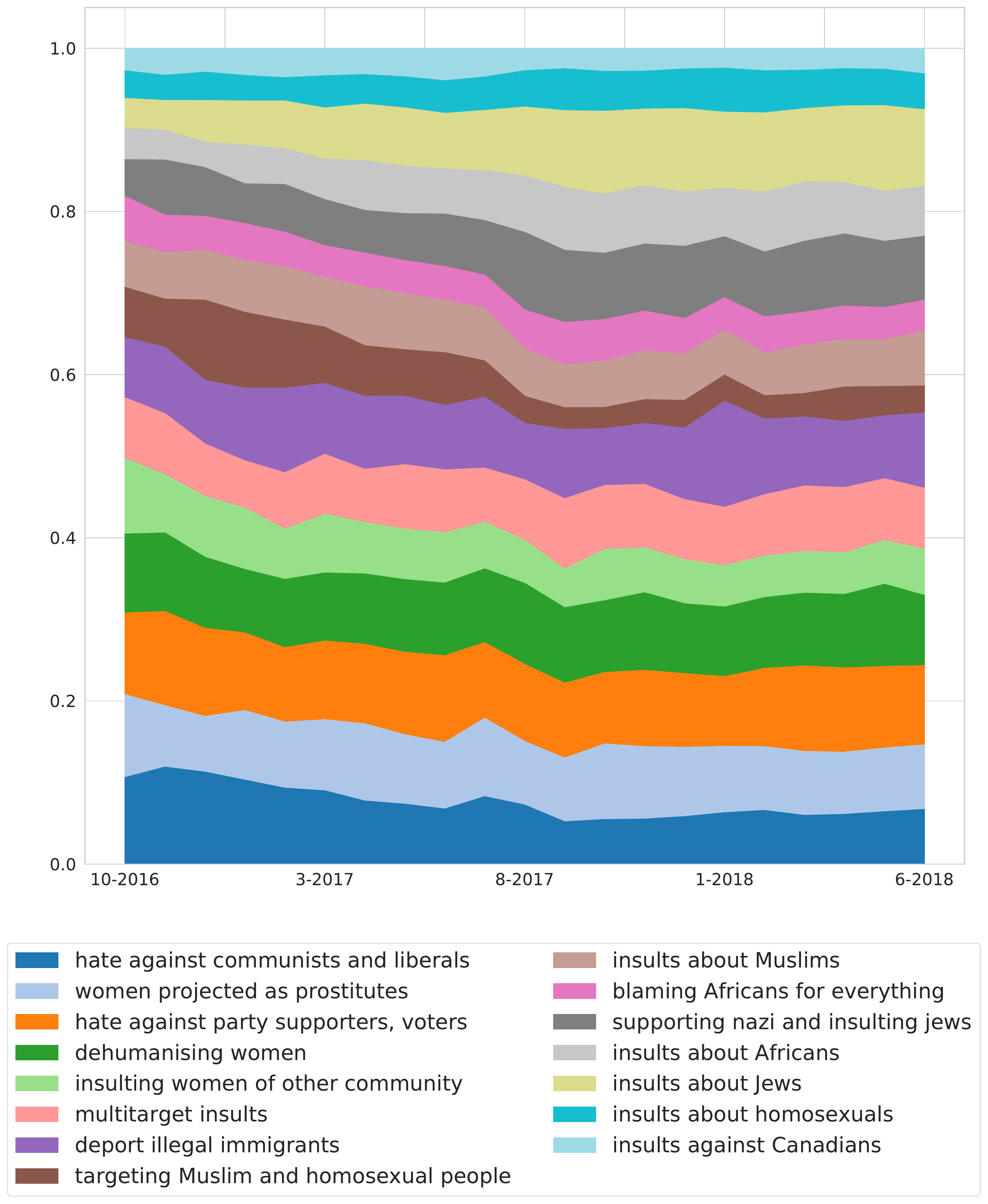}
    \caption{Top 10 topics and their normalised distribution per month for hate speech posts.}
    \label{fig:topic_dist_hate}
\end{figure}

\subsubsection*{Reaction of normal users} A careful observations of the topics extracted from the fear speech posts shows that the arguments presented in these topics most often look quite acceptable and amenable to the normal users resulting in their direct involvement in re-posting of and replying to the messages corresponding to these topics. The topics in the fear speech category receive around 1000 reposts from normal users with the highest average reposts being received by the topic `violence by Muslim community' ($\sim$ 2500). On the other hand, for the topics extracted from the hate speech posts hurling direct attacks on different communities are usually found to be repulsive by normal users and are much less frequently re-posted or replied to. Average number of reposts per topic is about 500 for hate speech topics with the highest average reposts being received by the topic `deport illegal immigrants' ($\sim$ 1100). Note that, in general the average number of reposts for any post on the platform is around 2 per posts.

\subsection*{Hashtags and web domains}

\subsubsection*{Hashtags} Hashtags are an important component of the overall content of any social media post. We investigate how fast a hashtag originating from one form of speech is adopted to scribe another form of speech. A hashtag is considered to have originated in fear/hate/normal speech, if a fear/hate/normal user uses it for the first time in one of their posts. One of the most surprising findings is how fast hashtags originating from normal speech gets adopted to fear speech ($\sim$ 83 days); this is significantly less compared to the time needed by hashtags originating from normal speech and getting adopted to hate speech ($\sim$ 124 days) ($p<1e^{-6}$, M-W U test, one-sided). This suggests that users posting fear speech carefully craft their messages to include hashtags mainly used by normal users. Consequently, the visibility of the corresponding fear speech post gets enhanced among the normal users. In addition, another observation is that the median time for a hate speech hashtag to get adopted into a fear speech post ($\sim$  73 days) is significantly ($p<1e^{-6}$, M-W U test, one-sided) lower than a fear speech hashtag to get adopted into a hate speech post ($\sim$ 88 days). This once again shows that fear speech users cleverly include hashtags used by hate speech users in their posts.%

\subsubsection*{Web domains} We investigate the popular domains shared by the fear and hate speech users. Around $\sim$ 6000 unique URLs were shared by each of these type of users. We manually inspected some of the most frequent domains (top 20) that were shared (see Table \ref{tab:urls}). Many of the fear speech posts shared URLs of unconfirmed blogs on atrocities by the Muslim community --- \url{islamexposedblog.blogspot.com}, \texttt{\url{thereligionofpeace.com}} and \texttt{\url{counterjihad.com}}. Few domains were right biased media  having low credibility like \texttt{American Center for Law and Justice}\footnote{\url{http://www.aclj.org} as accessed on Mar 7, 2022} and \texttt{Sputnik news}\footnote{https://sputnik.com/}. Another website portrays the unconfirmed atrocities on the white community - \texttt{whitenationnetwork}\footnote{https://tinyurl.com/567n8rat}. In fact, this website has been currently shut down. Other forms of conspiracy theories like \texttt{coronavirus is a hoax} also showed up on some of these websites. Overall, majority of the URLs shared by the fear speech users have fake/unconfirmed content which, most often, make them highly believable to the benign social media users.

Popular domains in hate speech posts are quite different in nature. We find \texttt{pagesix}\footnote{\url{https://pagesix.com/} accessed on March 10, 2022}, an entertainment news website and \texttt{towleroad}\footnote{\url{https://www.towleroad.com/}} an entertainment website for Gay and LGBTQ+ community which are both authentic. Both these websites are benign in nature, but the hate speech posts referred to them to insult the celebrities mentioned in these platforms. We also find \texttt{dailystormer}\footnote{\url{https://en.wikipedia.org/wiki/The_Daily_Stormer}}, \texttt{godhatefags}\footnote{\url{https://en.wikipedia.org/wiki/Westboro_Baptist_Church}} etc. which are popular far-right websites.

\begin{table}[!htpb]
\centering
\caption{Some of top relevant URLs along with the number of fear/hate speech posts.}

\begin{tabular}{ll}
\textbf{Fearspeech}             & \textbf{Hatespeech} \\\hline
aclg (243)                      &  pagesix (65)          \\
whitenationnetwork (54)         & towleroad (68)       \\
islamexposedblog (72)           & dailystormer (63)  \\
thereligionofpeace (40)         & weaselzippers (45)  \\
sputniknews (37)                & godhatesfags (28) \\  
counterjihad (33)               & thesmokinggun (20)      \\\hline
\end{tabular}
\label{tab:urls}
\end{table}

\subsection*{Interaction of the users with the content} 
Interaction with a post can be an essential indicator of how the audience engages with the post. We measure this using the re-posts, replies, and likes frequency. Here, we compare these interactions for fear and hate speech posts. As a baseline, we also compare these with the overall level of interaction with all posts. 

\subsubsection*{\#Likes} Fear and hate speech posts taken together receive $\sim$ 65\% of likes while at the overall level, less than $\sim$ 60\% posts receive one or more likes. As illustrated Figure~\ref{fig:mean_likes} we find that the average number of likes for fear speech is around $\sim$ 7 per post, which is significantly more ($p < 1e^{-6}$, M-W U, one-sided) than that of hate speech. We have shown the examples of the highly liked fear speech posts in Table ~\ref{table:samples_engaged}. %

\subsubsection*{\#Replies} Fear and hate speech posts taken together receive one or more replies in $\sim$ 16\% cases while at the overall level, less than $\sim$ 10\% posts receive one or more replies. Once again, as shown in Figure~\ref{fig:mean_replies} the mean number of replies per post is higher for fear speech as compared to hate speech ($p < 1e^{-6}$, M-W U, one-sided). We have shown the examples of the highly replied fear speech posts in Table ~\ref{table:samples_engaged}. Manual analysis revealed interestingly the post receiving higher reposts usually had less replies and likes. Further, around 0.3\% of the replies of the fear speech are from normal users, whereas 0.2\% of the replies of the hate speech are from normal users. %

\subsubsection*{\#Reposts} In terms of reposts, we observe that more number fear speech posts ($\sim $18\%) are reposted as compared to hate speech and overall posts ($\sim$ 11-13\%). The average number of reposts per post is significantly ($p < 1e^{-6}$, M-W U, one-sided) higher for fear speech (5 per post) than hate speech (3 per post (see Figure~\ref{fig:mean_reposts}). We have shown the examples of the highly reposted fear speech posts in Table ~\ref{table:samples_engaged}. Further, around 6\% of the reposts of the fear speech are from normal users, whereas 3\% of the reposts of the hate speech are from normal users. %

In summary, we observe that the average level of engagement of users with fear speech posts is much higher than hate speech posts which we believe is another reason for their prevalence.

\begin{figure}[h]
     \centering
     \begin{subfigure}[h]{0.3\linewidth}
         \centering
         \includegraphics[width=\linewidth]{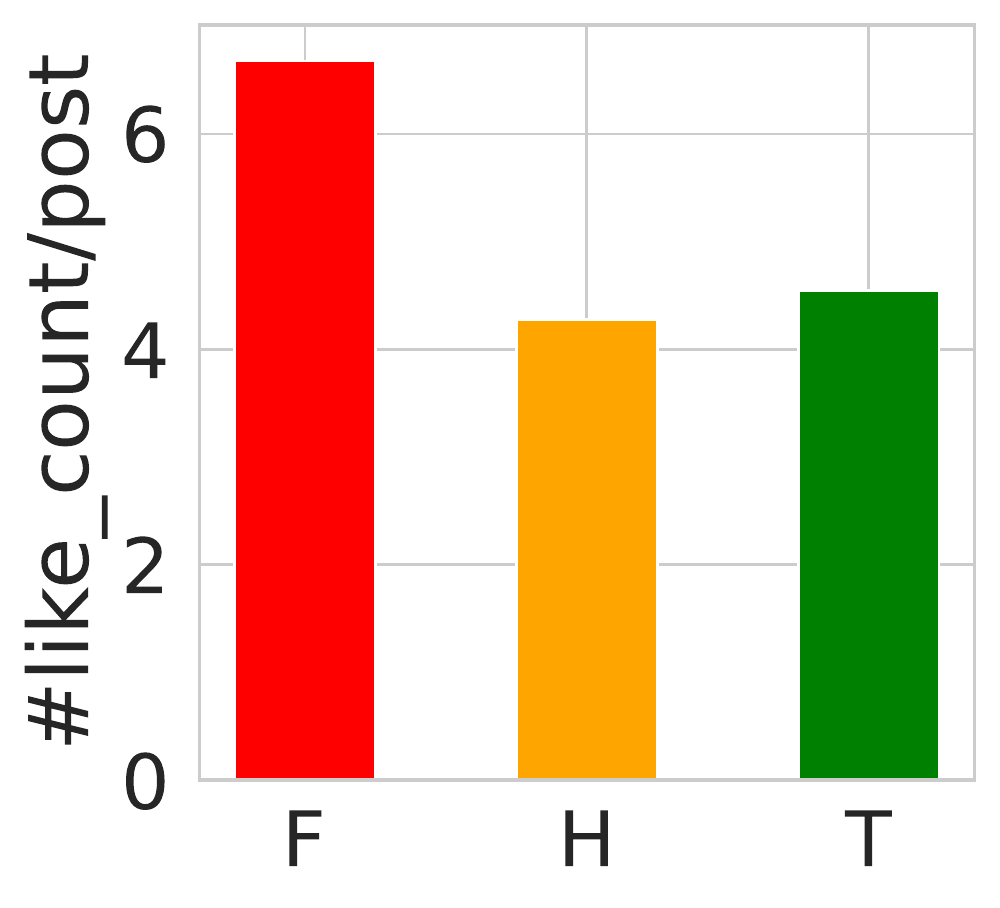}
         \caption{Mean likes per post.}
         \label{fig:mean_likes}
     \end{subfigure}
       \hfill
     \begin{subfigure}[h]{0.3\linewidth}
         \centering
         \includegraphics[width=\linewidth]{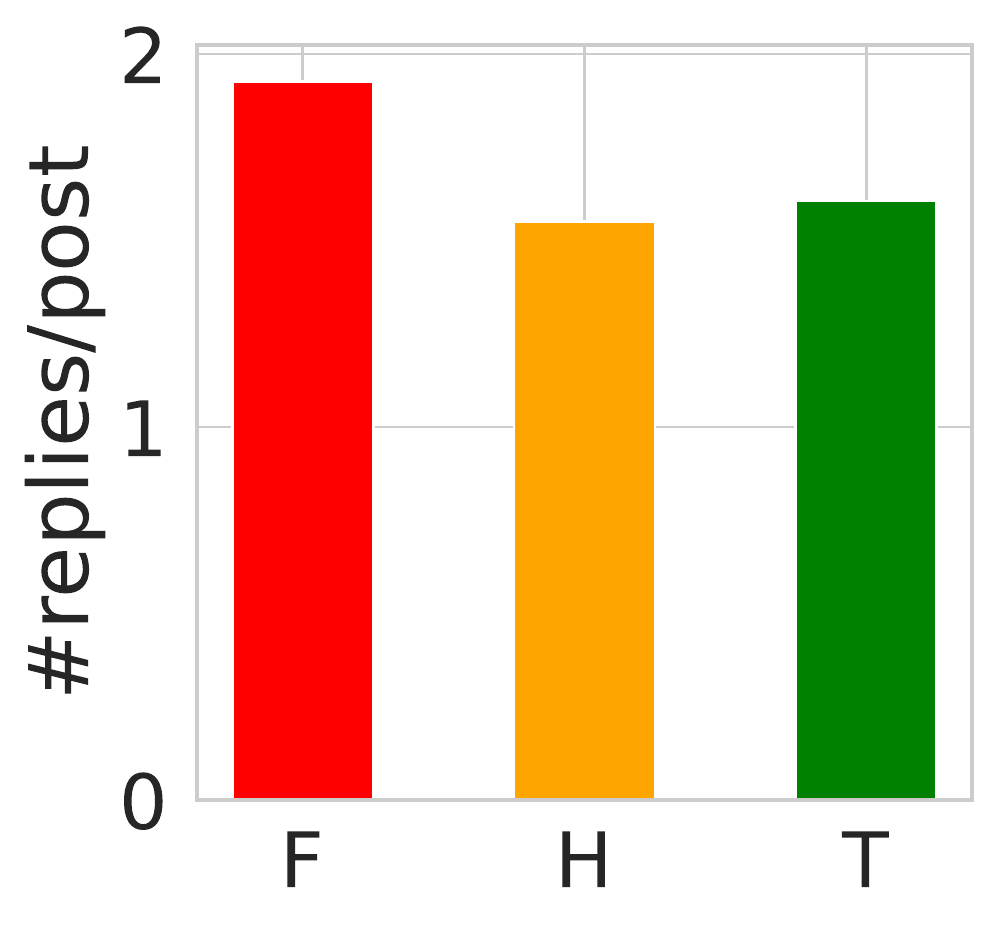}
         \caption{Mean replies per post.}
         \label{fig:mean_replies}
     \end{subfigure}
     \hfill
     \begin{subfigure}[h]{0.3\linewidth}
         \centering
         \includegraphics[width=\linewidth]{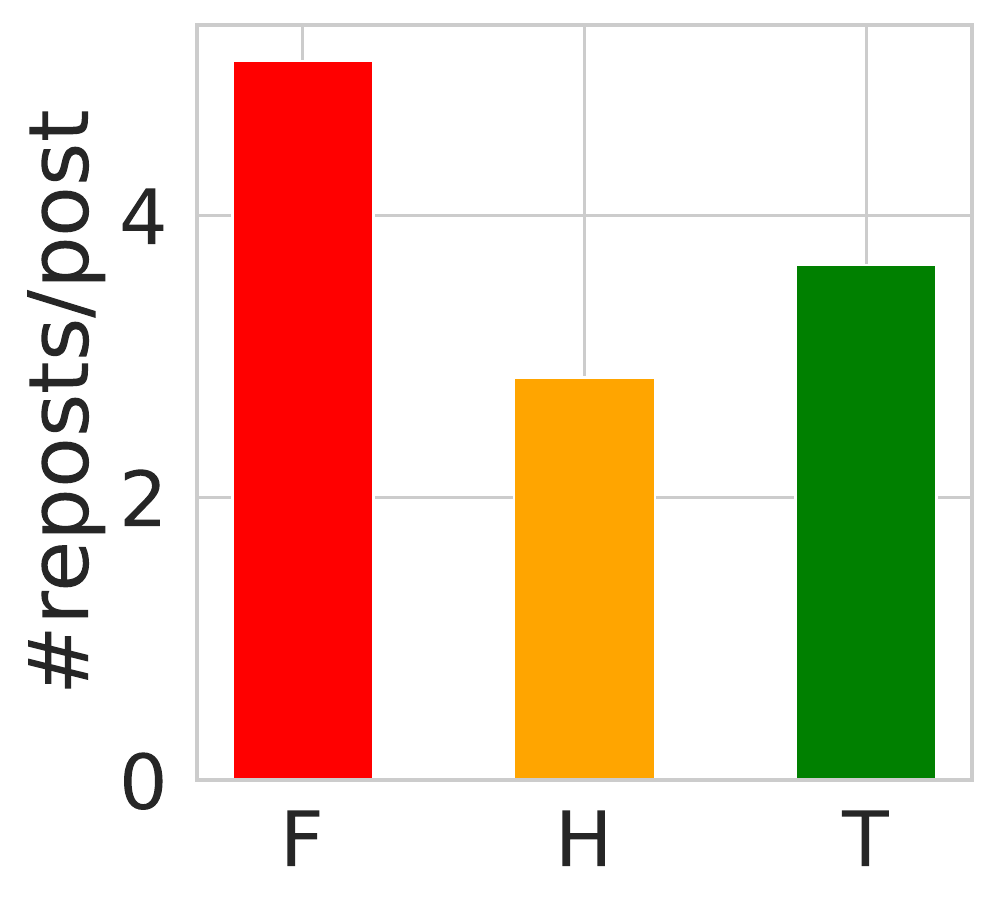}
         \caption{Mean re-posts per post.}
         \label{fig:mean_reposts}
     \end{subfigure}
        \caption{Interaction of users with posts. Here in the x-axis, we show the type of posts where F, H and T denotes fear speech, hate speech and total(overall) posts.}
        \label{fig:engagement}
\end{figure}

\subsection*{Pervasive impact of fear speech transcending to other social media platforms}

In this section, we demonstrate that the problem of fear speech is of significant general interest as it also prevails in other extensively moderated social media platforms, e.g., Twitter and Facebook. Note that the choice of these two platforms is motivated by the fact that both of them have their own strict hate speech policies in place and are constantly vigilant to remove harmful contents. We crawl large chunks of data from both these platforms and classify them as fear, hate or normal speech using our prediction model discussed earlier. Once again, we use the same confidence value based thresholding as used for the Gab dataset to designate a post to be fear/hate speech.

\subsubsection*{Twitter}

\begin{figure}
    \centering
    \includegraphics[width=\linewidth]{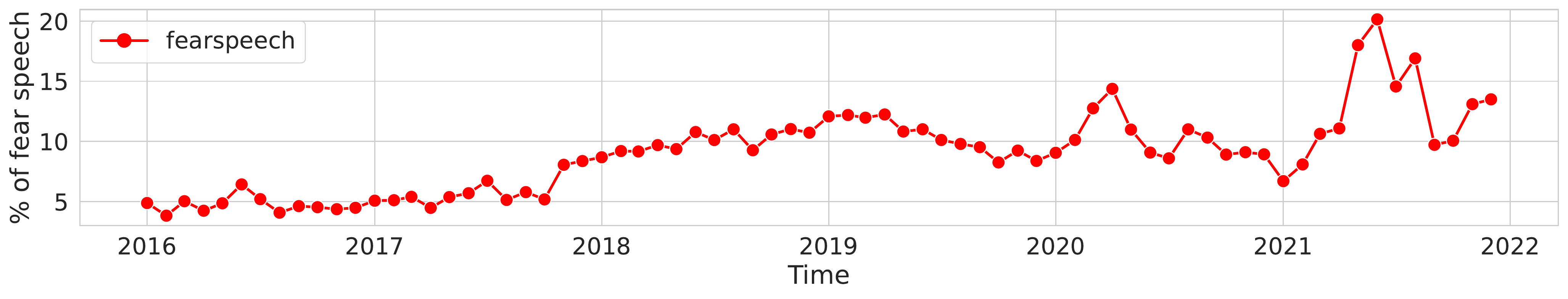}
    \caption{This figure shows the percentage of posts that were fear speech per month in the Twitter data.}
    \label{fig:temp_twitter}
\end{figure}

For Twitter, we use the topical keywords (the exact list will be shared in the repository) from the topics in Figure~\ref{fig:topic_dist_fear} and the academic research API to search through the history of tweets having those keywords. This way, we collect around 4,103,145 tweets over six years (2016-2022). We find that out of the entire dataset of around 4 million tweets, there are around 400k tweets ($\sim10\%$) were marked as fear speech by our model (examples in Table \ref{tab:twitter-fear-examples}). We further plot the timeline of the posts and find that there is an increasing trend in the number of fear speech posts (see in Figure \ref{fig:temp_twitter}) over time. The presence of such a huge volume of fear speech and its increasing temporal trend is alarming and should be analyzed by moderation policy experts. Not surprisingly, our model could only predict around 31,000 posts as hate speech which shows Twitter is quite active in moderating such hateful content.

\begin{table}[]
\begin{tabular}{p{7cm}l}
\textbf{Text}                    & \textbf{Date}      \\\hline
@AmosPosner Christians left tons of time for Jews to control media in th silence b/w the beat \&amp; when ppl yell ``Santa Claus is comin to town''                                                                              & 5/12/2016 \\\hline
MIGRANT SCANDAL: 200 illegals a DAY caught sneaking into UK - and that’s in just... https://t.co/avZNtrJyXk by \#rvaidya2000 via @c0nvey                                                                                       & 10/2/2017 \\\hline
@JudgeJeanine QUESTION PATRIOTS? ARE OUR OFFICIALS BREAKING THE LAW BY NOT UPHOLDING THE LAWS THEY WROTE, ALLOWING ILLEGAL IMMIGRATION TO OVERRUN OUR COUNTRY? IF YOU THINK SO.. SCREW BEING FIRED! HOW ABOUT CITIZENS ARREST? & 10/4/2019 \\\hline
@FBI San Diego Antifa leader calling for the killing of white men and raping white women. https://t.co/rqHhL5pxD2                                                                                                              & 28/6/2020 \\\hline
\end{tabular}
\caption{This table shows some examples of fear speech from the data collected from Twitter along with their dates.}
\label{tab:twitter-fear-examples}
\end{table}

\subsubsection*{Facebook}

\begin{figure}
    \centering
    \includegraphics[width=\linewidth]{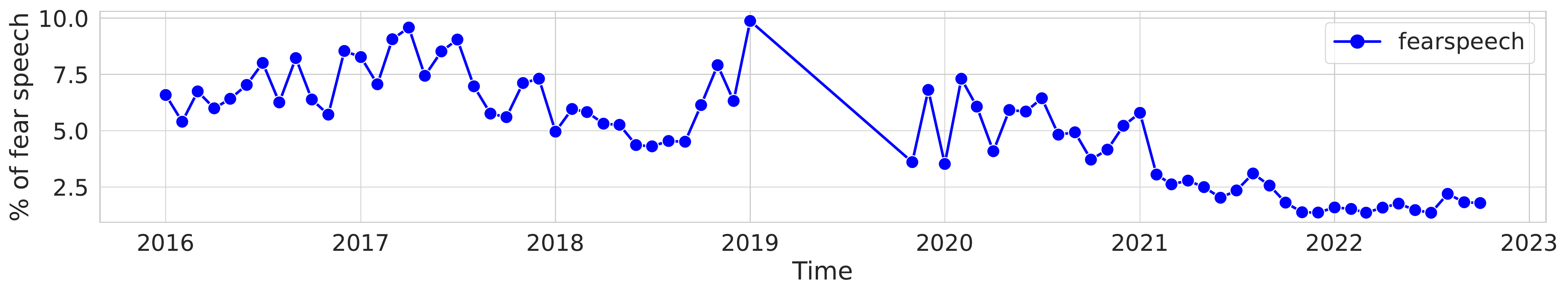}
    \caption{This figure shows the percentage of posts that were fear speech per month in the Facebook data.}
    \label{fig:temp_facebook}
\end{figure}

For Facebook, we use the historical search of CrowdTangle\footnote{\url{crowdtangle.com}} and use 73 public white supremacist pages used in the previous literature. We collect all the posts from these pages~\cite{10.1145/3313831.3376456}. This way, we obtain around 191,666 posts over six years (2016-2022). Our model predicts around 10k posts (around 4\%) are marked as fear speech (examples in Table \ref{tab:facebook-fear-examples}). We plot the timeline of the posts and find that there is a slightly decreasing trend in the number of fear speech posts (See in Figure \ref{fig:temp_facebook}). This decrease could possibly be because of the overall moderation of the white supremacist pages and not specifically the individual fear speech posts. Once again our model marked only 196 posts as hate speech pointing toward the strict hateful content moderation on Facebook.

\begin{table}[]
\begin{tabular}{p{7cm}l}
\textbf{Text}                 & \textbf{Date}      \\\hline
Have you noticed Islam is growing stronger? The "girl next door" is even jumping on the jihad train.                                                                                                                 & 18/5/2017 \\\hline
\#Turkey says its released 47,000 migrants into \#Europe. That us 47,000 on a \#hijrah most are men of military age. https://www.trtworld.com/turkey/number-of-migrants-leaving-turkey-reaches-47-113-minister-34211 & 29/2/2020 \\\hline
(Bangladesh: Muslims threaten to murder atheist blogger for criticizing political Islam, defending Buddhists) has been published on Jihad Watch                                                                      & 26/8/2020 \\\hline
Most attention goes to illegal aliens crossing by land, but data shows rising numbers trying to come by water.                                                                                                       & 14/1/2021 \\\hline
\end{tabular}
\caption{This table shows some examples of fear speech from the data collected from Facebook along with their dates.}
\label{tab:facebook-fear-examples}
\end{table}

We believe that these results together point to the pervasive nature of the problem and the necessity for special all-round attention from the community.

\section*{Discussion}

In this study, we aimed to understand what role fear plays in the polarized conversations and how it differs from the traditional form of polarized content --- hate speech. We find significant difference in how extreme hate speech (ExHate) and fear speech users (ExFear) exist and interact with other users in the network. ExFear users have more followers and can effectively interact with the general audience than the ExHate users. This indicates out that even within a polarized, hateful context, fear has a different reach in the audience. In the correct context, such type of polarized content may act as tipping points~\cite{buyse2014words,macy2021polarization} during some event. This is especially so when there are groups of coordinated actors who are interested in propagating an agenda~\cite{saha2021short}. Hence, it becomes important for the research community to understand how to moderate such different forms of extreme content. It is also interesting to think about some prioritization when moderating different forms of extreme content.

One of the main reasons why these difference exists is the language of the text used. While fear speech uses arguments and subtle ways to show some community as a threat, hate speech~\cite{hateoffensive} uses slurs, insults to dehumanize the community. There is a huge body of work for how hate speech spreads in social media and can be analysed~\cite{mathew2019spread,founta2018large,ribeiro2018characterizing}, detected in mono-lingual~\cite{mathew2020hatexplain,caselli-etal-2021-hatebert,koufakou-etal-2020-hurtbert,davidson2017automated} and multilingual scenarios~\cite{aluru2020deep,wang-etal-2020-detect,ousidhoum2019multilingual,ranasinghe-zampieri-2020-multilingual} and mitigated using suspension~\cite{ali2021understanding} and counter speech~\cite{chung-etal-2019-conan,qian2019benchmark,fanton2021human}. The presence of fear speech will create problem while deciding about the moderation policies because we might not be able to directly ban or suspend fear speech. The paper introducing the `fear speech' concept~\cite{buyse2014words} suggests creation of alternative arguments to the arguments given in fear speech. These alternative arguments should aim at diffusing the violence potential of fear speech. Since many instances of fear speech may also contain misinformation to exaggerate their arguments, researchers in misinformation~\cite{muhammed2022disaster} domain, news media and fact-checking organisations can play an important role. However, even such measures might not be effective unless the end user is aware. Hence, awareness events, similar to the ones done for hate speech~\cite{Raisinga44:online}, should be conducted to make the users question the content they are receiving.

Past research in this community has focused on the role of social media in polarization~\cite{asimovic2021testing,huszar2022algorithmic} and the role of user accounts in spreading such content~\cite{simchon2022troll}. Our research takes a step back and tries to understand the types in which polarization happens and whether there exists a difference in the audience using/perceiving it.\\
One limitation of our study is that our detection model is trained on Gab data. Further, most of the prevalence analysis is also based on the Gab data. Our choice of Gab as a social media platform is motivated by its unmoderated nature, which makes studying hate speech easier. Further obtaining certain nuanced data such as the time-varying structure of the followership network is easily possible for Gab. To compliment this study further, we perform some basic prevalence analysis on Twitter and Facebook as well and find significant amount of fear speech in these platforms. Our study renders hope that the investigation of fear speech can be easily extended to other platforms. Nevertheless, it seems as these platforms continue moderating hate speech, actors spreading such content might shift to more subtle ways like fear speech. Moreover, the content on `fringe' platforms does not stay only on those platforms anymore, and we have seen instances of seemingly fringe platforms affect main stream conversations~\cite{zannettou2017web}. Second, in an effort to scale up our findings to millions of posts, our study relies on the performance of automated classifiers on an inherently difficult task. While we have taken additional care while deciding the category of the posts, we might be missing some form of fear speech/hate speech. Section~\ref{sec:allmodels} provides robustness checks on our models.

\matmethods{
This section provides details on data collection, annotation and labeling, and user-level classification. 
Statistical tools used in this analysis are noted in section~\ref{sec:stats}.
\subsection*{Data Annotation}

There are different forms of toxic speech on social media. In this work, we primarily target hate speech and fear speech. For each post shown to the annotators, the annotator has to mark whether the speech is fear speech or hate speech, or normal. Further, they also need to mark the target communities towards which the particular posts are targeted. See \appendixnewname~in section \ref{sec:anno-guide} for more details.

To annotate the posts, we follow a hybrid strategy comprising both expert and crowd annotators. The expert annotators are a group of 4 undergraduate students who were trained using gold label annotations and detailed discussion sessions. The crowd workers were recruited from Amazon Mechanical Turk. We use a multi-label annotation framework, where a post can be assigned to both fear speech and hate speech.

To finalize the annotation guidelines and difficulty of the annotations, we first annotated a set of 1000 posts using the expert annotators. The expert annotators achieve a set of 0.51 Krippendorff's alpha. Next, we created using a pilot study to select the crowd workers. The pilot study is a set of 15 gold annotated posts from these 1000 data points used to test the Mechanical Turk workers who agreed to take part in our study. Out of 400 interested crowd workers, we selected 192 annotators. Of these annotators, 103 participated in the study. 

The annotation process comprised 24 rounds. In each round, we gave a fixed number of posts to annotate. The number of posts per round was kept low, around 150 initially and finally increased to 500 as the annotators became more familiar with the task at hand. For sampling the posts, we employed different strategies. Initially, our strategy revolved around using community-based keywords. In each round, we removed the keywords that gave more normal samples in order to retrieve more fear speech/hate speech posts.

\if{0}\subsection*{Human evaluation}

\updated{We select around 100 fear speech posts and 100 hate speech posts from the annotated dataset where all the annotators agree on the final label. We created a survey by posing pairs of fear speech and hate speech and asking the annotator to select the post they believed in more.  Each pair of posts was annotated by 9 annotators from Amazon mechanical Turk. In total, 243 unique annotators participated in the task. For each such pair we find how many annotators believed in fear speech and how many in hate speech. The information about the label of the posts were hidden from the annotators. Further, the pairs were sufficiently randomized so that the annotators cannot guess the labels of the posts.}\fi

\subsection*{Post classification}

We develop a bunch of classification models for this task. As baselines, we use three different feature extraction techniques -- bag of words vectors (\textbf{BoW}), GloVe word embeddings (\textbf{WE}) and \textbf{TF-IDF} features. We then use two one-vs-rest classifier -- logistic regression (\textbf{LR}), support vector classifier (\textbf{SVC}) as well as \textbf{XGBoost}. Additional details of the baseline models are noted in section~\ref{sec:allmodels}. 

Transformers are a recent NLP architecture formed using a stack of self-attention blocks having superior performance across a lot of benchmarks. We use several variations of the transformer models --- (i) pretrained models like \texttt{bert-base-uncased}, roberta-base (ii) models which are fine-tuned using data from hate speech related tasks like HateXplain~\cite{mathew2020hatexplain}, Twitter-roberta-hate~\cite{barbieri2020tweeteval} etc. (iii) models which are pretrained using social media dataset - \updated{HateBERT~\cite{caselli2021hatebert}}. In the category (iii), we also use a filtered out version of the Gab dataset to pretrain a \texttt{bert-base-uncased} model further and name it Gab-BERT~\footnote{We shall release this model upon acceptance of the paper}. All these models are added with a classification head. Gab-BERT is the best model among all others with a macro F1 score of 0.62 (the full set of results for all the models are presented in section~\ref{sec:allmodels}). 

We further hypothesize that hate speech and fear speech might show different forms of emotions. We use an emotion vector predicted using the model used in previous research work~\cite{demszky-etal-2020-goemotions}. This additional input vector increases the performance of the Gab-BERT model by 1 point for F1 score and 4 points for accuracy. %

\subsection*{User analysis}
\label{sec:user}
To conduct the user analysis, we wanted to understand the characteristics of the extreme fear and extreme hate users. To do this we find the users in the top 10\% percentile in terms of number of fear speech posts and hate speech posts separately. We remove the intersection of the users in both these sets. Finally, we end with 476 extreme fear speech (ExFear) and 478 extreme hate speech users (ExHate). We further perform the study on an extended set of users as well (noted in \appendixnewname~section~\ref{sec:user-analysis}).

\subsection*{Temporal movement of users}
\label{sec:temporal}

To understand the temporal influence of the users over the entire timeline, we utilise the follower-followee network per month which was referred to in~\cite{10.1145/3415163}. Then for each month we calculate the $k$-core or coreness metric~\cite{batagelj2011fast} to identify the influential users in the undirected version of the follower-followee network. Next we subdivide the nodes into 10 buckets based on their percentile ranks in terms of $k$-core value, i.e., the bottom 10\% percentile to the top 10\% percentile. Following this, we measure the time in months for a user to reach the inner core (core-0) in the network (further referred to as \textbf{time-to-reach-core}) from the time they join the network.

\subsection*{Data Availability Statement}
A repository of the data necessary to reproduce, analyze, and interpret all findings in this paper is available \url{https://osf.io/dc7vu/?view_only=8144833546e54a399ab883f0b0e3e7f7}. The code (including software information) for all studies and the analysis is available \url{https://github.com/punyajoy/Fearspeech-project}.

}

\showmatmethods{}

\if{0}

\begin{figure}
    \centering
    \includegraphics[width=0.5\linewidth]{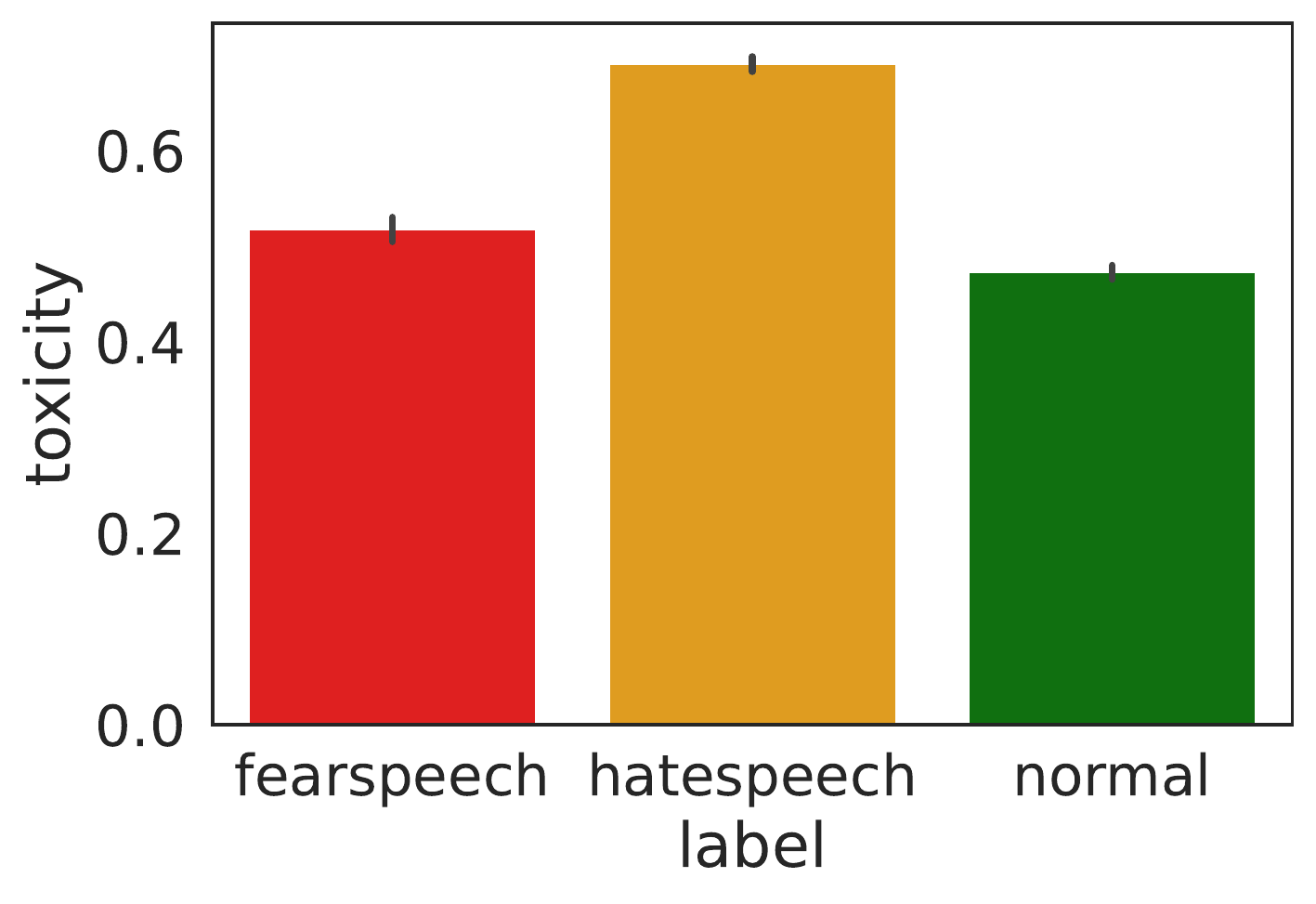}
    \caption{Perspective API based toxicity scores for fear speech, hate speech and normal posts. Error bars indicate  95\% confidence interval}
    \label{fig:toxicity}
\end{figure}

\if{0}
\begin{equation}
    \label{eq:1}
    var = \alpha*(\# fear speech) + \beta*(\# hate speech) + \gamma * (\# total)
\end{equation}
At first, we look at the users' positions in the follower-following network created at the end of the timeline (June 2018). In total, the graph had 279,961 nodes and 1.960,869 edges. Next, we calculate the \textbf{betweenness} and \textbf{eigen-vector} centrality of the undirected version of the former graph. While eigenvector centrality helps us understand the influence of the nodes, betweenness centrality helps us understand the degree to which nodes stand between each other. We further look at the number of \textbf{followers} and \textbf{followings} for each user. To check the influence on the normal users, we also noted the number of \textbf{normal followers} for each user. 

Since the following-followee network might not reveal everything about the users, we also make a hybrid interaction network considering three types of edges, i.e., reposts, replies, and mentions. A repost edge from A to B refers to B reposting a post of A. Similarly, a reply edge from A to B refers to B replying to A's posts, and finally, a mention edge from A to B refers to A mentioning B in their post. The weight of each of these edges notes the number of times such interaction took place. In total, the graph had around 109,202 nodes and 8,061,653 edges.

Next, we consider the users' interactions selected initially with the normal users. For each node in the set of 9,200 users,  we note the \textbf{average number of normal users} to who the users have interacted at least once. We also consider the \textbf{total interaction with the normal users/user} i.e., by taking a weighted sum of the interactions with normal users. We note these metrics for the reposts edges, mention edges, and reply edges. 

Apart from the overall user analysis, we also study the hate speech and fear speech extreme users. To find the users, we first find the 90\% percentile users in terms of the number of fear speech posts and hate speech posts. We remove the intersection of the users in both these sets. Finally, we end with 476 extreme fear speech (ExFear) and 478 extreme hate speech users (ExHate). We compare the former variables for this set of users as well.\fi

\section*{Results}

\subsection*{User analysis} 

At first, we analyse how using fear speech/hate speech influence the characteristics of the users. We inspect the coefficients of the linear regression model as described in the Materials and Methods section. We note the coefficients in the table~\ref{tab:coeff}. In terms of position in the follower-followee network, we find that number of fear speech posts shows a slight positive correlation with the centrality measures. In contrast, the number of hate speech posts shows a slight negative correlation. The correlation trend is similar for followers and following, although the coefficients are larger. In terms of time to reach the core~\kg{isnt this out of place?}, we notice a positive correlation with the number of hate speech posts. 

In terms of the number of normal followers, we find the number of fear speech posts correlates positively while the number of hate speech. Next, we consider the interaction with the normal followers. We find that the number of fear speech posts positively correlates with the number of normal reposters and total interaction with normal users per user. In contrast, the number of hate speech has a negative correlation. Similar results are found for replies from normal users. We see a negative correlation between hate speech and fear speech in terms of normal users mentioned. The total number of normal mentions per user is again positively correlated with the number of fear speech posts and negatively correlated with the number of hate speech posts~\kg{are we not including the prevalence over time figures? of hate and fear? like why. whats the mechanism, whats happening underneath the hood}.

\begin{table*}[!htpb]
\centering
\begin{tabular}{llllll}\hline
\textbf{Variable} & \textbf{Coeff of \#fear} & \textbf{Coeff of \#hate} & \textbf{Coeff of \#total} & \textbf{Obs} & \textbf{Mean Dep Var} \\\hline
\multicolumn{6}{c}{\textbf{Basic network properties}}                                                  \\\hline
\textbf{Eigen Vector}                    & 4.12e-06***       & -1.743e-06***  & 4.319e-07***  & 9245 & 0.004    \\
\textbf{Betweenness}                     & 1.668e-07***      & -2.019e-07*    & 2.442e-08***  & 9245 & 5.27e-05 \\
\textbf{Followers}                       & 0.0012***         & -0.0006***     & 0.0002***     & 9245 & 1012     \\
\textbf{Following}                       & 0.0010***         & -0.0005***     & 0.0002***     & 9245 & 926      \\
\textbf{Time to reach core 0}            & -8.849e-05        & 0.0002**       & -5.117e-05*** & 6700 & 3.52     \\\hline
\multicolumn{6}{c}{\textbf{Effect on normal users}}                                                    \\\hline
\textbf{Num of normal followers}         & 0.0009***         & -0.0008***     & 0.0002***     & 9170 & 204      \\
\textbf{Unique normal reposters}         & 0.0030***         & -0.0009***     & 0.0003***     & 8772 & 25       \\
\textbf{\# reposts by normal users}   & 0.0036***         & -0.0013***     & 0.0003***     & 8772 & 43       \\
\textbf{Unique normal mentions}          & -0.0004***        & -0.0013***     & 0.0003***     & 8756 & 9        \\
\textbf{\# mentions of normal users}  & 0.0022***         & -0.0015***     & 0.0004***     & 8756 & 24       \\
\textbf{Unique normal repliers}          & 0.0013***         & -0.0008***     & 0.0003***     & 9233 & 13       \\
\textbf{\# replies of normal users} & 0.0007***         & -0.0008***     & 0.0003***     & 9237 & 19       \\\hline
\end{tabular}
\caption{This table notes the coefficient values for number of hate speech, number of fear speech and number of total posts per user for different dependent variables. We also note the number of observations and average value of the dependent variable for the entire dataset.}
\label{tab:coeff}
\end{table*}

\if{0}
To further establish our results, we look into the extreme users in terms of fear speech (\textit{ExFear} users) and hate speech (\textit{ExHate} users) posts (Described in section~\ref{sec:user}). We find that the mean eigen vector centrality is higher for the \textit{ExFear} users compared with \textit{ExHate} users (MW U, $p < 0.001$). Similarly, we find that mean betweenness centrality is higher for the \textit{ExFear} compared to \textit{ExHate} users (Figure \ref{fig:betweenness}). More fine-grained analysis on the follower-followee distirbution shows that both the followers and followings are higher for \textit{ExFear} and \textit{ExHate} users  (MW U, $p<0.0001$ , Figure ~\ref{fig:follower_following}).

\begin{figure}[!htpb]
     \centering
     \begin{subfigure}[h]{0.45\linewidth}
         \centering
         \includegraphics[width=\textwidth]{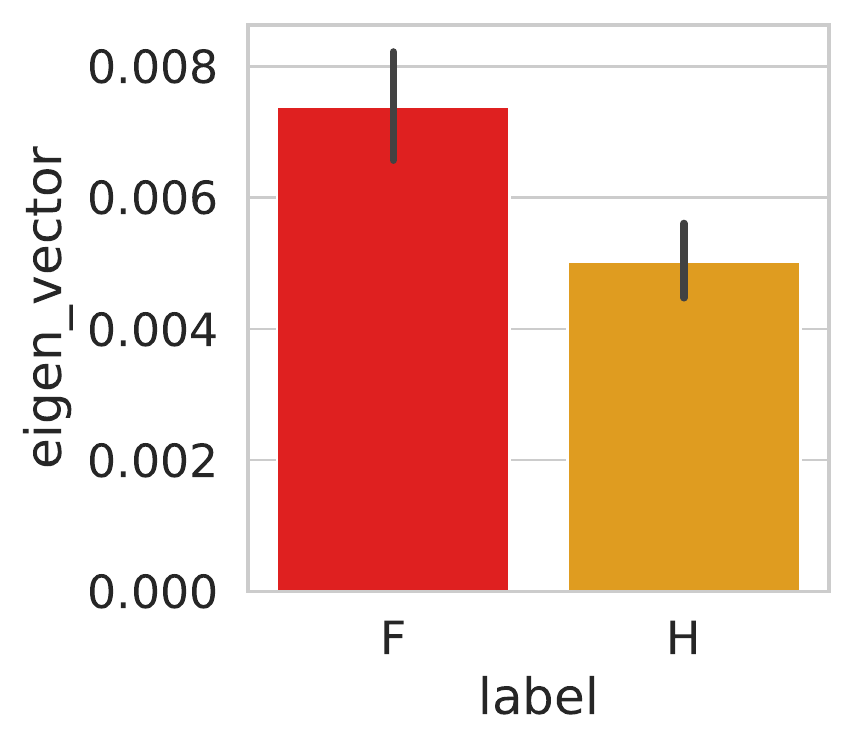}
         \caption{Eigen vector centrality of users}
         \label{fig:eigen_vector}
     \end{subfigure}
     \hfill
     \begin{subfigure}[h]{0.45\linewidth}
         \centering
         \includegraphics[width=\textwidth]{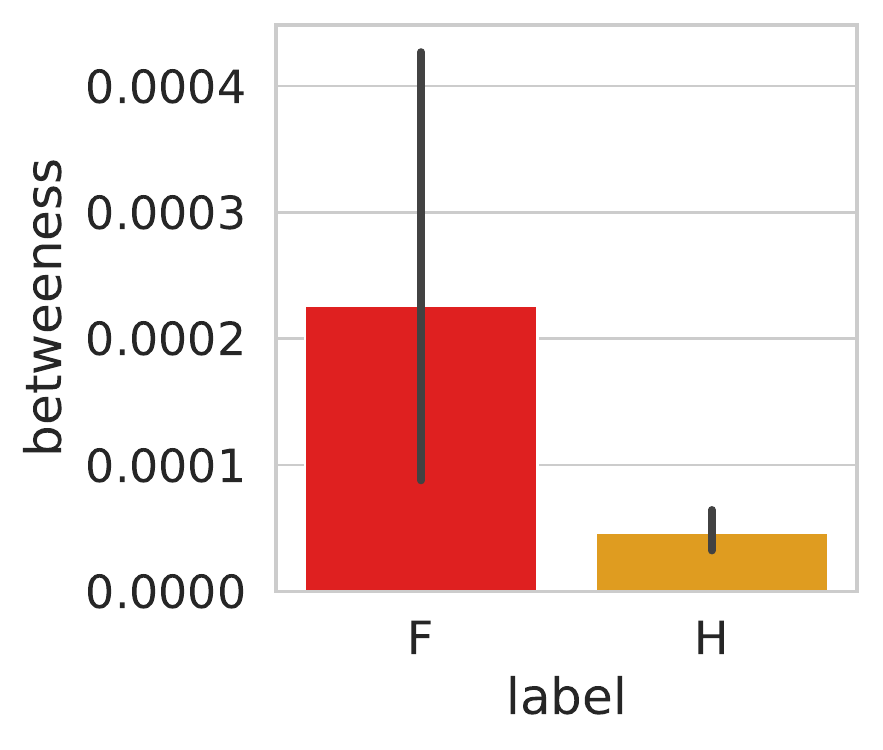}
         \caption{Betweenness centrality}
         \label{fig:betweenness}
     \end{subfigure}
    \caption{Centrality measures of ExFear (F) and ExHate (H) users  }
    \label{fig:network_metrics}
\end{figure}

\begin{figure}[!htpb]
     \centering
     \begin{subfigure}[h]{0.45\linewidth}
         \centering
         \includegraphics[width=\textwidth]{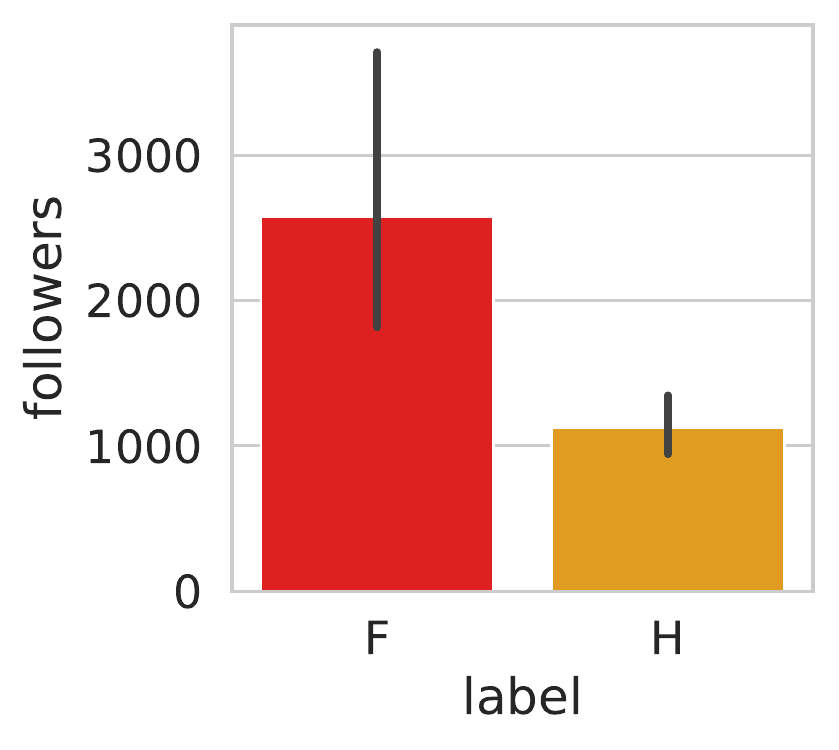}
         \caption{Average number of followers for different groups of users}
         \label{fig:followers}
     \end{subfigure}
     \hfill
     \begin{subfigure}[h]{0.45\linewidth}
         \centering
         \includegraphics[width=\textwidth]{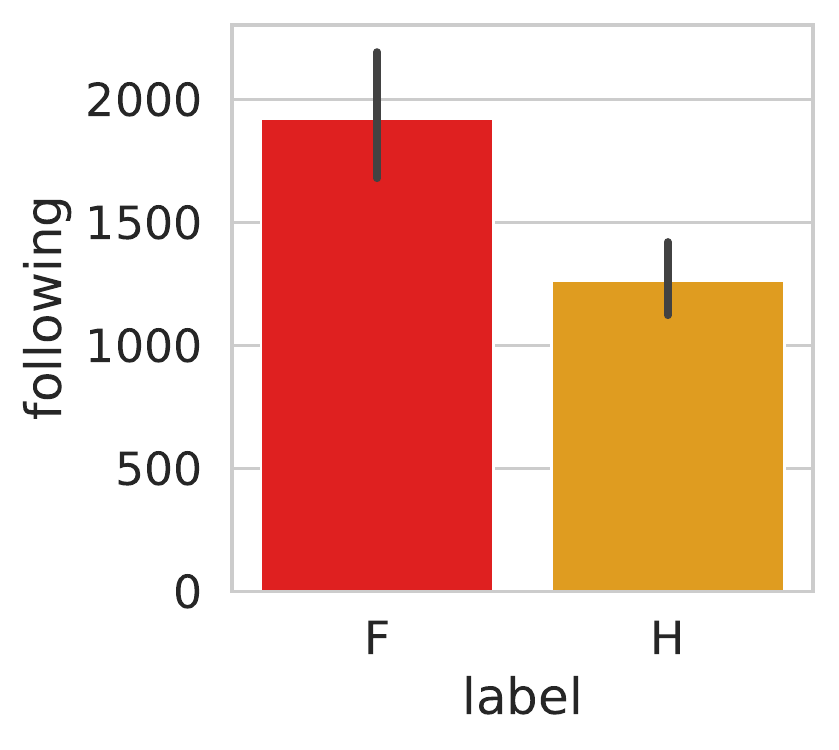}
         \caption{Average number of followings for different groups of users}
         \label{fig:following}
     \end{subfigure}
    \caption{Plots denoting follower-following properties for ExFear (F) and ExHate (H) users}
    \label{fig:follower_following}
\end{figure}

\fi

\if{0}Next, we look into how \textit{ExFear} and \textit{ExHate} users are affecting the normal users. We find the average ratio of normal followers for \textit{ExFear} users (21\%)  is higher than \textit{ExHate} users (18\% ,MW U, $p < 1e^{-6}$). More number of normal users reposts posts of fear speech users as compared to hate speech users. Further total number of reposts per user is also higher for the fear speech users. We observe a similar trend for the normal users mentioned by the \textit{ExFear} and \textit{ExHate} users and normal users replying to the posts of \textit{ExFear} and \textit{ExHate} users.%

Overall, from both these experiments, we notice that users using more fear speech cover more central positions than users using more hate speech. Further, they have a more substantial influence on normal users. This is established by the higher number of normal followers for users using more fear speech. Users using more fear speech can interact with more normal users, have more interactions from normal users, and more normal users mentioned in their posts~\kg{are we not including the prevalence over time figures? of hate and fear?}.

\begin{figure}[!htpb]
     \centering
     \begin{subfigure}[h]{0.45\linewidth}
         \centering
         \includegraphics[width=\textwidth]{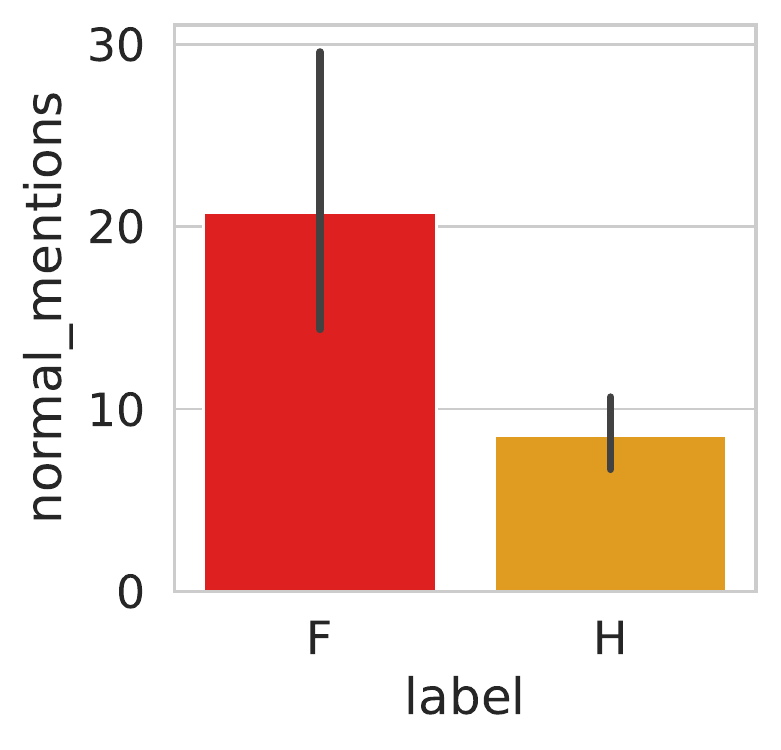}
         \caption{number of normal mentioned users per user}
         \label{fig:normal_mentions}
     \end{subfigure}
     \hfill
     \begin{subfigure}[h]{0.45\linewidth}
         \centering
         \includegraphics[width=\textwidth]{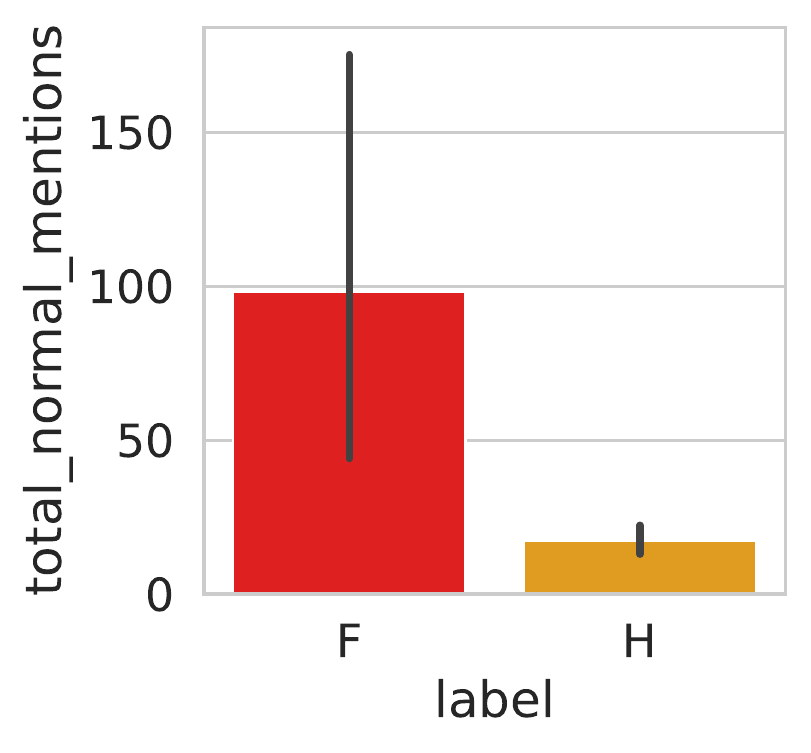}
         \caption{number of normal mentions per user}
         \label{fig:total_normal_mentions}
     \end{subfigure}
    \caption{distribution of normal mentions for ExHate (H) and ExFear (F)}
    \label{fig:mentions}
\end{figure}

\begin{figure}[!htpb]
     \centering
     \begin{subfigure}[h]{0.45\linewidth}
         \centering
         \includegraphics[width=\textwidth]{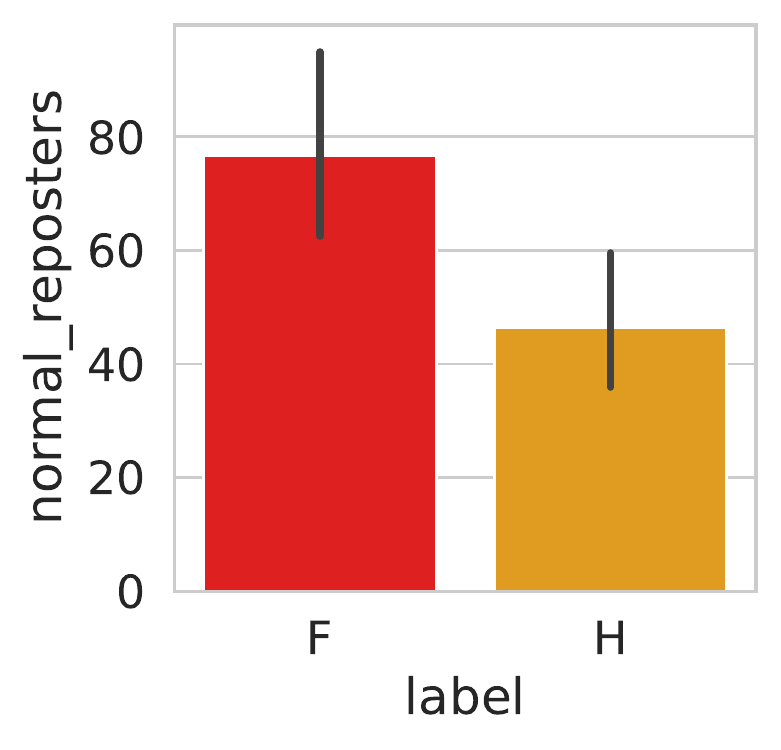}
         \caption{number of normal reposters per user}
         \label{fig:normal_reposts}
     \end{subfigure}
     \hfill
     \begin{subfigure}[h]{0.45\linewidth}
         \centering
         \includegraphics[width=\textwidth]{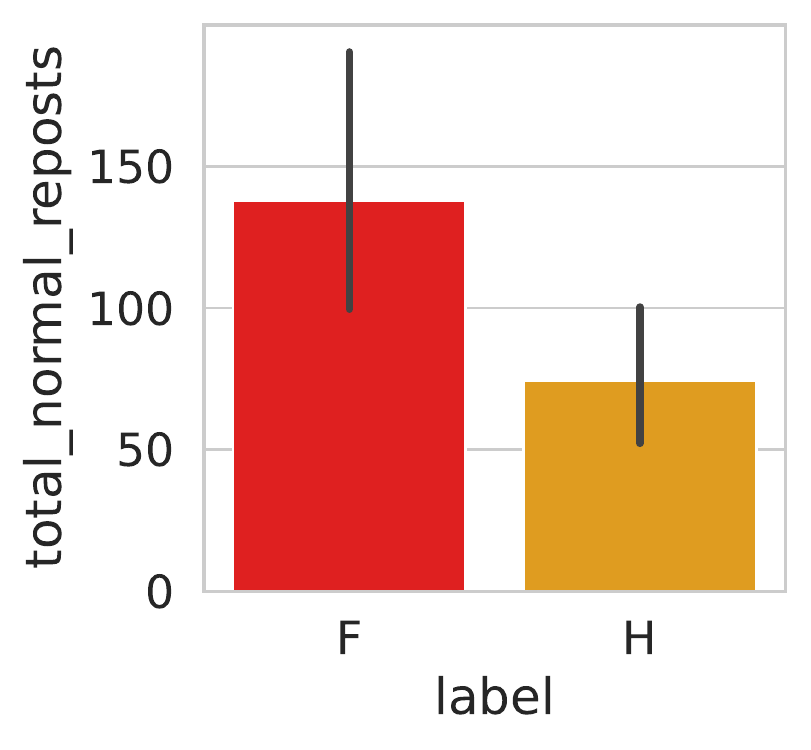}
         \caption{\# of reposts from normal users per user}
         \label{fig:total_normal_reposts}
     \end{subfigure}
    \caption{Distribution of reposts from normal users for ExFear (F) and ExHate (H) users}
    \label{fig:reposts}
\end{figure}

\begin{figure}[!htpb]
     \centering
     \begin{subfigure}[h]{0.45\linewidth}
         \centering
         \includegraphics[width=\textwidth]{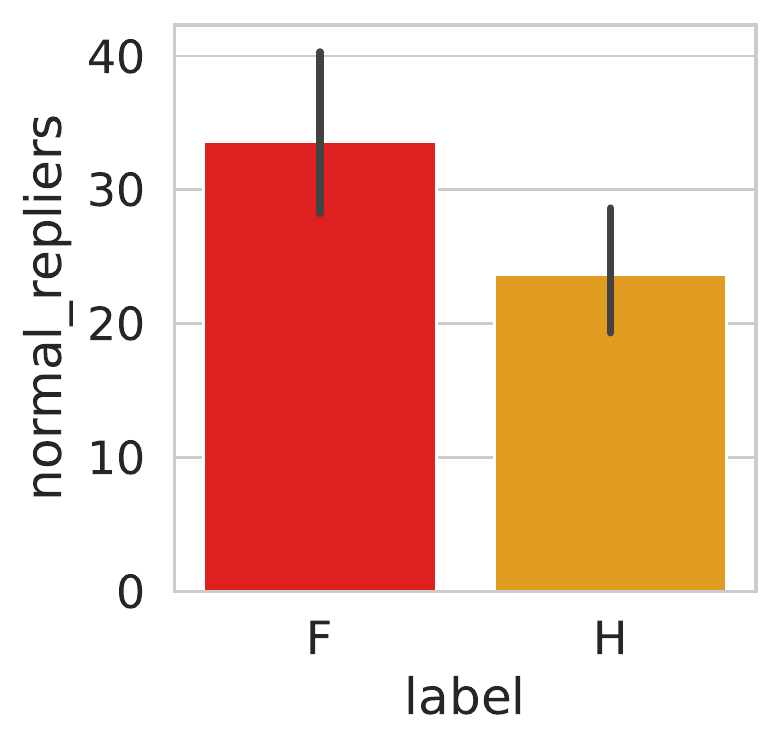}
         \caption{number of normal repliers per user}
         \label{fig:normal_replies}
     \end{subfigure}
     \hfill
     \begin{subfigure}[h]{0.45\linewidth}
         \centering
         \includegraphics[width=\textwidth]{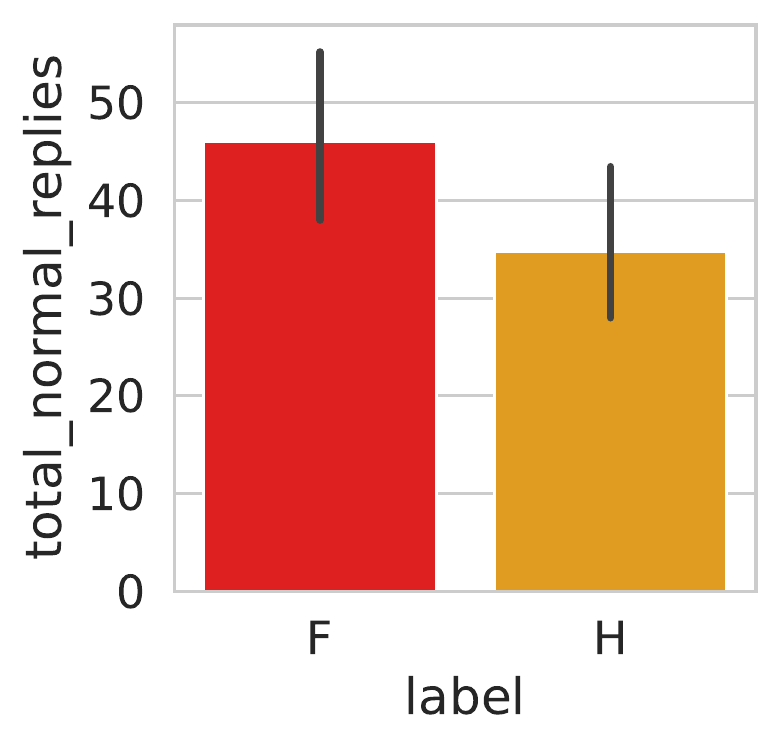}
         \caption{\# of replies from normal users / user}
         \label{fig:total_normal_replies}
     \end{subfigure}
    \caption{Distributions of replies from normal users for ExFear (F) and ExHate (H) users}
    \label{fig:replies}
\end{figure}

\fi

\subsubsection*{Interaction with the posts} Interaction with any post can be an essential indicator of how the audience is engaging with the posts. We measure this using the retweets, replies, and likes frequency. Here, we compare these interactions for the fear speech and hate speech posts. To set a baseline, we also compare them with the overall level of interaction for all the posts. 

In terms of \textbf{likes}, we find that $\sim$ 65\% of posts are liked by at least one user out of the posts labeled as fear speech and hate speech. This is slightly more than overall posts, where at least one user likes less than $\sim$ 60\% posts. Further, we find that the average number of likes for fear speech is around $\sim$ 7 per post, which is much more than hate speech (MW U, one-sided, $p < 1e^{-6}$). This is illustrated in Figure~\ref{fig:mean_likes}.

In terms of \textbf{replies}, we find that around $\sim$ 16\% of the posts have at least one reply for fear speech and hate speech. Again this is slightly higher than the overall posts ($\sim$ 10\%). The mean number of replies per post is again higher for fear speech as compared to hate speech (MW U, one-sided, $p < 1e^{-6}$). This is illustrated in Figure~\ref{fig:mean_replies}.

In terms of \textbf{reposts}, we observe that more number fear speech posts ($\sim $18\%) are reposted as compared to hate speech and overall posts ($\sim$11-13\%). The average number of reposts per post is much higher for fear speech (5 per post) than hate speech (3 per post, MW U test, one-sided,  $p < 1e^{-6}$). This is illustrated in Figure~\ref{fig:mean_reposts}.

In summary, we find that the percentage of posts getting replies, reposts, and likes for fear speech and hate speech is much higher than posts in the total Gab posts. The average engagement of fear speech posts is much higher than hate speech posts.

\subsection*{Post Level Analysis} In order to understand why such difference exists in terms of user characteristics and interactions with the posts, we try to analyse difference between the fear speech and hate speech using different linguistic experiments.

\begin{figure}[h]
     \centering
     \begin{subfigure}[h]{0.3\linewidth}
         \centering
         \includegraphics[width=\linewidth]{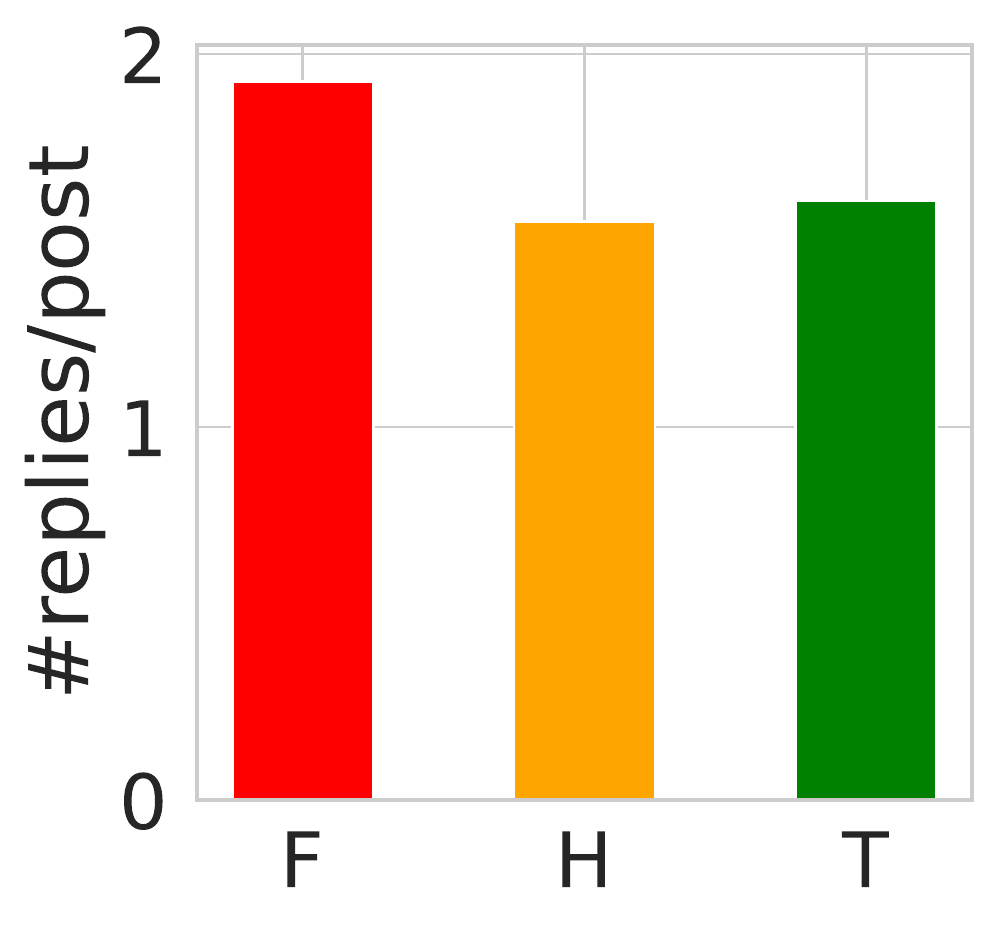}
         \caption{mean replies per post}
         \label{fig:mean_replies}
     \end{subfigure}
     \hfill
     \begin{subfigure}[h]{0.3\linewidth}
         \centering
         \includegraphics[width=\linewidth]{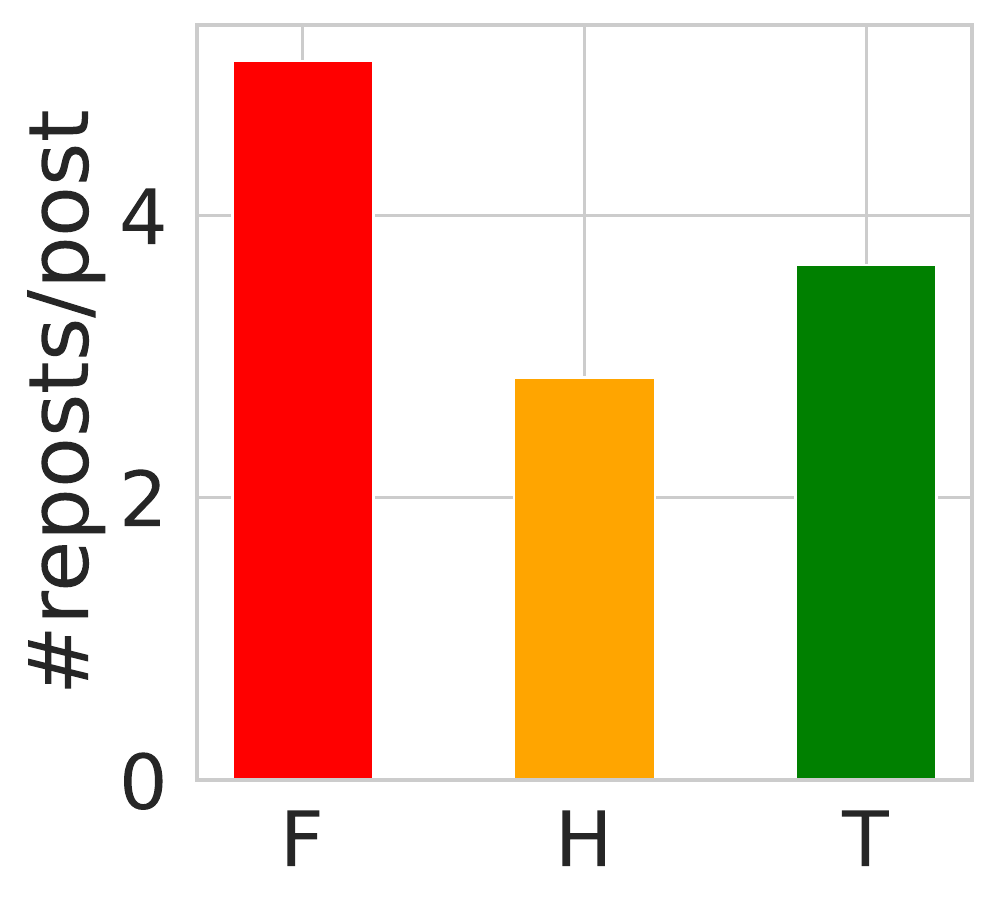}
         \caption{mean reposts per post}
         \label{fig:mean_reposts}
     \end{subfigure}
     \hfill
     \begin{subfigure}[h]{0.3\linewidth}
         \centering
         \includegraphics[width=\linewidth]{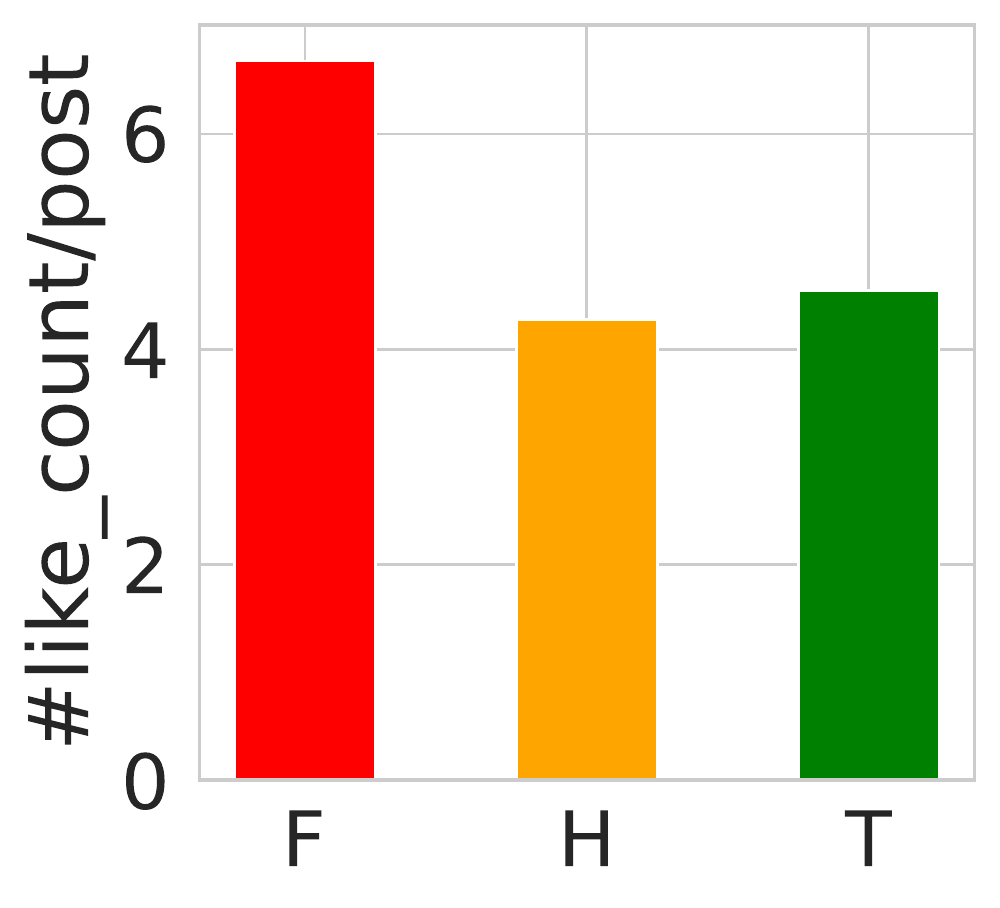}
         \caption{mean likes per post}
         \label{fig:mean_likes}
     \end{subfigure}
        \caption{Average engagements of posts belonging to fearspeech(F), hatespeech(H) and overall posts(T).}
        \label{fig:engagement}
\end{figure}

\subsection*{Textual properties} Next, we examine the linguistic properties of hate and fear speech posts. Understanding the linguistic properties can help us understand how these posts differ on a larger scale~\kg{I am not sure if this section is needed. If we want to keep it, we can remove the emoji analysis (our previous findings could be specific to WhatsApp), and dig deeper into understanding why these differences exist and what they mean in practice. not just presenting the stat}. 

As noted in the paper~\cite{saha2021short}, fear speech often contains \textbf{emojis} with hidden meanings to target the target communities. We extracted the emojis using the emoji package by detecting the Unicode of the emojis. After analyzing the emojis in the posts, we observed that emojis are used less in the Gab posts (less than 3\%). The use of emojis in fear speech is further less, around 1.7\% of posts containing emojis compared to normal and hate speech posts (around 3.5\% posts). The average number of emojis for the posts having emojis are noted in Figure \ref{fig:mean_emojis}. 

The average number of \textbf{web links} used in fear speech posts or any other posts is found to be similar and is closed to 1. But $\sim$ 30\% of fear speech posts have web links in it compared to $\sim$ 36\% of overall posts. The percentage of hate speech posts with a web link is further less at around $\sim$ 15\%. This clearly shows that the presence of web links is more important for fear speech posts (as noted in Figure \ref{fig:mean_urls}).

The usage of \textbf{hashtags} also varies across different types of posts. Hashtags encourage people to take part in conversations on the same topic, and people engage with one another even if they are not following each other~\cite{enli2018social}. Around 25\% of fear speech posts have at least one hashtags which is much higher than hate speech($\sim$11\%) and overall posts ($\sim$16\%). Among the posts having at least one hashtags (noted in \ref{fig:mean_hashtag}), a fear speech post has around 3.5 hashtags per post compared to about 2.5 hashtags per post (Mann-Whitney U test $p<0.0001$). Hence, there is more significant usage of hashtags in fear speech posts compared to hate speech posts.

\begin{figure}[h]
     \centering
     \begin{subfigure}[h]{0.3\linewidth}
         \centering
         \includegraphics[width=\linewidth]{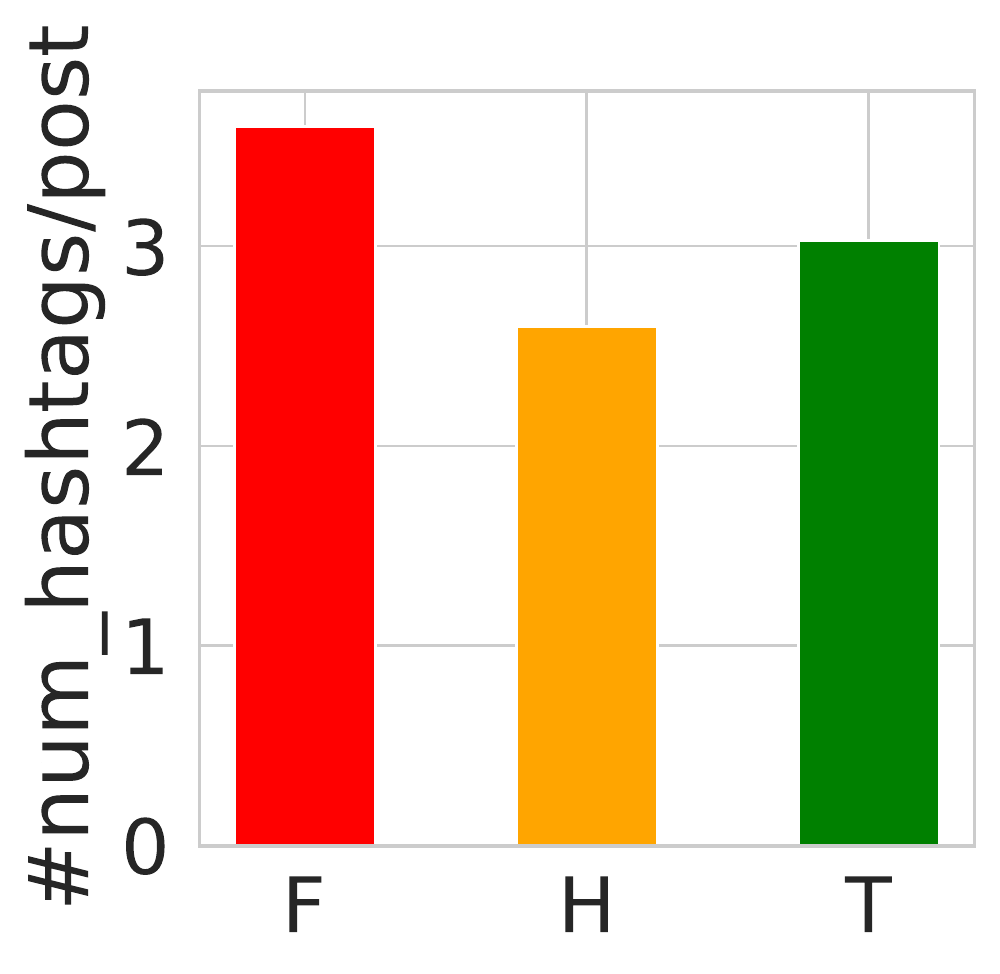}
         \caption{mean hashtags / posts}
         \label{fig:mean_hashtag}
     \end{subfigure}
     \hfill
     \begin{subfigure}[h]{0.3\linewidth}
         \centering
         \includegraphics[width=\linewidth]{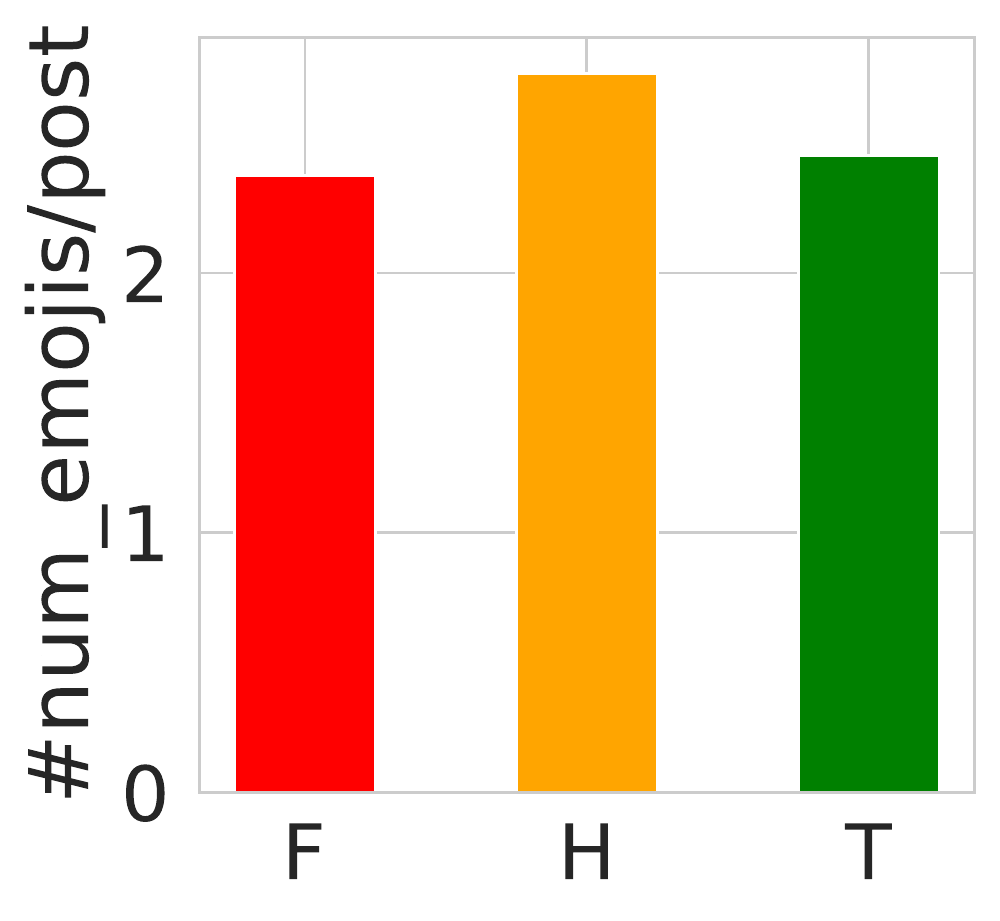}
         \caption{mean emojis / posts}
         \label{fig:mean_emojis}
     \end{subfigure}
     \hfill
     \begin{subfigure}[h]{0.3\linewidth}
         \centering
         \includegraphics[width=\linewidth]{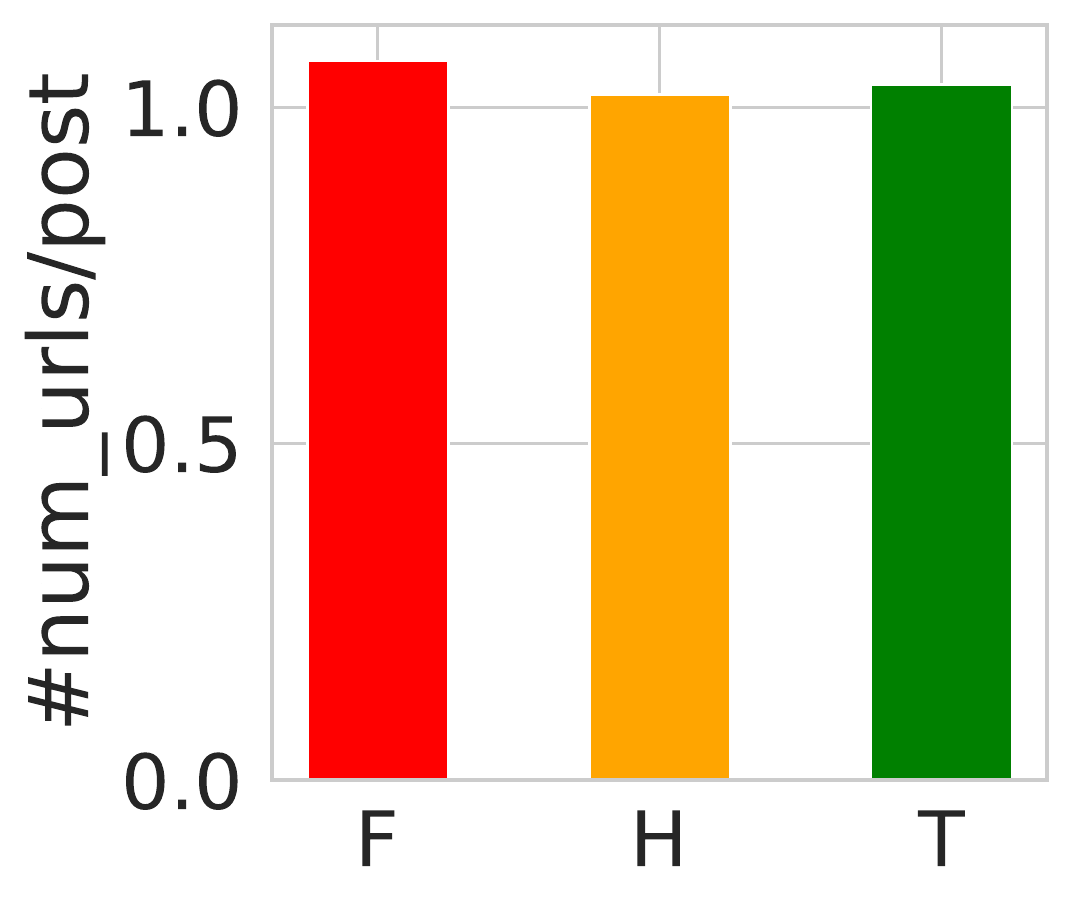}
         \caption{mean web links / posts}
         \label{fig:mean_urls}
     \end{subfigure}
     
        \caption{Average textual properties of fear speech(F), hate speech(H) compared to overall posts(T).}
        \label{fig:textual_properties}
\end{figure}

\subsubsection*{Hashtags} To explore the hashtags further, we explore how fast they flow between different types of speech, i.e., hate speech to fear speech. Essentially, we note in what kind of speech it originates in~\footnote{A hashtag originates in normal speech if a normal labeled person is using it.}, and the time it takes to transfer to some other kind of speech. The figure \ref{fig:hashtag_shift} shows the overall statistics for each type of transfer. Considering the hashtag transfer between fear and hate speech, we find that the median time required for a hashtag in hate speech to transfer to a hashtag in fear speech ($\sim$ 73 days) is less than the vice-versa, i.e., $\sim$ 88 days (MW U, one-sided, $p<1e^{-6}$). Secondly, the transfer from fear speech to normal ($\sim$ 107) takes slightly more than hate speech to normal($\sim$ 121, MW U, one-sided, $p<1e^{-4}$). One surprising thing is how fast hashtags in normal gets transferred to fear speech ($\sim$ 83 days); this is significantly less compared to hashtags transferred from normal to hate speech ($\sim$ 124 days) (MW U, one-sided, $p<1e^{-6}$). This suggests that users writing fear speech are utilising the 

\begin{figure}[ht]
    \centering
    \includegraphics[width=\linewidth]{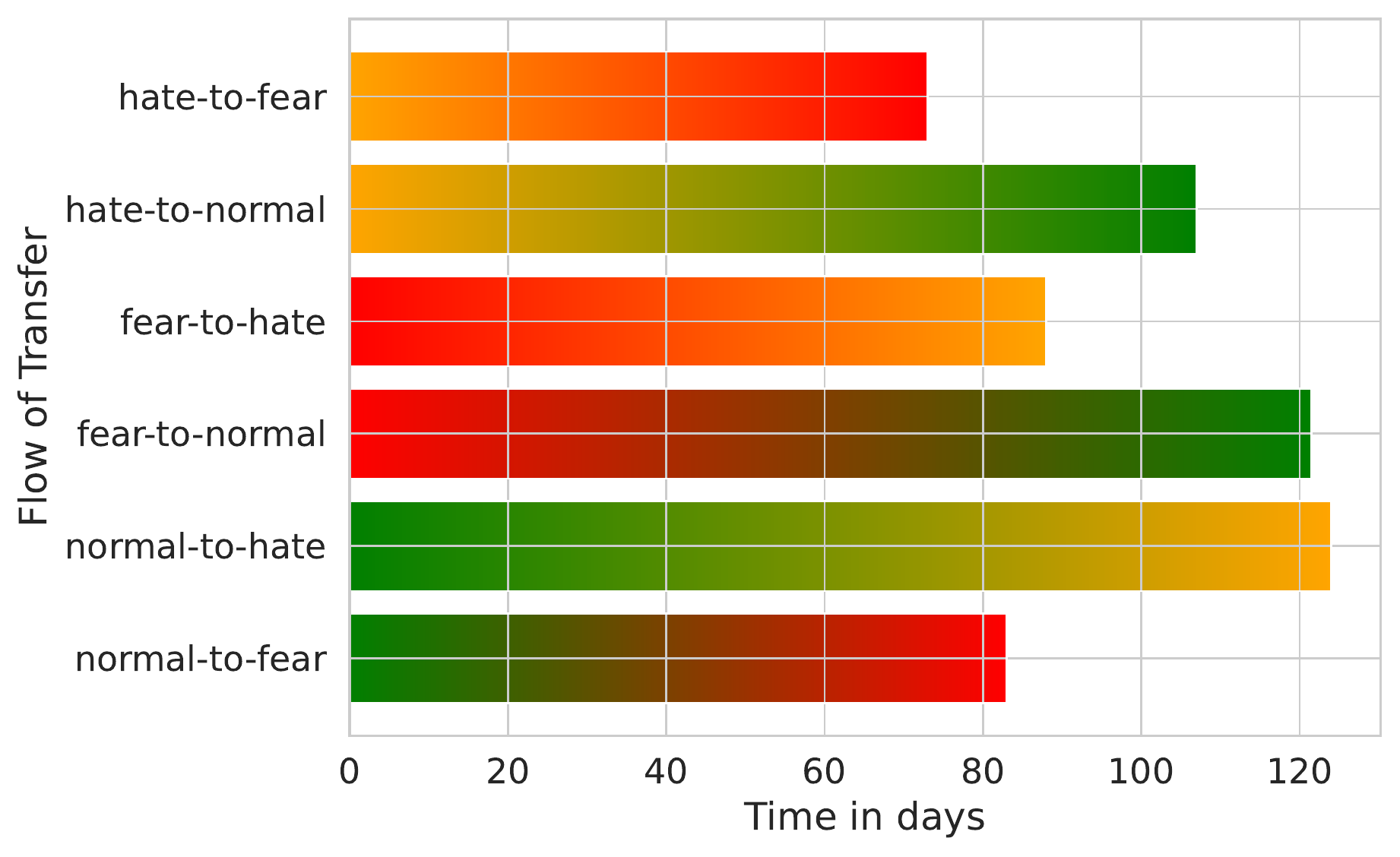}
    \caption{Median time required to transfer the hashtags from one type of speech to another. The name A-to-B refers to the hashtag first originating in A type of speech and transfers to B type of speech.}
    \label{fig:hashtag_shift}
\end{figure}

\subsubsection*{Website links} To explore the website links (URLs) further, we try to analyze URLs in fear speech posts and hate speech posts. In both these cases, we found $\sim$ 6000 unique websites. We further extract and manually analyze the top 20 domains based on their sightings. Few of the relevant website domains from both categories are noted in the Table \ref{tab:urls}. For the popular domains in fear speech category, we find that domains like `islamexposedblog.blogspot'~\footnote{https://islamexposedblog.blogspot.com/}, `thereligionofpeace'~\footnote{https://www.thereligionofpeace.com/} and `counterjihad'~\footnote{\ps{give link}} contained several unconfirmed blogs about atrocities by Muslim community. Few domains were right biased media and having low credibility like `American Center for Law and Justice'~\footnote{\url{http://www.aclj.org} as accessed on Mar 7, 2022} and Sputnik news~\footnote{\ps{add}}. Another website shows the atrocities on the white community - whitenationnetwork~\footnote{https://tinyurl.com/567n8rat}. This website is currently shut down. Other forms of conspiracy theories also showed up on these platforms, like ``coronavirus is a hoax''. 

Popular domains in hate speech posts are slightly different in nature. We find pagesix \footnote{\url{https://pagesix.com/} accessed on March 10, 2022}, an entertainment news website and towlerroad \footnote{\url{https://www.towleroad.com/}} an entertainment website for Gay and LGBTQ+ community. Both these websites are benign in nature, but the hate speech posts were using them by insulting the celebrities in these platforms. Apart from that we also find dailystromer~\footnote{\url{https://en.wikipedia.org/wiki/The_Daily_Stormer}}, godhatefags\footnote{\url{https://en.wikipedia.org/wiki/Westboro_Baptist_Church}} which are popular far-right websites.

\begin{table}[!htpb]
\centering
\begin{tabular}{ll}
\textbf{Fearspeech}             & \textbf{Hatespeech} \\\hline
aclg(243)                      &  pagesix(65)          \\
whitenationnetwork(54)          & towleroad(68)       \\
islamexposedblog(72)            & dailystormer(63)  \\
thereligionofpeace(40)         & weaselzippers(45)  \\
sputniknews(37)                & godhatesfags(28) \\  
counterjihad(33)               & thesmokinggun(20)      \\\hline
\end{tabular}
\caption{Few of the relevant urls (number of posts) from the top 20 urls in fear speech and hate speech posts}
\label{tab:urls}
\end{table}

\subsubsection*{Empath analysis}
To further characterize the text, we use Empath~\cite{fast2016empath}, a tool to perform large-scale textual analysis. We select a set of 15 categories that might show the difference between hate and fear speech posts~\cite{buyse2014words} and report the normalized values along with their significance in table ~\ref{fig:empath}. We find that categories like `aggression', `kill' , `horror', `violence', `crime' and `war' were more significant for fear speech posts. Hate speech scores are higher for categories such as `negative emotion’, `swearing words’ and `hate’~\kg{gvr kiran: we need to explain what this means. not just a description. why would kill, violence be significant for fear?}. 

\begin{figure}[!htpb]
    \centering
    \includegraphics[width=0.8\linewidth]{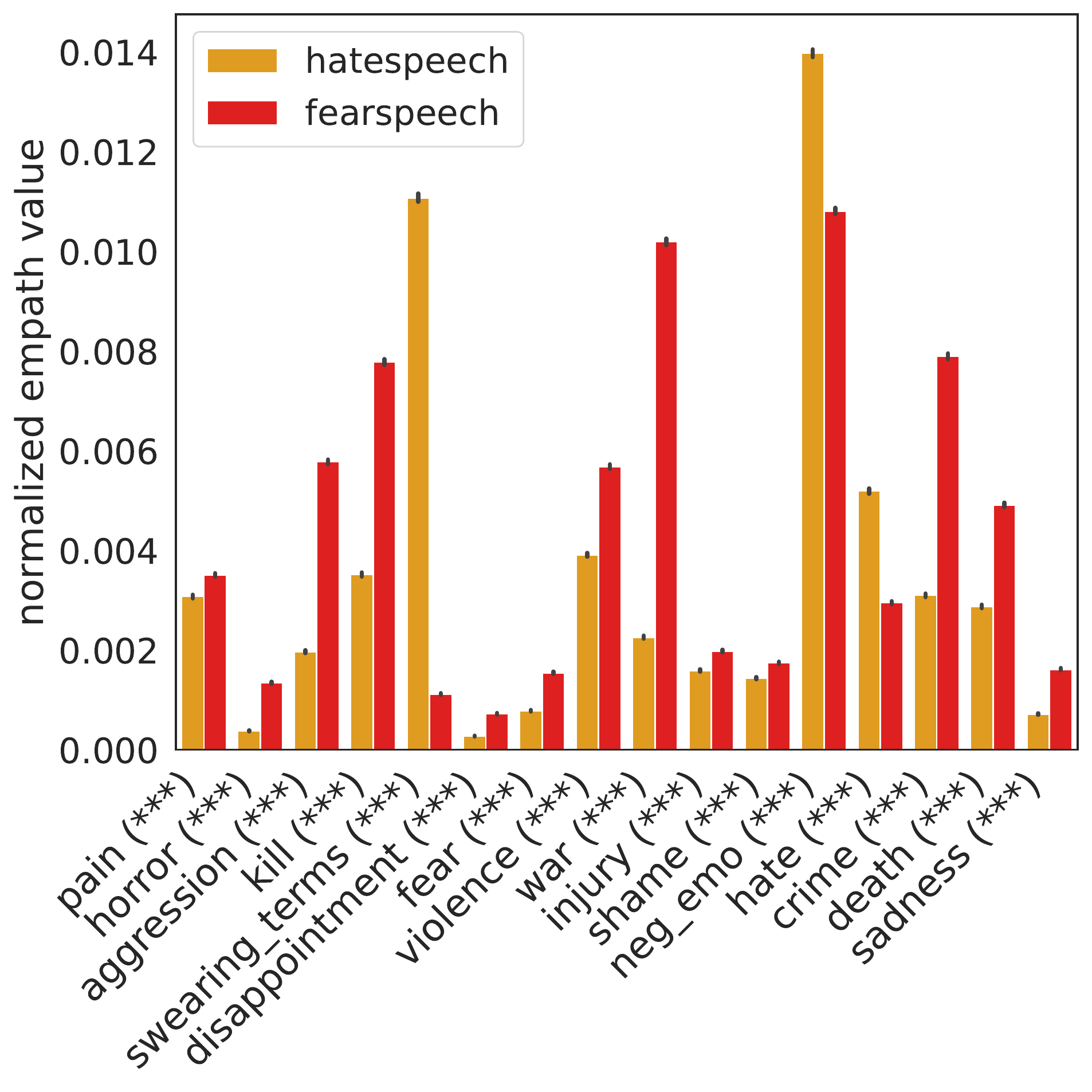}
    \caption{We report the mean values for 15 chosen categories of Empath for fear speech and hate speech posts.}
    \label{fig:empath}
\end{figure}

\subsubsection*{Topic modelling} We use LDA modelling~\cite{hoffman2010online} to model the topics in the fear and hate speech posts (More details in the \appendixnewname~section~\ref{sec:topic_modelling}). Then for each month, we plot the top 10 topics and their normalized distribution considering the total posts in that month. Overall we notice one broad level differences in fear speech topics (refer figure~\ref{fig:topic_dist_fear}) and hate speech topics (refer figure~\ref{fig:topic_dist_hate}). Topics in the fear speech mostly portrayed other communities as perpetrators, while topics were dehumanizing or insulting the target communities. 

Considering the fear speech topics, we find that topics like `America needs to wake up' and `Ideology of Islam is dangerous' have a more prominent presence across all the months. Here the topic `America needs to wake up' makes implicit calls to Americans to see the atrocities by other communities. The topic `violence by Muslim communities' notes the various unconfirmed violent activities by the Muslim communities. It had a tiny percentage initially (Oct and Nov 2016) but increased to a significant ratio afterward. On the other hand, the topic -``immigrants manipulating elections'' was prominent during the initial time periods but died out after April 2017. Another interesting topic was `jews controlling media' - which points out how Jews control media platforms. Apart from that, illegal immigration as a problem was shown in topics like ``illegal immigration in Europe, `illegal immigration in the USA' etc. 

Among the hate speech topics, three of the most consistent topics are `multi-target insults' --- where a single hate post targeted multiple communities, `women being projected as prostitutes', and hate against voters from different demography. Other topics like insults about Muslims and Canadians occur rarely and have smaller ratios. Insults for the jewish community rose after August 2017. This might be an effect of the influx of users during that time period. The topic, which has post targeting both homosexuals and Muslims reduced after March 2017, might have merged with the multi-target insults. The topics `support for Nazi', `insulting and blaming Africans' all increased significantly in terms of ratio after August 2017.

\begin{figure}
    \centering
         \includegraphics[width=\linewidth]{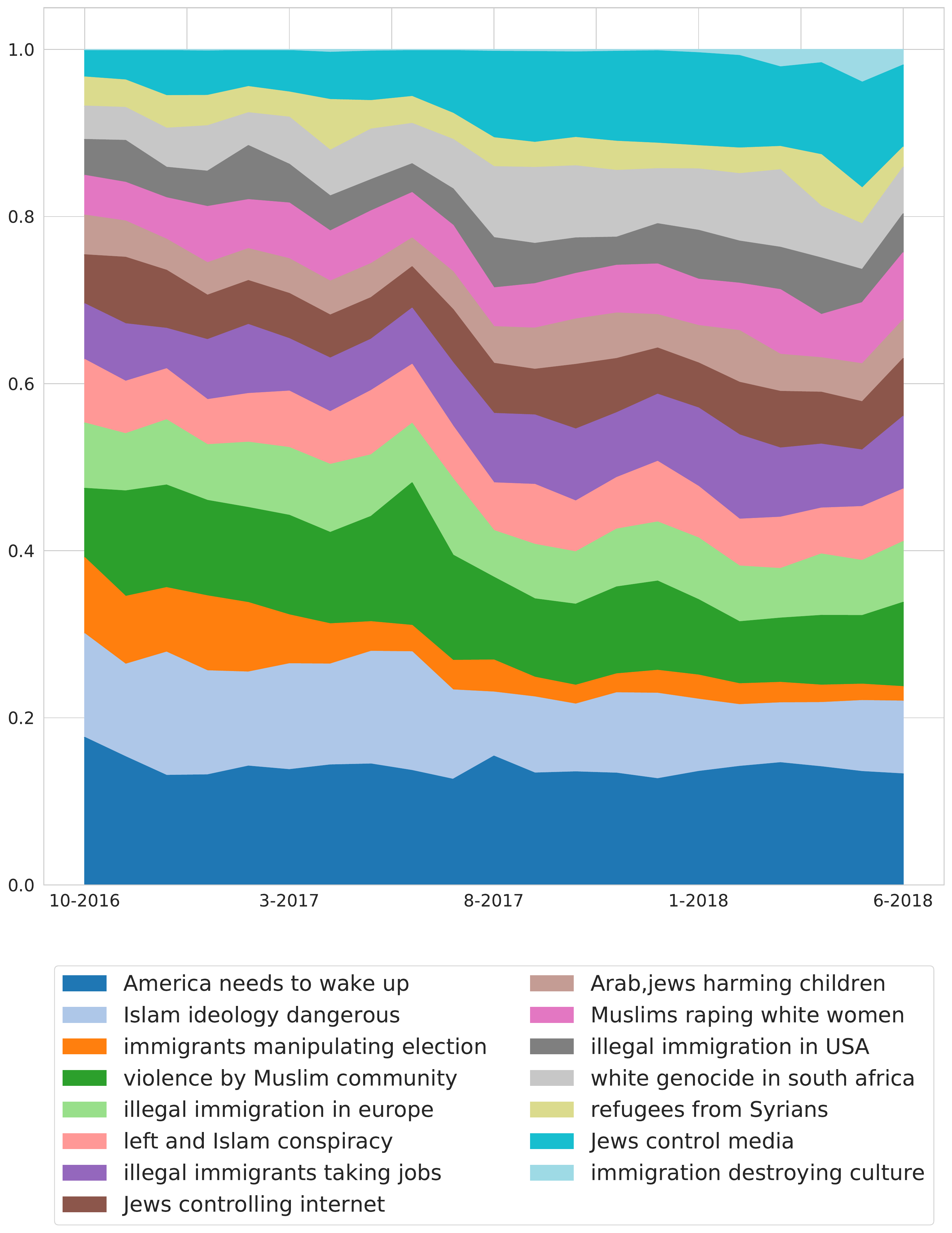}
    \caption{Top 10 topics and their normalised distribution per month for fear speech posts~\kg{problems with the figure}}
    \label{fig:topic_dist_fear}
\end{figure}

\begin{figure}
    \centering
         \includegraphics[width=\linewidth]{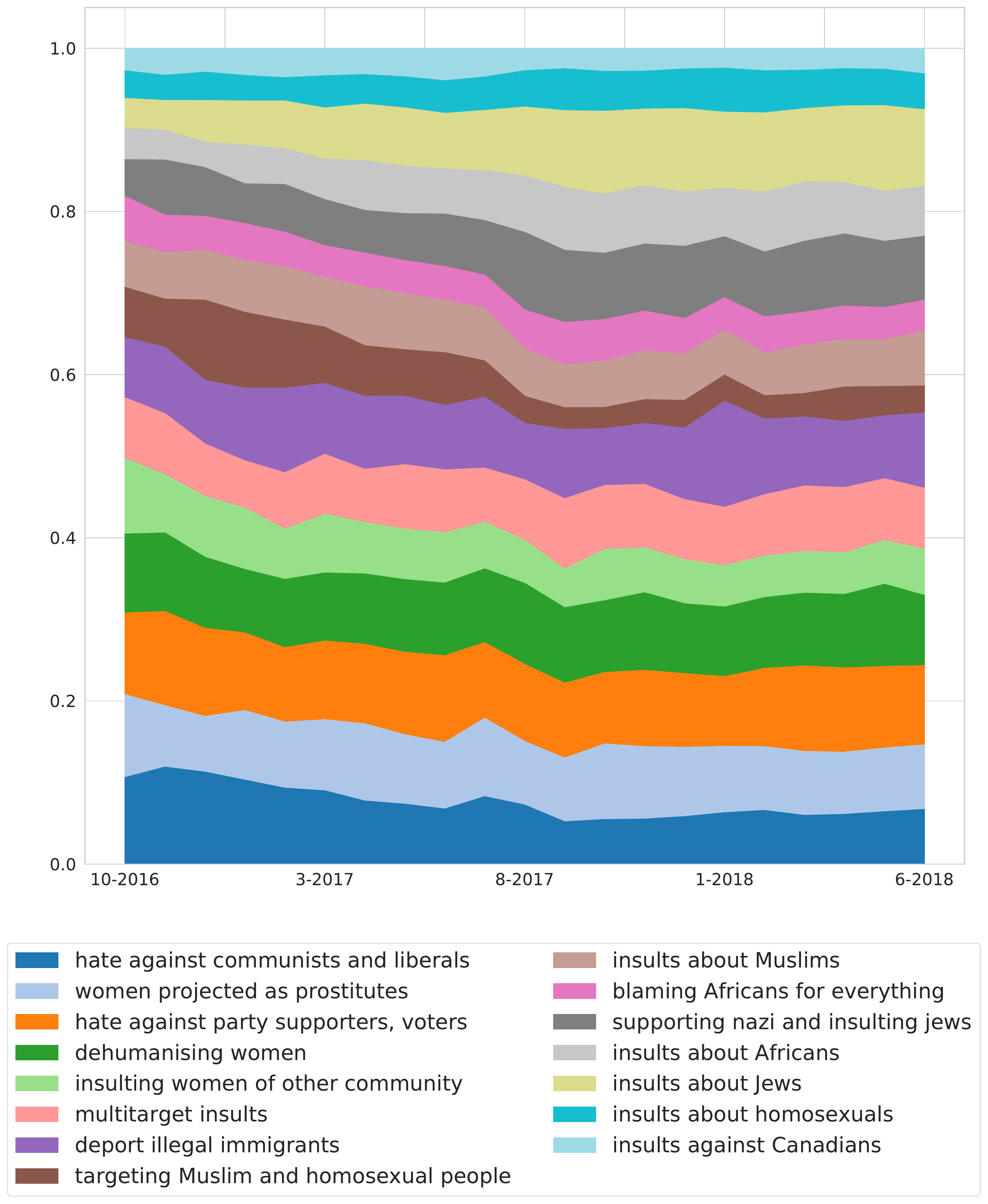}
    \caption{Top 10 topics and their normalised distribution per month for hate speech posts}
    \label{fig:topic_dist_hate}
\end{figure}

\fi

\bibliography{pnas-sample}

\appendix
\setcounter{table}{0}
\renewcommand{\thetable}{S\arabic{table}}
\setcounter{figure}{0}
\renewcommand{\thefigure}{S\arabic{figure}}
\setcounter{section}{0}
\renewcommand{\thesection}{S\arabic{section}}

\section{Statistical tools}
\label{sec:stats}
\subsection{Mann-Whitney U (MW U) test}
Mann-Whitney U test is a non-parametric test which is an alternative to the independent sample $t$-test. It compares two sample means that come from the same population, and is used to test whether the two sample means are equal or not. Usually, the Mann-Whitney U test is used when the data is ordinal or when the assumptions of the $t$-test are not met. For our purpose we use a python package~\cite{shier2004statistics} with two-sided hypothesis and `asymptomatic' method.

\subsection{Matching users}
\label{sec:matching}
To understand if the distinction between fear vs. hate is the sole reason for the results rather than some other confounding factor, we perform matching between ExFear and ExHate users using propensity score matching. Propensity score based matching (PSM)~\cite{10.1145/3178876.3186162} is a quasi-experimental method in which we use statistical techniques to construct an artificial group by matching treated unit with a non treated unit of similar characteristics. For this, we use a python library \footnote{\url{https://github.com/benmiroglio/pymatch}} to obtain the propensity score-based matching. To this purpose, we use the average number of messages per month, the standard deviation of the number of messages per month, and the number of months they were active (how many months they had at least one post). Next we calculate the propensity scores using logistic regression. Using these scores the users were matched using 0.0005 as threshold between the matched propensity scores. After matching, we measure significance with M-W U test between the propensity scores of the treatment and the matched control set and find the differences in the propensity of the two sets to be non significant at $p=0.499$. We find 476 matched users out of 479 ExHate and 483 ExFear users. Since the fraction of matched users is close to 100\%, our results remain unaffected by whether we choose only the matched users or all the users in both the sets.

\section{Additional details for annotation}
\label{sec:anno-guide}
\noindent\textbf{Hate speech:} is a language used to express hatred toward a targeted individual or group, or is intended to be derogatory, to humiliate, or to insult the members of the group, on the basis of attributes such as race, religion, ethnic origin, sexual orientation, disability, or gender~\cite{mathew2020hatexplain}.
A post is a hate speech if one or more of the following are true --
\begin{itemize}
  \item it is targeted against a person or group of persons,
  \item it uses derogatory or racial slur words within the post
  \item it makes use of disparaging terms with the intent to harm or incite harm,
  \item it refers to and supports other hateful facts, hate posts and organization,
  \item it refers to the other group as inferior as cultural superiority,
  \item it makes use of idiomatic, metaphorical, collocation or any other indirect means of expressions that are harmful or may incite harm,
  \item it expresses violent communications.
\end{itemize}

\noindent\textbf{Fear speech} is an expression aimed at instilling (existential) fear of a target group on the basis of attributes such as race, religion, ethnic origin, sexual orientation, disability, or gender~\cite{buyse2014words}.

A post is a fear speech if it creates fear about a target group using one of the following notions.
\begin{itemize}
  \item Something done by the target group in the past (and the possibility of that happening again) - \textbf{historical domination}.
  \item Some tradition of the target group which is shown to take precedence over in-groups - \textbf{cultural domination}.
  \item The target group taking over jobs or education institutes - \textbf{economic domination}.
  \item The target group taking over land/ living places - \textbf{geographic domination}.   
  \item The target group killing people of the in-group - \textbf{existential domination}.
  \item Speculation that the target group would take over and dominate in the future over the in group - \textbf{future domination}.
\end{itemize}

\noindent\textbf{Normal speech} will be the label of the posts which are neither hate speech nor fear speech. Note that this class is not equivalent to unbiased or polite language. There may be political propaganda, use of slur words but since our aim is to identify fear and hate speech we should annotate them as normal class.

\noindent In addition to the above definition, we further add specific guidelines so that the annotators avoid some common mistakes. These are noted below. 

\begin{itemize}
    \item The presence of certain words does not justify a text being hateful or fearful. The annotators should look into the context of the post.
    \item Since some of the fear speech heavily used common news sources, we asked the annotators to look into the post if it is a news headline or if it is trying to create or spread fear.
    \item A post might be both partially hateful and partially fearful. The annotator should carefully read the whole post before finalising the labels.
    \item Fear and hate speech both should be about some target community (explicit or implicit). Fear/hate speech about media or government institutions are not considered in this task. Such cases should be marked as normal.
    \item Finally, the annotators should mark the targets present in the posts.
\end{itemize}

Some of the annotated examples from our dataset are shown in Table~\ref{table:samples}.

\begin{table}[!htbp]
\centering
\caption{This table notes the annotation for some of the sample posts}

\begin{tabular}{|p{4.8cm}|>{\centering\arraybackslash}p{1cm}|>{\centering\arraybackslash}p{1.2cm}|}
\hline
\multicolumn{1}{|c|}{{ \textbf{Example}}} &
  { \textbf{Class}} &
  { \textbf{Targets}} \\ \hline
{I call bullshit on Google.How does a search for "white parents with white children" return these images?} &
  { Normal} &
  { Other} \\ \hline
{ Thousands of Americans have been killed by illegal aliens. Illegals break up American families forever. Go home. Stay home. Stop breeding out of control.} &
  { Hate Speech, Fear Speech} &
  { Refugee} \\ \hline
{ Toledo Tree-Dweller Guns Down White Man After an Argument at a Bar. Blacks routinely use homicide as their go-to for conflict resolution.} &
  { Fear speech} &
  { African} \\ \hline
{ @user Wouldn't you save a dog before you'd save an illegal or a Muslim? Or Hellary? Every sentient being with a central nervous system feels pain. Animals don't build mosques or vote Demonrat. They really ARE better than a lot of people!} &
  { Hate speech} &
  { Islam, Refugee} \\ \hline
{ Today, do something for the environment and kill your local queer} &
  { Hate Speech} &
  { Homosexual} \\ \hline
{ Don't really care if a fighters black, white, yellow or Dalmatian. If he is good he is good.} &
  { Normal} &
  { African, Asian, Caucasian} \\ \hline
{ "I call BS! NO muslim woman would ever accept gays, any other form of religion, free speech (she has no clue what it is after all) or ever adapt yo the way of life of the country her family invaded!"} &
  { Fear speech} &
  { Islam, Women} \\ \hline
\end{tabular}
\label{table:samples}
\end{table}

\section{Models}\label{sec:allmodels}
In this section we present additional details about the classification models we utilised in this paper.

\subsection{Baseline models}
\label{sec:baseline}
Baseline models can be categorised based on what kind of feature extractor and classifier model are being used. We note three different types of features and classifications and report the performance of all of their combinations in Table~\ref{tab:baseline_models}.

\noindent\textit{Feature extraction}: In case of BOW vectors, we create a vocabulary vector of all unique words in the training set and for each post a binary vector is created based on the presence of words (1 if a word is found in the vocabulary else 0). In case of GloVe word embeddings, it takes into consideration local as well as global context of words and we use a word vector pretrained with 840B tokens, 2.2M vocab, cased, 300 dimensional vectors using common crawl. In case of TF-IDF feature extraction method, we use tfidf-vectorizer module from scikit-learn which initializes a vectorizer and uses its fit\_transform method which gives us the product of term-frequency and inverse-document frequency of each word in the text.

\noindent\textit{Classification model}: Logistic regression is a machine learning method used to predict the outcome of a dependent variable based on previous observations. Support vector machines (SVMs) are a set of supervised learning methods used for classification and regression problems and they are effective in high dimensional spaces. XGBoost (extreme gradient boosting) is a library for developing fast and high performance gradient boosting tree models.

\subsection{Transformer model variations}
\label{sec:transformer}
Transformers are a recent architecture in the NLP literature~\cite{devlin2018bert} which utilises self-attention blocks to create contextual embeddings in a self-supervised manner. These embeddings are further used with a classification head in this task. Here, we detail the different forms of transformer models used and the results are noted in Table~\ref{tab:transformer_models}.

\subsubsection{Pretrained models} These models are pretrained on a huge corpus collected from web using various pretrained strategies. With these models, we add a linear layer to classify the posts into fear speech, hate speech and normal class in a multilabel fashion. We note the models below.

\noindent\textit{BERT base} (\texttt{bert-base-uncased}): BERT is a transformer model pretrained in a self-supervised fashion on a large corpus of raw English data including the entire Wikipedia (2,500 million words) and Book Corpus (800 million words), with no human labelling. 

\noindent\textit{RoBERTa base} (\texttt{roberta-base-uncased}): RoBERTa is a transformer model in which the texts are tokenized using a byte version of byte-pair encoding (BPE)~\cite{shibata1999byte} %
and a vocabulary size of 50,000 for training it. Unlike BERT, the masking is done dynamically during pretraining (e.g., it changes at each epoch and is not fixed).

\subsubsection{Fine-tuned models} These models are variants of the pretrained models which are further fine-tuned using a classification head on some related task, i.e., offensive post detection, hate speech detection. Fine-tuning on a related dataset can help the model learn additional semantic features required for the task. We describe the models below.

\noindent\textit{HateXplain model} (\texttt{bert-hatexplain}): This model is used for classifying a text as abusive (hate speech and offensive) or normal. The model is trained using data from Gab and Twitter and human rationales were included as part of the training data to boost the performance~\cite{mathew2020hatexplain}. The model also has a rationale predictor head that can predict the rationales given an abusive sentence.

\noindent\textit{BERT latent hatred} (\texttt{bert-latent-hatred}): While many hate speech datasets are highly explicit and overt, this model is used for understanding implicit hate speech. This model was fine-tuned on implicit hate data~\cite{elsherief2021latent} which contains 22,056 tweets from the most extremist groups in the US where 6,346 of these tweets contain implicit hate speech. We train a \texttt{bert-base-uncased} model to fine-tune on this dataset considering train (8) : val (1) : test (1) stratification. The model reaches a macro F1-score of 0.60 and an accuracy of 0.74 for the test dataset.

\noindent\textit{BERT dynamic hate} (\texttt{bert-dynamic-hate}): Here a \texttt{bert-base-uncased} model is fine-tuned on a dataset~\cite{vidgen2020learning} of nearly 40,000 entries, generated and labelled by trained annotators over four rounds of dynamic data creation. In specific, we train a \texttt{bert-base-uncased} model to fine-tune on this dataset considering train (8) : val (1) : test (1) stratification. The model reaches a macro F1-score of 0.80 and accuracy of 0.80 for the test dataset.

\noindent \textit{BERT Gab hate} (\texttt{bert-gab-hate}):  The model used is \texttt{bert-base-uncased} fine-tuned on Gab hate speech corpus~\cite{kennedy2022introducing}, consisting of 27,665 posts from the social network service gab.ai, each annotated by a minimum of three trained annotators. Once again we train a \texttt{bert-base-uncased} model to fine-tune on this dataset considering train (8) : validatopm (1) : test (1) stratification. The model reaches a macro F1-score of 0.32 and accuracy of 0.94 for the test dataset. The lower performance is due to the huge imbalance in the dataset.

\noindent\textit{Twitter base RoBERTa hate} (\texttt{twi-rob-hate}): This is a RoBERTa-base model trained on \~58M tweets and fine-tuned for hate speech detection using 2 labels - hate/non hate with the TweetEval benchmark~\cite{barbieri2020tweeteval}. For hate speech detection task, a popular dataset~\cite{basile2019semeval} was used with 9000, 1000, 2970 as train, validation and test dataset. The model achieves a macro-F1 score of 0.46 with the RoBERTa-base model and a score of 0.52 with the RoBERTa-base model (retrained on Twitter) on the test dataset.%

\noindent\textit{Twitter base RoBERTa offensive} (\texttt{twi-rob-offensive}): This is also a RoBERTa-base model trained on \~58M tweets and fine-tuned for offensive language identification with the TweetEval benchmark~\cite{barbieri2020tweeteval}. For hate speech detection task, a popular dataset~\cite{basile2019semeval} was used with  11,916, 1,324 and 860 as train, validation and test dataset. It achieves a macro F1-score of 0.78 with the RoBERTa-base model and a score of 0.81 with the RoBERTa-base model (retrained on Twitter) on the test dataset.%

\subsubsection{MLM pretrained models} These models are essentially pretrained models which are further pretrained using the masked language modelling framework (MLM)~\cite{taylor1953cloze}.%
In this framework, we randomly mask 15\% of the words in the input, the model has to then predict the masked words. We describe the two models used here.

\noindent\textit{HateBERT} (\texttt{hatebert}): HateBERT~\cite{caselli-etal-2021-hatebert} is an English pretrained BERT model obtained by further training the English \texttt{bert-base-uncased} model with more than 1 million posts from banned communities from Reddit considering train 1,478,348 and 149,274  as train and test data points respectively. 

\begin{figure}
    \centering
    \includegraphics[width=\linewidth]{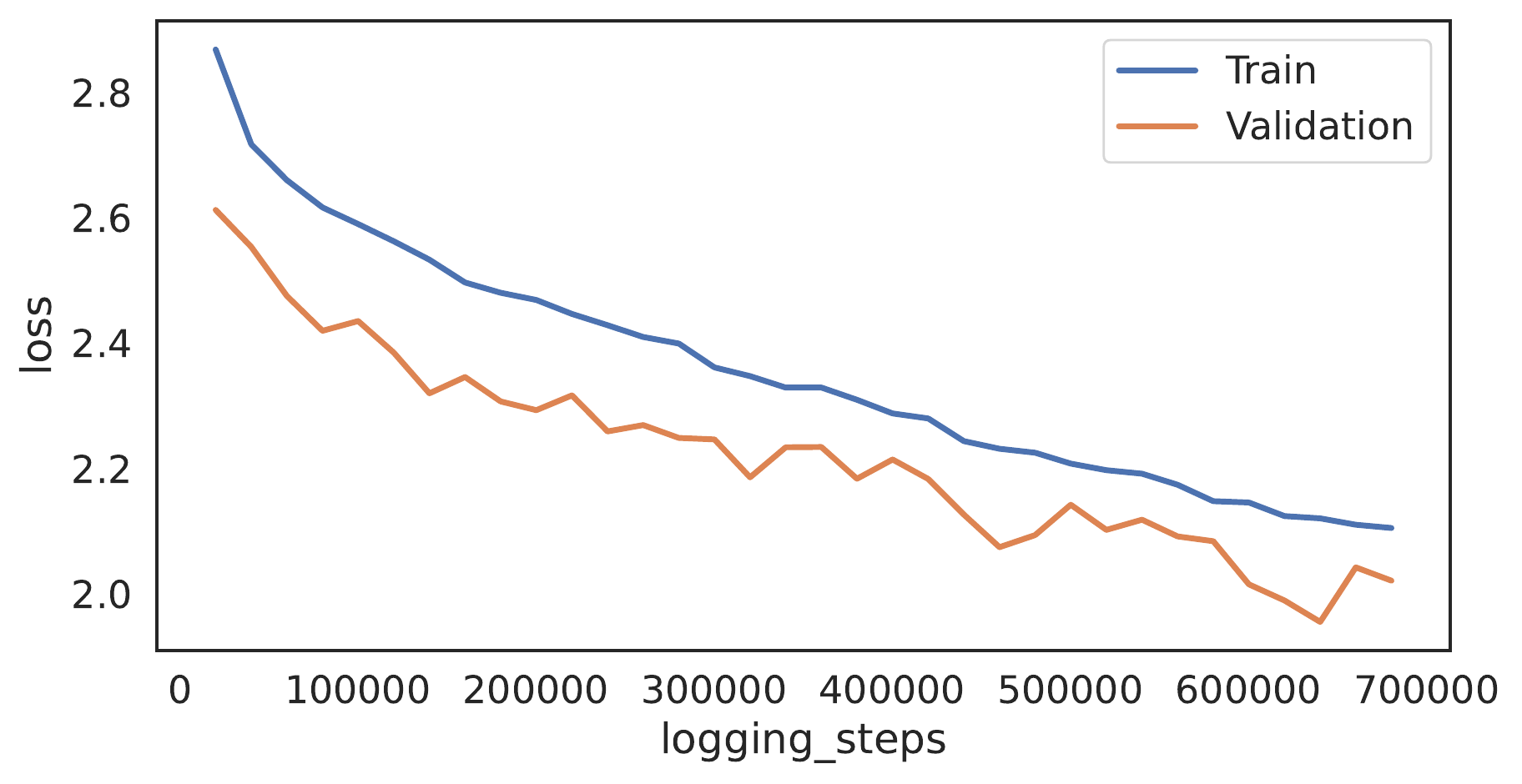}
    \caption{Cross entropy loss for train and validation set.}
    \label{fig:loss}
\end{figure}

\begin{figure}
    \centering
    \includegraphics[width=\linewidth]{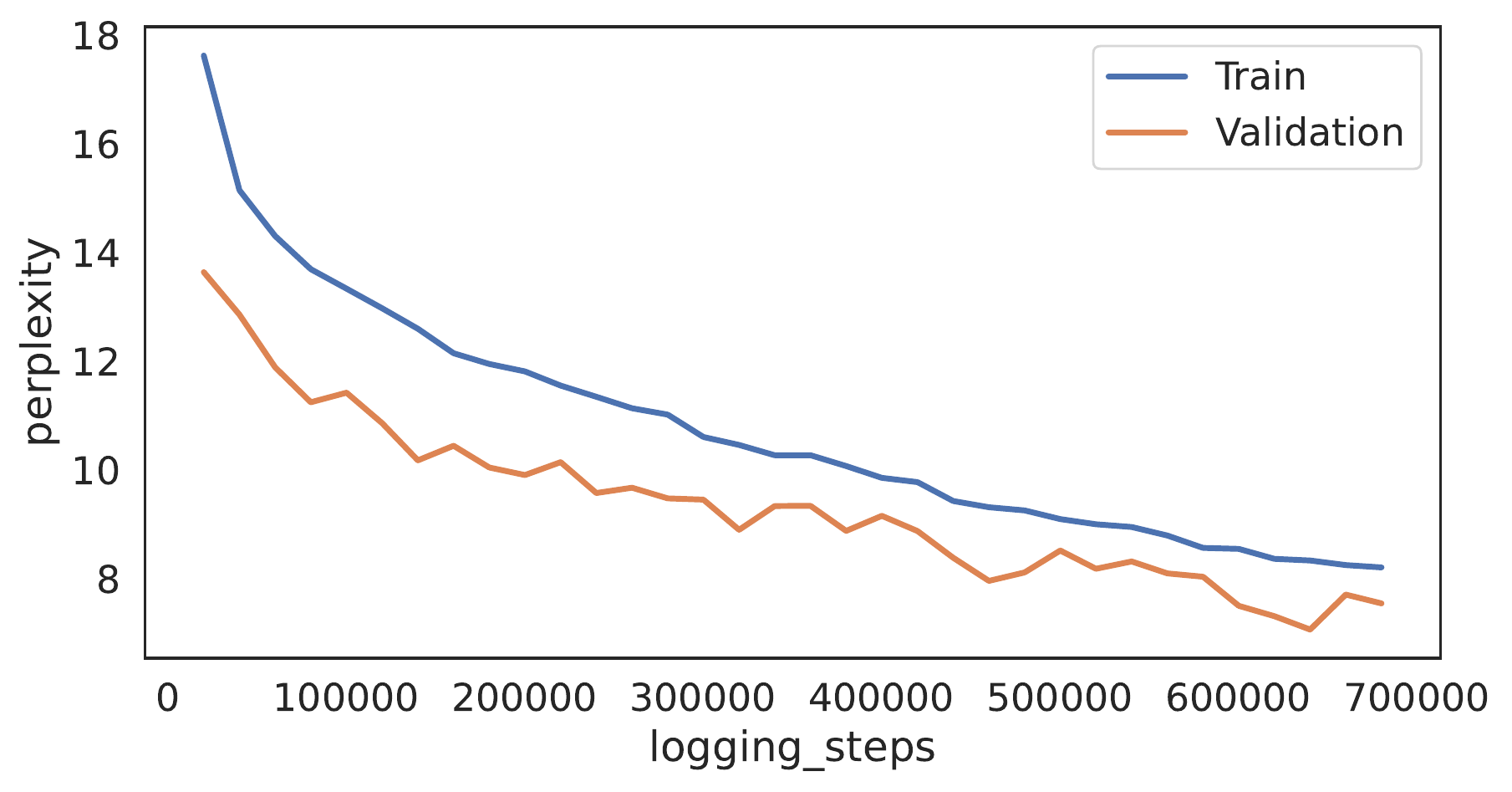}
    \caption{Perplexity for train and validation set.}
    \label{fig:perplexity}
\end{figure}

\noindent\textit{Gab-BERT} (\texttt{gab\_bert}): We further collect posts from the Gab data using a filtered set of keywords~\footnote{\url{https://www.dropbox.com/s/gjhs4s6bfa9l4i0/slur_keywords.json}}. Our total dataset for pretraining contains 1,393,504 posts; we pretrain the \texttt{bert-base-uncased} model further on this dataset considering 1,392,504 and 1,000 as train and test data points respectively. We use block size of 128, batch size of 10 and 5 as the number of epochs for the continued pretraining. We note the loss and perplexity curve in the Figures~\ref{fig:loss} and~\ref{fig:perplexity}. Here, we observe that the final perplexity is around 7.54 and we get a final loss of 2.02 on the validation set. %

After pretraining with this dataset we add a classification head to train it on the labeled dataset. We further calculate an emotion vector using the model trained with an emotion classification data~\cite{demszky-etal-2020-goemotions}. We consider 28 different emotions each representing one unit of the emotion vector. Later this emotion vector is appended to the end of the pooled output for the forward direction and the training continues similar to a generic transformer model. This model is referred to as \texttt{gab\_bert+emotion} in the Table~\ref{tab:transformer_models}.

\subsection{Performance comparison across model} We use all the models from former sections and train them to classify the posts into fear speech, hate speech and normal. We split our dataset of $\sim$ 10000 points into train, validation and test in the ratio of 8:1:1. We report results on the test dataset as shown in table~\ref{tab:baseline_models} and~\ref{tab:transformer_models}. We find that among the baseline models in table~\ref{tab:baseline_models}, the highest macro F1 score (0.56) is achieved by the XGBoost and BOW vectors. The highest accuracy (0.42) is achieved by the SVM with TFIDF vectors.

Among the transformers models as shown in ~\ref{tab:transformer_models}, the highest accuracy (0.56) is achieved by \texttt{twi-rob-hate}. In terms of macro F1 score, \texttt{gab\_bert+emotion} achieves the highest F1 score. Interestingly, the same model is either highest or second highest across all the different metrics used. Hence, we use this 
model for the large scale dataset extraction.

\begin{table}[!htbp]
\centering
\caption{\footnotesize{This table shows the evaluation of the baseline models. The first column shows the machine learning models and second columns shows the features used. The rest of the columns represents Accuracy (Acc), Macro F1 score (F1), Precision (Pre), Recall (Rec) and Hamming score (Ham) }
\label{tab:baseline_models}}
\begin{tabular}{|c|c|c|c|c|c|c|}
\hline
\textbf{Model}   & \textbf{Features}                      & \textbf{Acc} & \textbf{F1} & \textbf{Pre} & \textbf{Rec} & \textbf{Ham}(↓) \\ \hline
LR      & \multirow{3}{*}{BOW} & \underline{0.40}     & 0.51     & \textbf{0.52}      & 0.50   & \textbf{0.32}       \\ \cline{1-1} \cline{3-7} 
SVM     &                               & 0.38     & 0.51     & \underline{0.50}      & 0.53   & \underline{0.34}       \\ \cline{1-1} \cline{3-7} 
{XGBoost} &                               & 0.10   & \textbf{0.56}     & 0.42      & \textbf{0.82}   & 0.43       \\ \hline
LR      & \multirow{3}{*}{TFIDF}        & 0.39     & 0.53     & \underline{0.50}      & 0.57   & \underline{0.34}       \\ \cline{1-1} \cline{3-7} 
SVM     &                               & \textbf{0.42}     & 0.47     & \textbf{0.52}      & 0.51   & \textbf{0.32}       \\ \cline{1-1} \cline{3-7} 
XGBoost &                               & 0.13     & 0.54     & 0.42      & \underline{0.76}   & 0.42       \\ \hline
LR & \multirow{3}{*}{\begin{tabular}[c]{@{}c@{}}WE\\ (google)\end{tabular}} & 0.36 & 0.54 & 0.49 & 0.64 & 0.36 \\ \cline{1-1} \cline{3-7} 
{SVM}     &                               & 0.37     & \underline{0.55}    & \underline{0.50}      & 0.64   & 0.35       \\ \cline{1-1} \cline{3-7} 
XGBoost &                               & 0.35     & 0.48     & 0.47      & 0.51   & \underline{0.34}       \\ \hline
\end{tabular}
\end{table}

\begin{table}[!htbp]
\centering
\caption{\footnotesize{This table shows the evaluation of the transformers models. The first column shows the different variation of transformers models. Here, the input is text, except the last row where we also pass a emotion vector along with the text. The rest of the columns represents Accuracy (Acc), Macro F1 score (F1), Precision (Pre), Recall (Rec) and Hamming score (Ham)}}
\label{tab:transformer_models}
\begin{tabular}{|c|c|c|c|c|c|}
\hline
\textbf{Model}                  & \textbf{Acc}  & \textbf{F1}   & \textbf{Pre} & \textbf{Rec} & \textbf{Ham}(↓) \\ \hline
\texttt{bert-base-uncased}    & 0.51 & 0.60 & 0.61 & 0.59 & 0.27   \\ 
\texttt{roberta-base-uncased}          & 0.51 & 0.60 & 0.60 & \underline{0.62} & 0.28   \\\hline
\texttt{twi-rob-offensive}      & 0.54 & 0.61 & 0.60 & \textbf{0.63} & 0.27   \\ 
\texttt{twi-rob-hate}           & \textbf{0.56} & 0.61 & 0.61 & 0.61 & \underline{0.26}  \\ 
\texttt{bert-gab-hate}          & 0.51 & 0.61 & \underline{0.63} & 0.58 & \underline{0.26} \\ 
\texttt{bert-hatexplain}        & 0.52 & 0.58 & 0.59 & 0.58 & 0.28   \\
\texttt{bert-dynamic-hate}      & 0.54 & 0.59 & 0.60 & 0.60 & 0.27   \\ 
\texttt{bert-latent-hatred}      & 0.51 & 0.59 & 0.61 & 0.57 & 0.27   \\ \hline
\texttt{hatebert}             & 0.47 & 0.57 & 0.59 & 0.56 & 0.29   \\ 
\texttt{gab\_bert}              & 0.51 & \underline{0.62} & \textbf{0.65} & 0.60 & 0.25   \\ 
\texttt{gab\_bert + emotions} & \underline{0.55} & \textbf{0.63} & \textbf{0.65} & \underline{0.62} & \textbf{0.25}   \\ \hline
\end{tabular}
\end{table}

\section{Large scale user analysis}
\label{sec:user-analysis}
To perform user analysis at a larger scale, we first select the users posting at least 10 fear/hate speech posts. This was considered to remove the confounding users from the dataset. Consequently, we have around 9,200 users. Next, we study the correlation of the number of fear speech posts, number of hate speech posts, and total posts (as a baseline) with several variables. For each variable, we build a generalized linear regression model, which is shown in equation \ref{eq:1}. When the dependent variables are counting variables, we use negative binomial regression~\cite{hilbe2011negative}. Otherwise, the default setting is considered~\cite{gill2019generalized}. We note the $\alpha,\beta$ and  $\gamma$ for different dependent variables in Table \ref{tab:coeff}.

\begin{equation}
    \label{eq:1}
    var = \alpha*(\# fear speech) + \beta*(\# hate speech) + \gamma * (\# total)
\end{equation}

At first, we inspect the coefficients of the linear regression model as described in the Materials and Methods section. We note the coefficients in the Table~\ref{tab:coeff}. In terms of position in the follower-followee network, we find that number of fear speech posts shows a slight positive correlation with the centrality measures. In contrast, the number of hate speech posts shows a slight negative correlation. The correlation trend is similar for followers and following, although the coefficients are larger. In terms of time to reach the core, we notice a positive correlation with the number of hate speech posts. 

In terms of the number of normal followers, we find the number of fear speech posts correlates positively while the number of hate speech posts correlate negatively. Next, we consider the interaction with the normal followers. We find that the number of fear speech posts positively correlates with the number of normal reposters and total interaction with normal users per user. In contrast, the number of hate speech has a negative correlation. Similar results are found for replies from normal users. We see a negative correlation between normal users mentioned and the number of hate speech as well as fear speech posts. The total number of normal mentions per user is again positively correlated with the number of fear speech posts and negatively correlated with the number of hate speech posts.

\begin{table*}[!htpb]
\centering
\caption{This table notes the coefficient values for number of hate speech, number of fear speech and number of total posts per user for different dependent variables. We also note the number of observations and average value of the dependent variable for the entire dataset.}
\label{tab:coeff}

\begin{tabular}{llllll}\hline
\textbf{Variable} & \textbf{Coeff of \#fear} & \textbf{Coeff of \#hate} & \textbf{Coeff of \#total} & \textbf{Obs} & \textbf{Mean Dep Var} \\\hline
\multicolumn{6}{c}{\textbf{Basic network properties}}                                                  \\\hline
\textbf{Eigenvector}                    & 4.12e-06***       & -1.743e-06***  & 4.319e-07***  & 9245 & 0.004    \\
\textbf{Betweenness}                     & 1.668e-07***      & -2.019e-07*    & 2.442e-08***  & 9245 & 5.27e-05 \\
\textbf{Followers}                       & 0.0012***         & -0.0006***     & 0.0002***     & 9245 & 1012     \\
\textbf{Following}                       & 0.0010***         & -0.0005***     & 0.0002***     & 9245 & 926      \\
\textbf{Time to reach core 0}            & -8.849e-05        & 0.0002**       & -5.117e-05*** & 6700 & 3.52     \\\hline
\multicolumn{6}{c}{\textbf{Effect on normal users}}                                                    \\\hline
\textbf{\# normal followers}         & 0.0009***         & -0.0008***     & 0.0002***     & 9170 & 204      \\
\textbf{unique normal reposters}         & 0.0030***         & -0.0009***     & 0.0003***     & 8772 & 25       \\
\textbf{\# reposts by normal users}   & 0.0036***         & -0.0013***     & 0.0003***     & 8772 & 43       \\
\textbf{unique normal mentions}          & -0.0004***        & -0.0013***     & 0.0003***     & 8756 & 9        \\
\textbf{\# mentions of normal users}  & 0.0022***         & -0.0015***     & 0.0004***     & 8756 & 24       \\
\textbf{unique normal repliers}          & 0.0013***         & -0.0008***     & 0.0003***     & 9233 & 13       \\
\textbf{\# replies of normal users} & 0.0007***         & -0.0008***     & 0.0003***     & 9237 & 19       \\\hline
\end{tabular}
\vspace{-\baselineskip}
\end{table*}

\section{Common users}
\label{sec:common_users}

We perform a separate analysis on the users who were common in the initial ExFear and ExHate set and were removed. We have around 445 users in this set (say ExCommon). We perform the same set of user-level analysis for the ExCommon users as have been done for the other two sets of users. In terms of the position in the network, we find that average \textit{eigen vector} centrality of the ExCommon users are similar to that of the ExFear users; the \textit{betweenness} centrality of the ExCommon users are however larger than that of the ExFear users. Further the ExCommon users are similar to the ExFear users in terms of the average number of followers and percentage of normal followers. ExCommon users have even better interaction with the normal users than the ExFear users, i.e., they get more reposted, mention more normal users and get replied by more normal users. All the results are statistically significant ($p<0.01$ M-W U test).

\section{Topic modelling}
\label{sec:topic_modelling}
Topic modeling is a type of statistical modeling for discovering the abstract `topics' that a collection of documents contain. Latent Dirichlet Allocation (LDA) is an example of topic model and is used to classify text in a document to a particular topic. Topic modelling is carried out separately for hate and fear speech posts. The posts are preprocessed to remove website links, stop words etc. The posts are then tokenized into list of words. A bag of words corpus is created using the dictionary generated by \textsc{gensim} library on the preprocessed posts~\cite{rehurek_lrec}. We use the multicore LDA model from the same repository to create the topics for fear speech and hate speech posts. 

\begin{figure}
    \centering
    \includegraphics[width=0.87\linewidth]{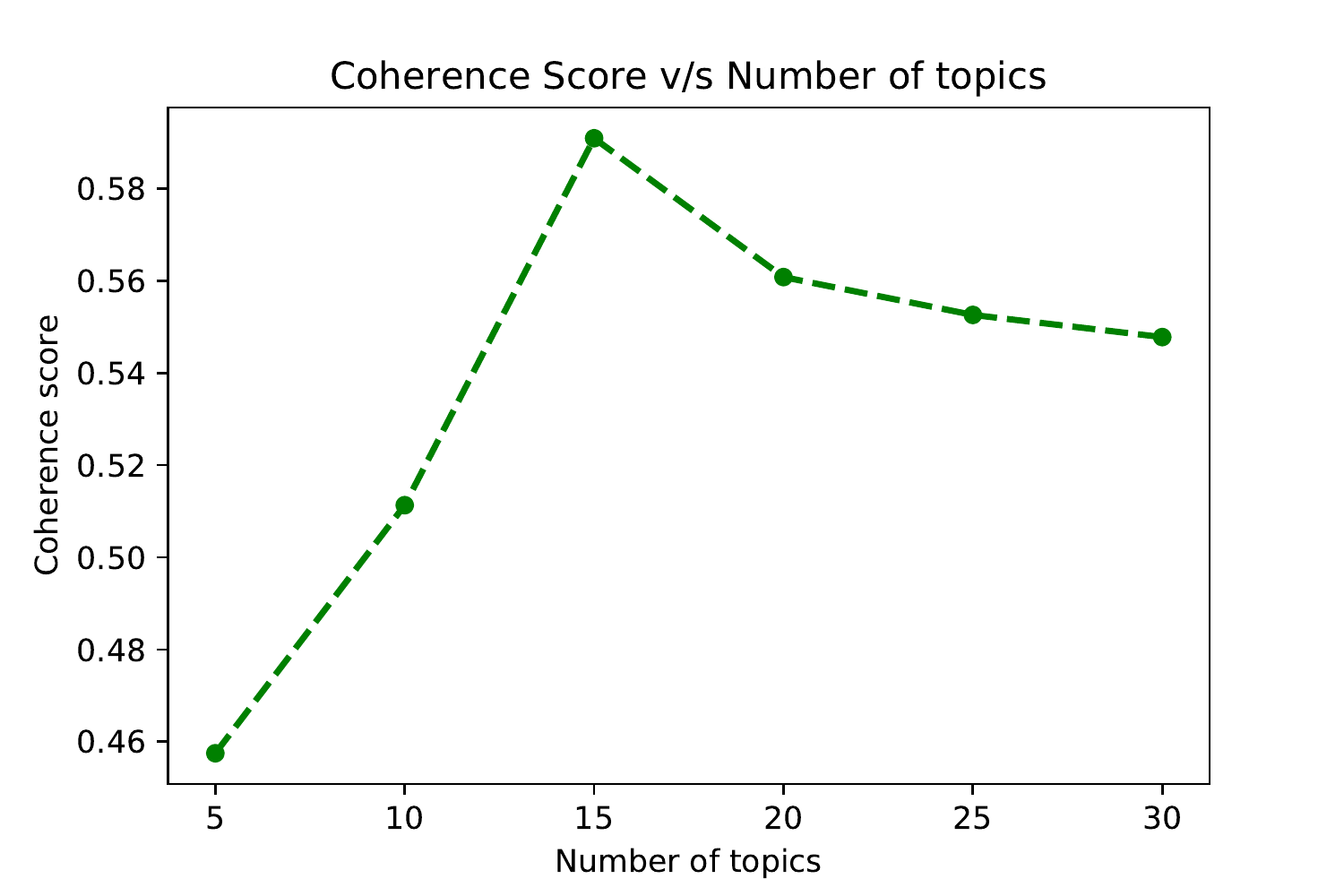}
    \caption{Coherence score vs number of topics for topic modelling in hate speech posts.}
    \label{fig:coh_fear}
    \vspace{-\baselineskip}
\end{figure}

\begin{figure}%
    \centering
    \includegraphics[width=0.87\linewidth]{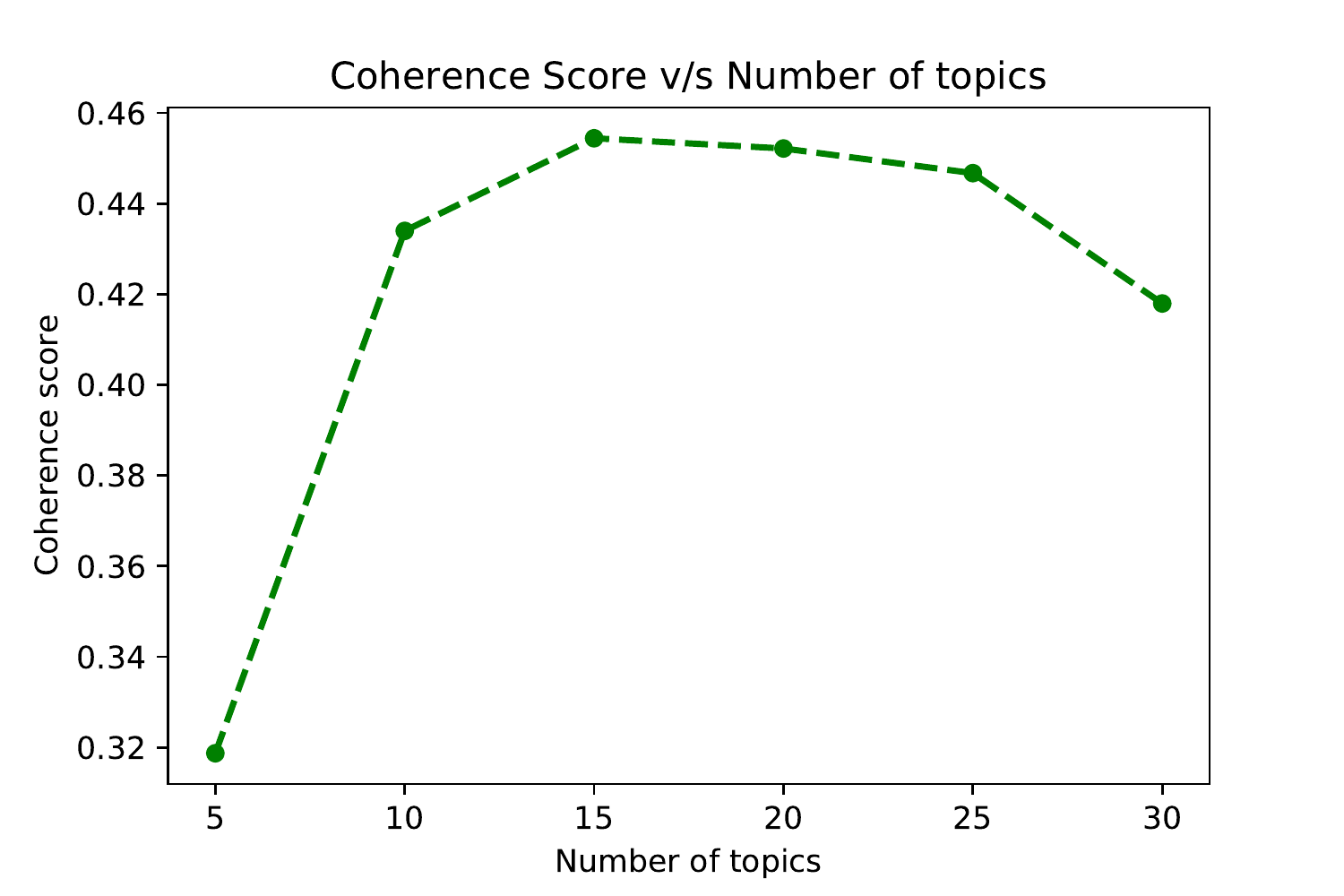}
    \caption{Coherence score vs number of topics for topic modelling in fear speech posts.}
    \label{fig:coh_hate}
    \vspace{-\baselineskip}
\end{figure}

We decide the number of topics based on the coherence scores~\cite{roder2015exploring}. We vary the number of topics ranging from 5 to 30 topics at intervals of 5. For topics of both fear speech and hate speech posts, we find that the highest coherence score is achieved with 15 topics (see  Figures~\ref{fig:coh_fear} and \ref{fig:coh_hate}). For the LDA model with 15 topics, we note the topic representing words and their abbreviation in Table~\ref{tab:topic_names_fear} and \ref{tab:topic_names_hate} respectively. The abbreviations were manually ascribed by seeing topic words associated with a particular topic and 20 random sample posts which had that topic.

\begin{table}[ht!]
    \centering
            \caption{Topic words and the abbreviated topic names for fear speech posts.}
        \label{tab:topic_names_fear}
    \begin{tabular}{p{3cm}|p{5cm}}
            \textbf{Topic name} & \textbf{Topic words} \\\hline
            Jews control media & jews, jewish, israel, people, world\\\hline
            illegal immigration in europe & europe, invasion, immigration, middle, muslim\\\hline
            Arab, Jews harming children & kids, children, school, people, schools\\\hline
            illegal immigration in USA & state, illegal, aliens, california, police\\\hline
            violence by Muslim community & muslim, attack, attacks, police, muslims\\\hline
            left and Islam conspiracy & islam, left, refuse, society, admit\\\hline
            white genocide in South Africa & white, whites, black, genocide, people\\\hline
            Muslims raping white women & rape, women, children, muslim, girls\\\hline
            Islam ideology dangerous & islam, muslims, religion, sharia, people\\\hline
            immigration destroying culture & world, western, population, crime, immigration\\\hline
            refugees from Syrians & israel, syria, isis, terrorist, terrorists\\\hline
            immigrants manipulating election & trump, maga, news, speak, america\\\hline
            America needs to wake up & people, country, going, know, time\\\hline
            Jews controlling internet & evil, global, agenda, world, elite\\\hline
            illegal immigrants taking jobs & people, country, like, illegals, need\\\hline
        \end{tabular}

\end{table}

\begin{table}[ht!]
    \centering
            \caption{Topic words and the abbreviated topic names for hate speech posts. }
        \label{tab:topic_names_hate}
    \begin{tabular}{p{3cm}|p{5cm}}
            \textbf{Topic Name} & \textbf{Topic words} \\\hline
            multitarget insults & f*ggot, c*nt, f*cking, asshole, know\\\hline
            insults about Africans & n*ggers, africa, white, whites, hang\\\hline
            insults about Muslims & sh*t, piece, kill, f*cking, hole\\\hline
            insults about Jews & jews, hate, white, k*kes, race\\\hline
            insults about homosexuals & f*ggot, twitter, moron, nazi, f*cking\\\hline
            dehumanising women & like, good, time, know, look\\\hline
            hate against party supporters, voters & f*ck, b*tch, f*cking, little, dumb\\\hline
            hate against communists and liberals & trump, traitor, president, obama, vote\\\hline
            insults against Canadians & deport, illegal, b*stards, f*ckers, goat\\\hline
            deport illegal immigrants & country, want, send, need, sh*thole\\\hline
            blaming Africans for everything & white, black, people, free, racist\\\hline
            insulting women of other community & scum, disgusting, wh*re, tr*nny, ugly\\\hline
            targeting Muslim and homosexual people & muslim, muslims, islam, maga, news\\\hline
            supporting nazi and insulting jews & k*ke, f*ggots, right, people, hitler\\\hline
            women projected as prostitutes & women, like, d*ck, filthy, mouth\\\hline        
            \end{tabular}

\end{table}

\end{document}